\documentclass[twocolumn]{aastex62}
\usepackage{graphicx}
\usepackage{amssymb}
\usepackage{amsmath}
\usepackage{txfonts}
\usepackage[T1]{fontenc}
\newcommand{\mc}{\multicolumn}

\newcommand       \magni        {\,{\rm mag}}

\newcommand     \gtsim  {\lower.5ex\hbox{$\buildrel > \over \sim$}}
\newcommand     \ltsim  {\lower.5ex\hbox{$\buildrel < \over \sim$}}
\newcommand     \simgt  {\lower.5ex\hbox{$\buildrel > \over \sim$}}
\newcommand     \simlt  {\lower.5ex\hbox{$\buildrel < \over \sim$}}
\newcommand{\PreserveBackslash}[1]{\let\temp=\\#1\let\\=\temp}

\newcommand       \cm           {\,{\rm cm}}

\newcommand       \keV           {\,{\rm keV}}

\newcommand       \NH           {N_{\rm H}}
\newcommand       \rmH          {\,{\rm H}}

\newcommand{\AV}{A_V}

\shortauthors{Shan et al.}
\begin{document}

\title{Distances of the Galactic Supernova Remnants Using Red Clump Stars}


\author[0000-0002-4462-2341]{S.S. Shan}

\affil{National Astronomical Observatories, Chinese Academy of Sciences,\\ 20A
Datun Road, Chaoyang District, Beijing 100012, China\\}
\affil{School of Astronomy, University of Chinese Academy of Sciences,
Beijing 100049, China\\}

\author{H. Zhu}
 \affil{National Astronomical Observatories, Chinese Academy of Sciences,\\ 20A
Datun Road, Chaoyang District, Beijing 100012, China\\}
\affil{Harvard-Smithsonian Center for Astrophysics, 60 Garden Street, Cambridge, MA 02138, USA\\}

\author{W.W. Tian}
\affil{National Astronomical Observatories, Chinese Academy of Sciences,\\ 20A
Datun Road, Chaoyang District, Beijing 100012, China\\}
\affil{School of Astronomy, University of Chinese Academy of Sciences,
Beijing 100049, China\\}

 \author{M.F. Zhang}
\affil{National Astronomical Observatories, Chinese Academy of Sciences,\\ 20A
Datun Road, Chaoyang District, Beijing 100012, China\\}
\affil{School of Astronomy, University of Chinese Academy of Sciences,
Beijing 100049, China\\}

\author{H.Y. Zhang}
 \affil{National Astronomical Observatories, Chinese Academy of Sciences,\\ 20A
Datun Road, Chaoyang District, Beijing 100012, China\\}

\author{D. Wu}
\affil{School of Astronomy, University of Chinese Academy of Sciences,
Beijing 100049, China\\}

\author{A.Y. Yang}
\affil{National Astronomical Observatories, Chinese Academy of Sciences,\\ 20A
Datun Road, Chaoyang District, Beijing 100012, China\\}
\affil{Max-Planck-Institut f\"ur Radioastronomie, Aufdem H\"ugel 69. D-53121, Bonn, Germany}


\begin{abstract}
  We carry out a project to independently measure the distances of supernova remnants (SNRs) in the first quadrant of the Galaxy.
In this project, red clump (RC) stars are used as standard candles and extinction probes to build the optical extinction (A$_V$) - distance(D) relation in each direction of extinction-known SNRs. The distances of 15 SNRs are well determined. Among them, the distances of G65.8-0.5, G66.0-0.0 and G67.6+0.9 are given for the first time.
We also obtain 32 upper/lower limits of distances, and
 the distances to G5.7-0.1, G15.1-1.6, G28.8+1.5 and G78.2+2.1 are  constrained. Most of the distances measured by the RC method are consistent with previous results. The RC method provides an independent access to the distances of SNRs.
 
\end{abstract}


\section{Introduction}
\label{sec:intro}
Supernova remnants (SNRs) play key roles in the final evolution of stars,  reshaping and heating  the interstellar medium, and the birth of the high-energy cosmic rays.
 Reliable distances to SNRs are essential to constrain their physical parameters such as age,  physical size, expansion velocity and  explosion energy of the progenitor supernovae,
which reveal the evolutionary process of SNRs.
However, obtaining reliable distances of SNRs is a really challenging job. About 20\% of Galactic SNRs have distance measurements \citep{Green2014b}. 

There are several  popular methods to measure the distances to Galactic SNRs. Firstly, the kinematic method is based on the flat rotation curve of the Milky Way. By combining 21 cm HI absorption with CO emission, \citet{Tian2007} developed an improved way to measure the distances of the extended radio sources by minimizing the possibility of a false absorption spectrum. Their methods have been applied to several SNRs, e.g., SNRs Kes 69 \&75, Tycho's SNR \citep{Tian2008, 2011Tian}. Secondly, distance determinations to the shell-type SNRs can be inferred by the relation
 between the mean surface brightness ($\Sigma$) at a specific radio frequency and physical diameter (D) of an SNR,
$\Sigma$=aD$^{\beta}$. Distance is the ratio of physical diameter
 and the angular diameter \citep[e.g.][]{Clark1976,Milne1979,Case1998}. $\Sigma$-D relation is frequently used since $\Sigma$ is easy to be observed in radio bands for most radio SNRs. 
Thirdly, the distances can be accessible when SNRs are associated with the objects with known distances like OB associations \citep[e.g.][Vela remnant]{Cha1999} or pulsars \citep[e.g.][]{Cordes2002}. Additionally, the proper motion and the shock velocity can be used to calculate the distance \citep[e.g.][Kepler's SNR]{Vink2008,Katsuda2008}. For the shell-type SNRs in the adiabatic phase, distances can be calculated by the X-ray flux and thermal temperature of X-ray -emitting gas \citep{Kassim1994}. Finally, the extinction measurements  can also indicate distances \citep[e.g.][]{Chen2017,Zhao2018}, which this paper focuses on.

Red clump (RC) stars are characterized by an obvious concentration region in the colour-magnitude diagram (CMD). They are usually low-mass stars in the early stage of core He-burning. Their helium cores almost have the same mass. Meanwhile their absolute magnitude weakly depends on metal abundance and ages in the K band \citep{Alves2000}. Hence, RC stars are good enough to be standard candles in the infrared band. Assuming that intrinsic colour of RC stars is homogeneous, then their CMD spread along the colour is just caused by interstellar extinction traced by  RC stars. \citet{Zhu2015} applied a similar measurement to determine the distance of SNR G332.5-5.6.

 We closely follow the RC method and systematically measure the distances to 47 SNRs with known extinction in the first Galactic quadrant, with the aim of enlarging the reliable distance sample of SNRs. In Section 2, the method is described in detail. We summarize methods of measuring the optical extinction, the hydrogen column density and the distances of SNRs compiled from the literature in Section 3. The uncertainties of this method are analyzed  in Section 4. In Section 5, we discuss our results and make a comparison with distances measured by other methods. Finally, a brief summary is given.

\section{Build  A$\rm _v$-D Relation}

\begin{figure}
\centering
\includegraphics[angle=0,width=0.45\textwidth]{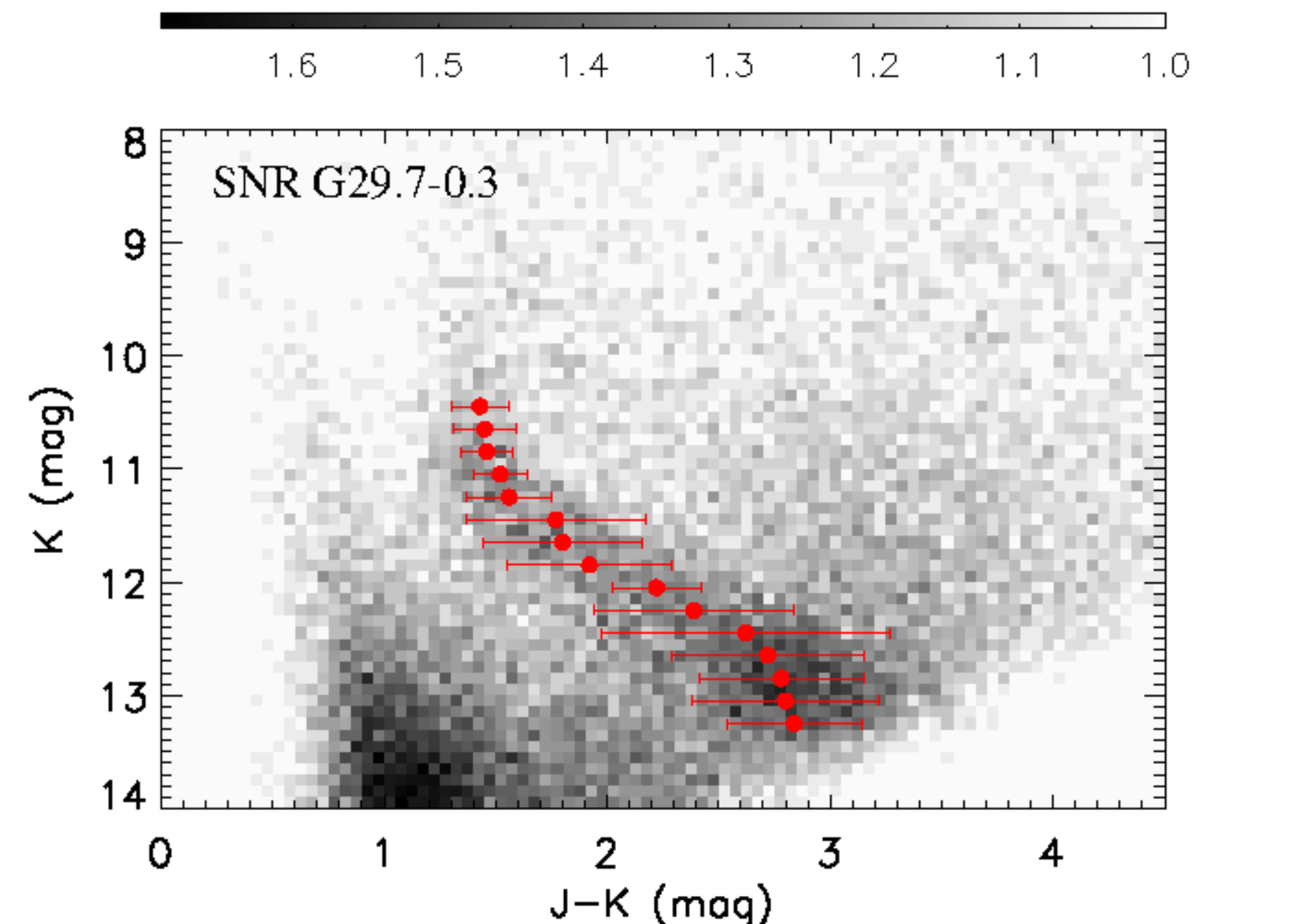}
\caption{Colour-magnitude diagram for 21032 stars within 0.5 deg$^2$ of G29.7-0.3, the grey colours denote stellar densities in the logarithmic scale.
The red dot and  lines show the fitted location of the RC peak density and its extent with $1\sigma$.}
\label{fig1}
\end{figure}
\begin{figure}
\centering
\hspace{-7mm}
\includegraphics[angle=0,width=0.45\textwidth]{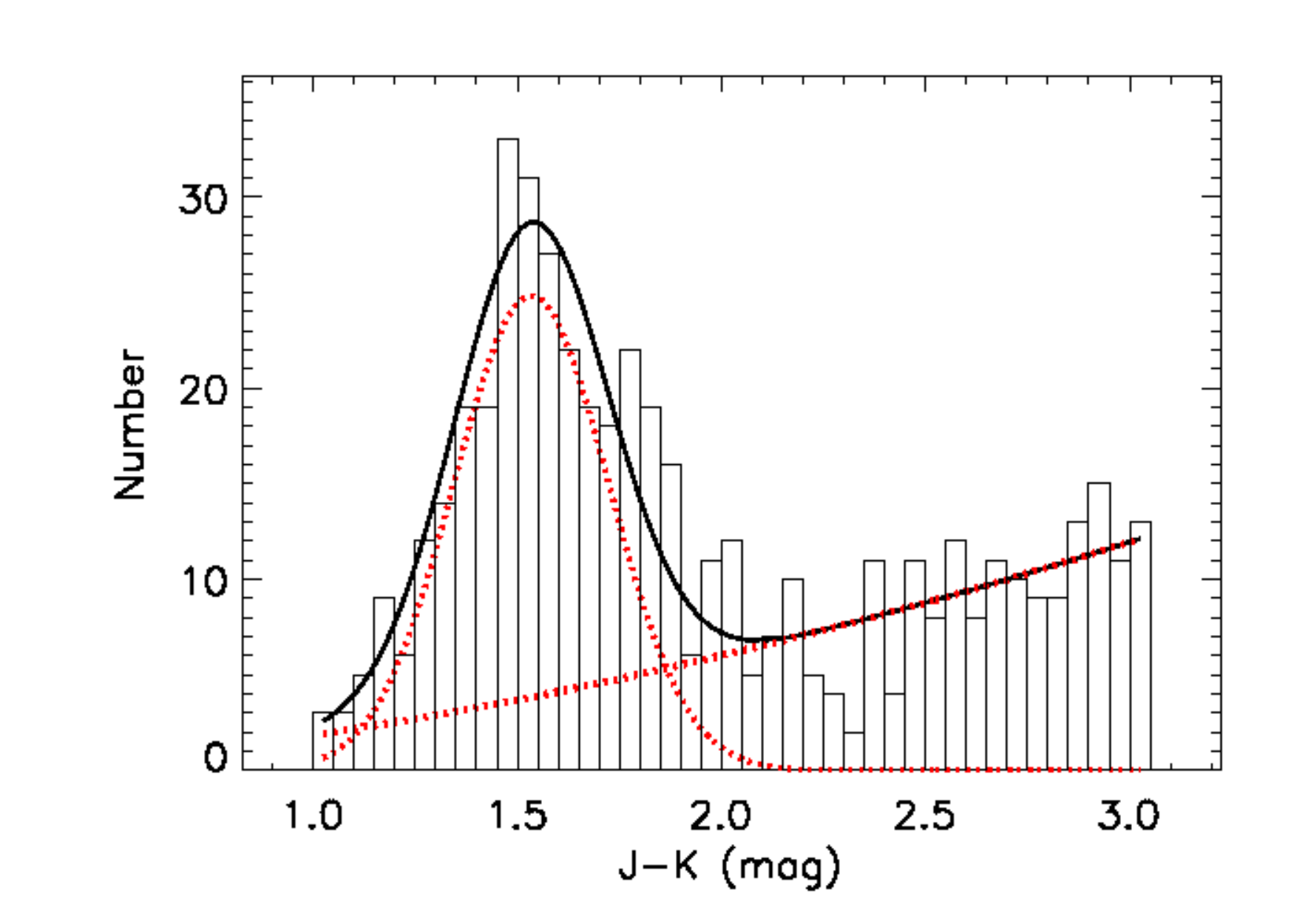}
\caption{Histogram of the $\rm J-K$ values of the selected stars
 in the $\rm 11.1 <K < 11.4$. The black curve is the best fit to this histogram
 to obtain  the value of $\rm J-K$  for the peak density of RC stars. The red dotted curves are the Gaussian and power-law components, respectively.}
\label{fig2}
\end{figure}

 To better illustrate this method, we start with G29.7-0.3 as an example.
We extract the stars from the 2MASS All-sky Point Source Catalog in the J and $\rm K_{ S}$ (hereafter K) bands \citep{Skrutskie2006} centered on the SNR in $ 1^{\circ}$
$\times$  $ 0.5^{\circ}$ ($\bigtriangleup l\,$ $\times$ $\bigtriangleup b \,$) area. The reason why we choose the size of 0.5 deg$^2$ will be discussed later in this section.
 Their magnitudes in J and K bands are used to construct the CMD (K vs. J-K) since the RC stars are easy to be identified on CMDs \citep[e.g.][]{Gao2009}. In Figure~\ref{fig1}, RC stars are concentrated in the middle of the CMD. The bulk of stars in the left region of the CMD are predominantly main-sequence stars; those in the right  are mainly dwarfs and red-giant-branch stars.

 In principle, there is a maximal density of RC stars in each range of K apparent magnitude. We divide the stars sample into a number of horizontal strips in K through the CMD. The locations of the RC stars  in different strips indicate different distances and reddening. The width of each strip is usually 0.3 mag and it will be extended to 0.5 mag or 0.7 mag when the counts of the RC's peak density  are less than 10. The length of each strip in $J-K$ is fixed by the RC's distribution in order to include most of the RC stars and minimize the contamination of the stars of other types. For each strip, we apply an empirical function to fit the histogram of star counts
\citep{Durant2006}:
\begin{equation}
\rm y=A_{RCs}exp\{ \frac{-[(J-K)-(J-K)_{peak}]^2}{2\sigma ^2}\}+A_{C}(J-K)^{\alpha}
\label{eq1}
\end{equation}
Where ($\rm J-K$) represents the stellar colour, $\rm A_{RCs}$ and $\rm A_{C}$ stand for
 the normalizations of the RC stars and the contaminant stars, respectively.
 The first term is a Gaussian of $(J-K)_{\rm peak}$ and the width $\sigma$  to fit the  RC stars distribution;
the second term is a power law to fit the contaminant stars. For instance,
 Figure~\ref{fig2} shows the best fit for the $\rm 11.1 < K < 11.4$ strip: the stellar colour ($\rm J-K$) at the peak density of the RC stars $\rm (J-K)_{peak}$ is 1.56 mag and the $\sigma$ is 0.19 mag.
The $(J-K)_{\rm peak}$  value is applied to calculate the average extinction of this field as equation (2).
We assume that the intrinsic colour $\rm (J-K)_0$ is  0.63 mag and
the mean absolute magnitude of  RC stars in K band is  $-1.61$ mag. We discuss further in section~\ref{sec:uncertain}.
Then the extinction and the corresponding distance are derived from the following functions\citep{Indebetouw2005}:
\begin{equation}
\rm A_K=0.67 \times[(J-K)_{peak}-(J-K)_{0}]
\end{equation}
\begin{equation}
\label{f2}
\rm \frac{A_K}{A_V}=0.1615-\frac{0.1483}{R_V}
\end{equation}

\begin{equation}
\rm D(kpc)=10^{[0.2(m_K-M_K+5-(0.1137\pm0.003) \times A_V)]}/1000
\end{equation}
Where $\rm A_{V}$ and $\rm A_K$ are extinction in V and K bands, $\rm R_V=3.1\pm0.18$ (we discuss the value of $\rm R_V$ in Section~\ref{rv} ).
In Equation \ref{f2}, the conversion from $\rm A_K$ to $\rm A_{V}$  follows the empirical relation of \citet{Cardelli1989}, which contributes an uncertainty of $\sim$3\% to the optical extinction.

This process is repeated for all strips until the 2MASS observational limit is reached.
 Since the extinction grows with the increasing distance in a given line of sight, 
 there is a one-to-one correspondence between the extinction and distance. Hence, 
 the distance of G29.7-0.3 is  obtained by overlapping its extinction value 
 on the $\rm A_V-D$ relation in its direction.

\begin{figure}
\centering
\includegraphics[angle=0, width=0.45\textwidth]{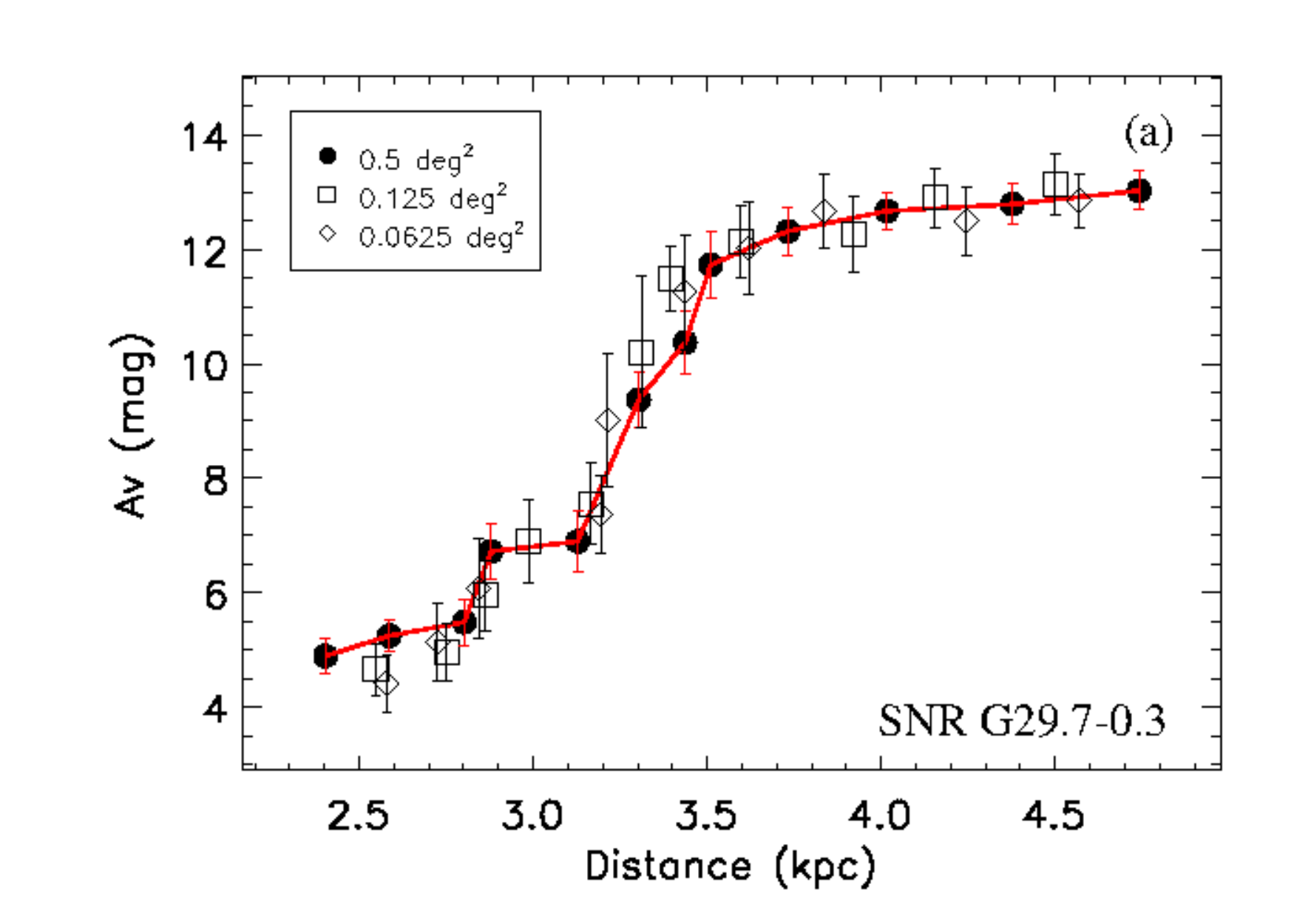}
\includegraphics[angle=0, width=0.45\textwidth]{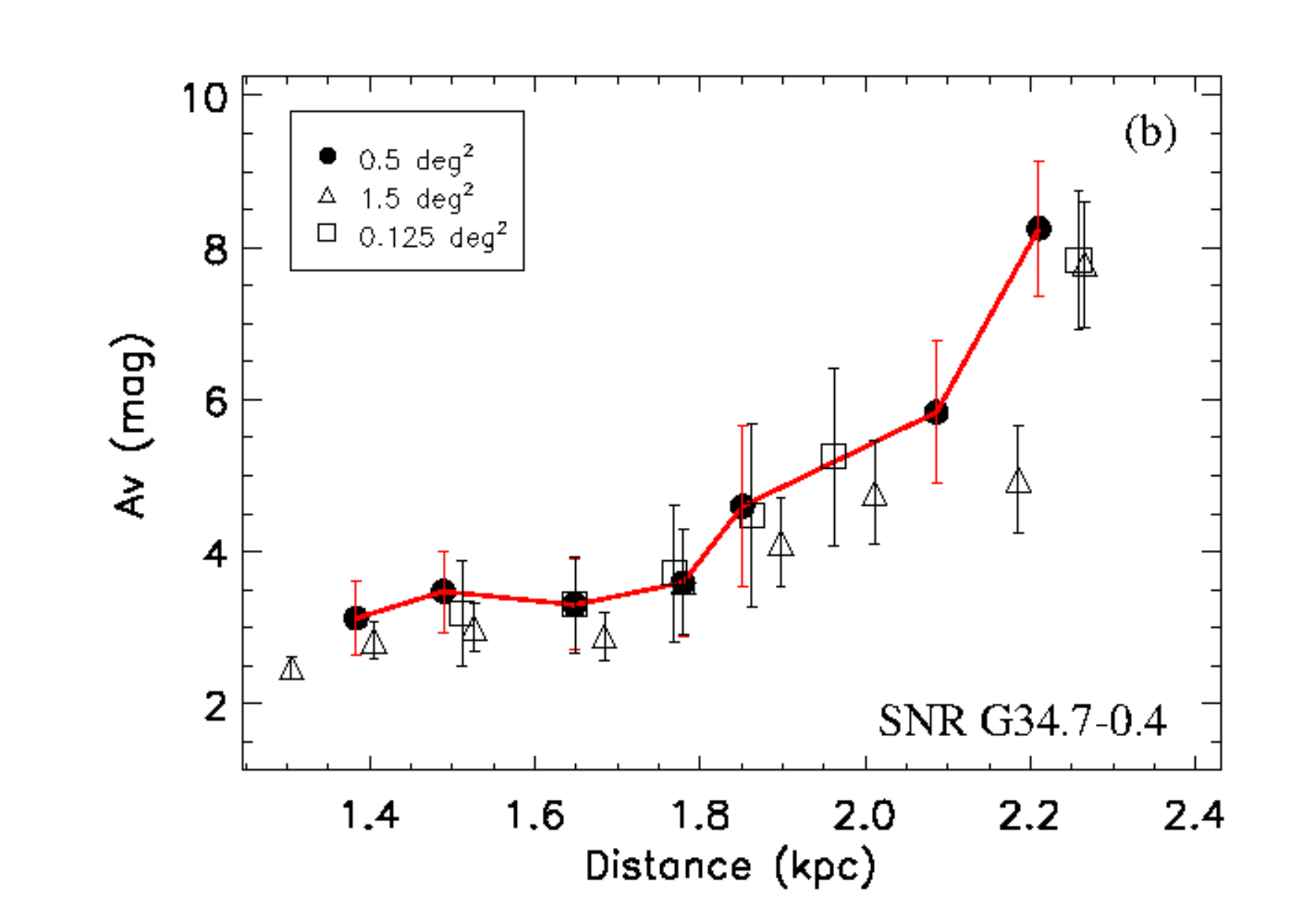}
\caption{The red curve associated with the extinction uncertainties is established by RC stars within 0.5 deg$^2$. (a) The $\rm A_V$-D relations in the direction of G29.7-0.3 within different bins. The black curve associated with the extinction uncertainties is established by stars within 0.0625 deg$^2$. 
(b) The $\rm A_V$-D relations in the direction of G34.7-0.4 within different bins.}
\label{fig3}
\end{figure}

 We tempt to find an optimal bin size for each SNR.
On the one hand,  a larger bin size can help increase the accuracy in determining the extinction. This would enlarge the amount of the RC stars, then decrease the uncertainty of RC's colour $\rm (J-K)_{peak}$. Meanwhile it would allow a narrow horizontal strip in $m_K$ that is used to perform the fits of Equation ~\ref{eq1}, then yield a better sampling of the pairs of extinction and distance. On the other hand, a smaller bin sizes can decrease the dispersion of extinction in the  line of sight.

\begin{figure*}
\centering
\includegraphics[angle=0, width=0.49\textwidth]{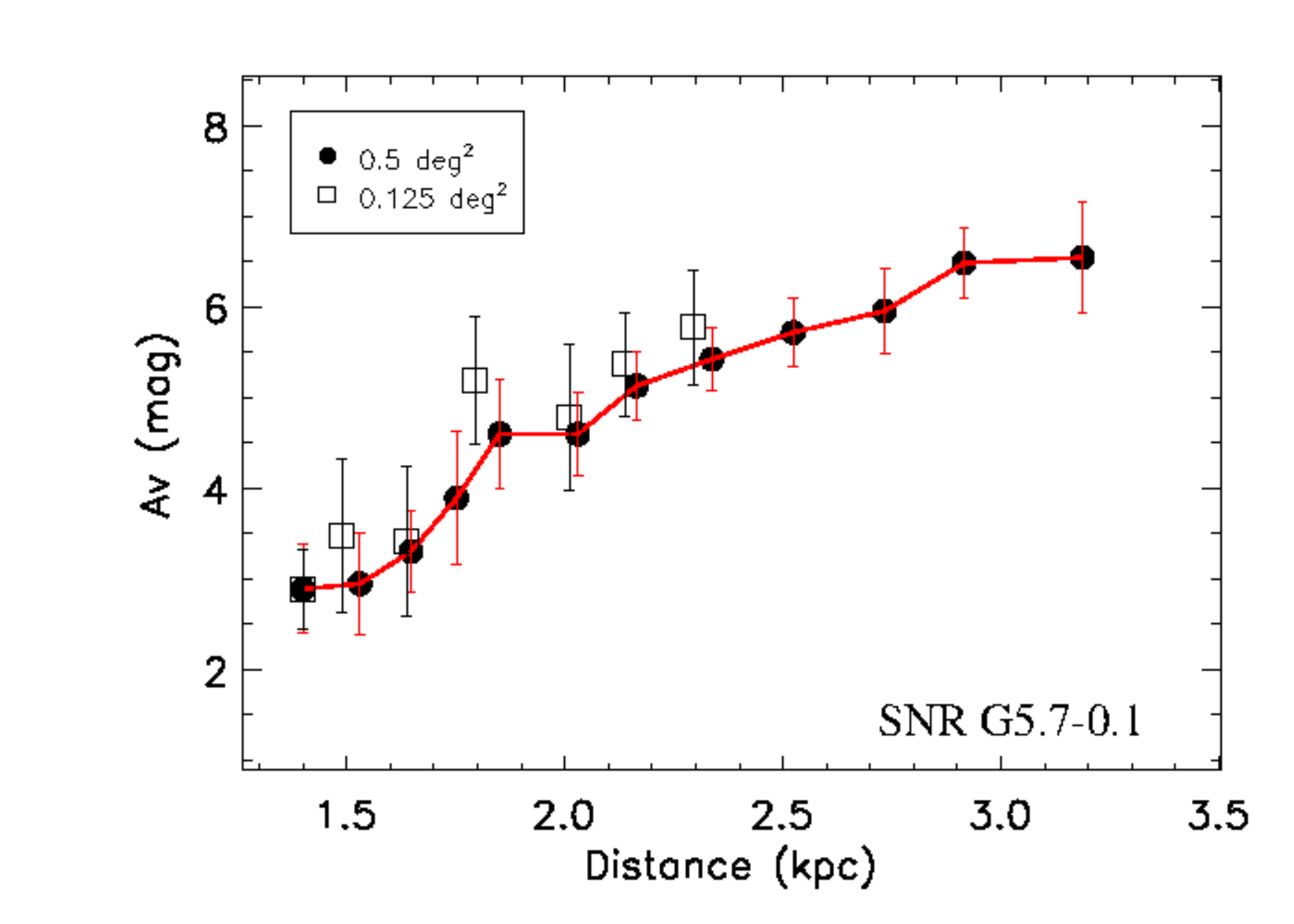}
\includegraphics[angle=0, width=0.49\textwidth]{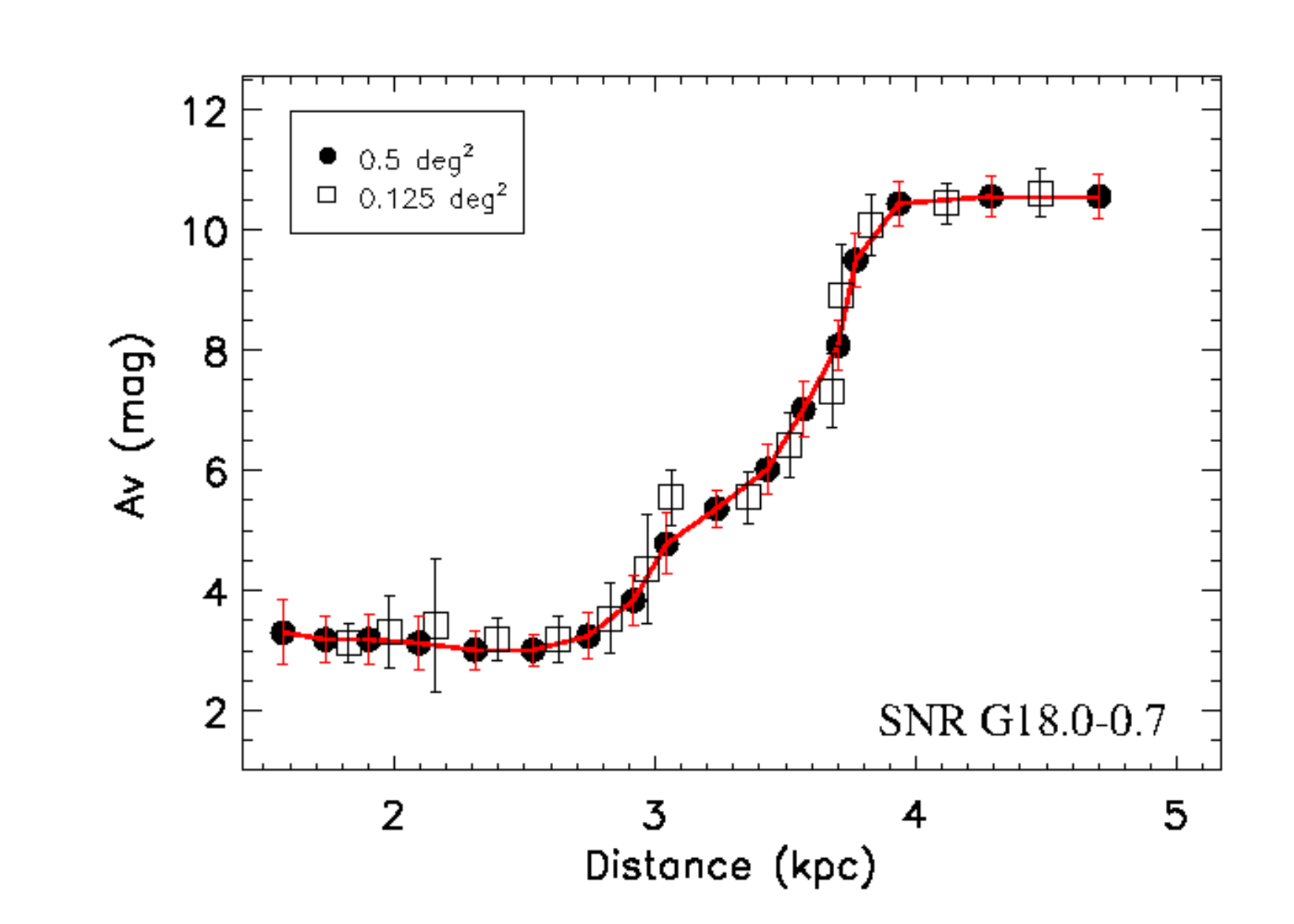}
\includegraphics[angle=0, width=0.49\textwidth]{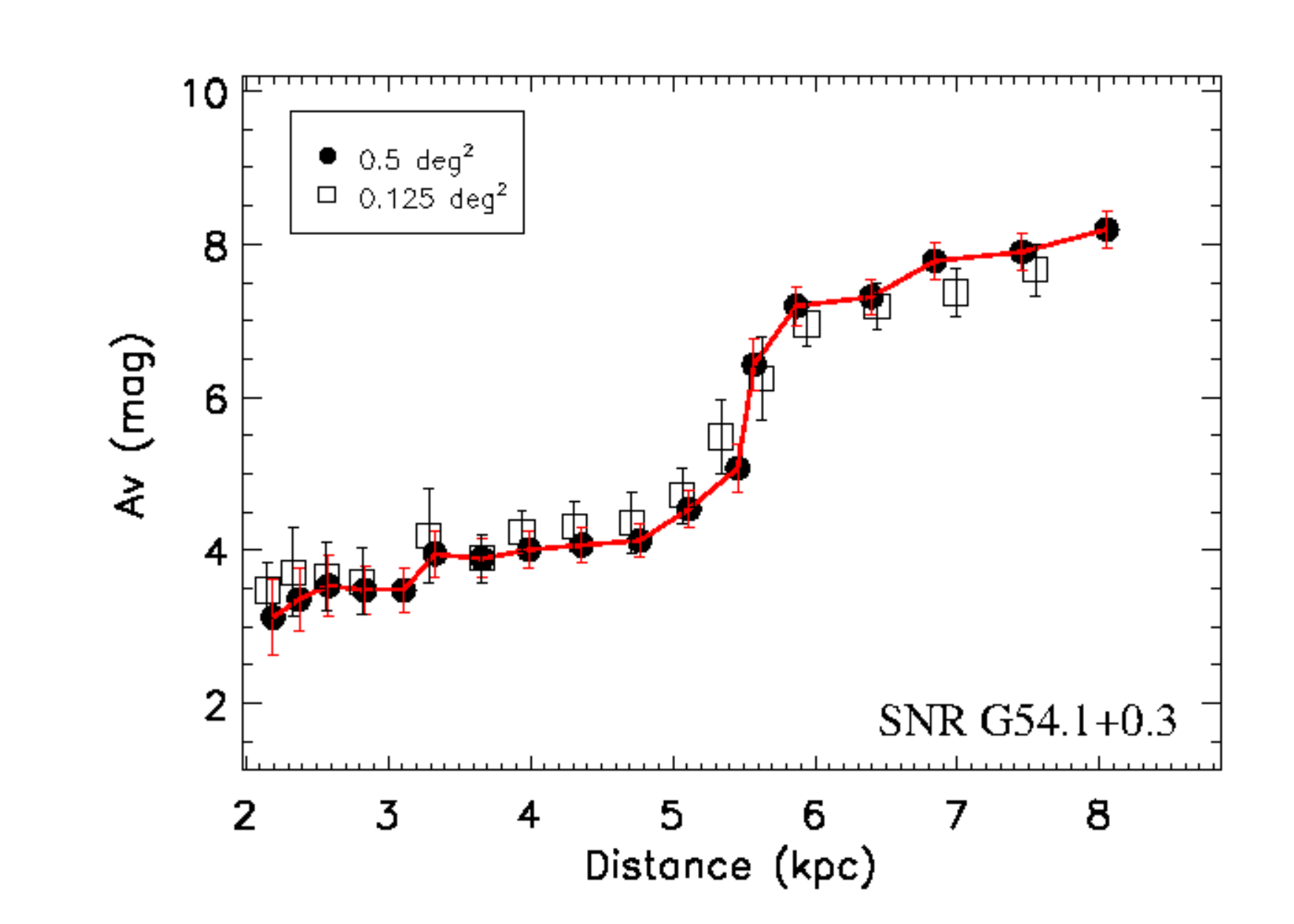}
\includegraphics[angle=0, width=0.49\textwidth]{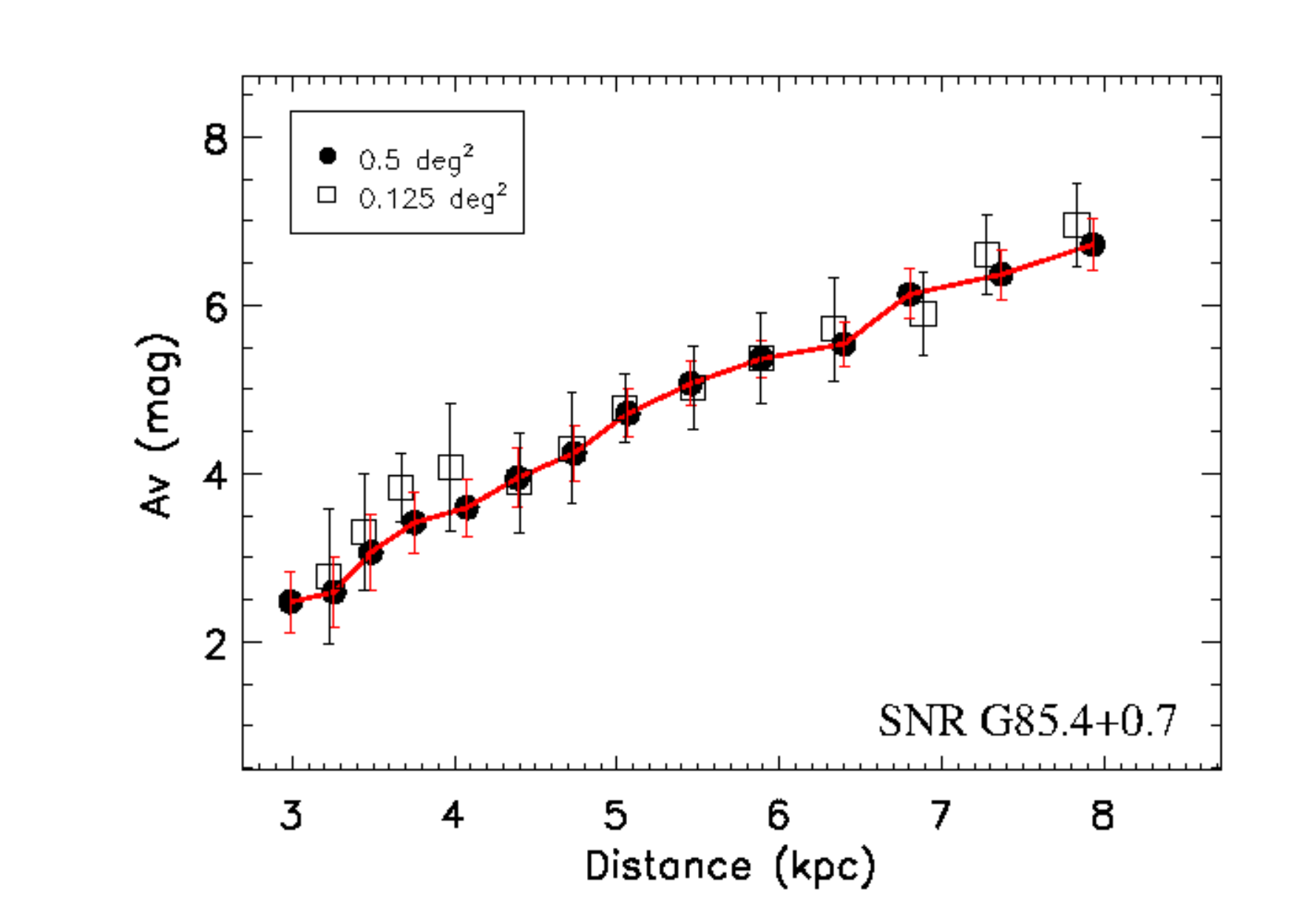}
\caption{ Comparison of the  $\rm A_V$-D curves within different bin size for different objects. The red curves associated with the extinction uncertainties are established by RC stars within 0.5 deg$^2$.}
\label{binsize}
\end{figure*}

We present two typical examples (one with high stellar density, the other one with low stellar density ) to illustrate how to select the bin sizes. For the SNR with high stellar density in the sightlines, the $\rm A_V$-D relations are constructed  within the bin sizes of 0.5 deg$^2$ ($1.0^{\circ}$
$\times$ $0.5^{\circ}$), 0.125 deg$^2$ ($0.5^{\circ}$
$\times$ $0.25^{\circ}$), 0.0625 deg$^2$ ($0.25^{\circ}$
$\times$ $0.25^{\circ}$, the smallest possible area around the target), respectively (See Figure~\ref{fig3} (a) ). 
It is found that  the three $\rm A_V$-D relations are fully consistent with each other. In this case, the effect of bin sizes can be neglected when deriving the  distance of SNRs.
 For the one with low stellar density , the $\rm A_V$-D relations are constructed  within the bin sizes of 1.5 deg$^2$ ($\bigtriangleup l\, 1.5^{\circ}$
$\times$ $\bigtriangleup b \, 1^{\circ}$), 0.5 deg$^2$ ($1.0^{\circ}$
$\times$ $0.5^{\circ}$), 0.125 deg$^2$ ($0.5^{\circ}$
$\times$ $0.25^{\circ}$,the smallest possible area around the target), respectively (See Figure~\ref{fig3} (b) ).
The numbers of sampling points are 9, 7, 6, respectively. The $\rm A_V$-D relations of 0.5 deg$^2$ and 0.125 deg$^2$ are almost the same while  the data points of 1.5 deg$^2$ are  systematically  lower. As equation (2) shows, $\rm A_V$ is determined by the stellar colours of  RC stars located in the peak density $\rm (J-K)_{peak}$. The lower $\rm A_V$  for 1.5 deg$^2$  is a result of 1.5 deg$^2$ concentrating in  stellar colour($\rm J-K)_{peak}$  lower than those of  0.5 deg$^2$ and 0.125 deg$^2$. The $\rm A_V$-D relations indicate that extending the area to 1.5 deg$^2$  might enlarge extinction dispersion. 
In principle, we choose a larger size of bin when the dispersion of extinctions is at the same level.  The  of 0.5 deg$^2$ is a good balance for the two examples. Then, we test the bin sizes of  
 0.5 deg$^2$ and 0.125 deg$^2$ for the objects almost at an interval of 15$^{\circ}$ in the longitude. In addition to Figure~\ref{fig3}, Figure~\ref{binsize} presents the results in the four typical directions. In each of the panels, the two curves are  consistent with each other, except for the number of sampling points within 0.125 deg$^2$ is less than that of  0.5 deg$^2$ for some objects. Hence, we usually build the  $\rm A_V$-D curve  within 0.5 deg$^2$  when deriving the distance of an SNR.

In recent years, both theory and observations  have shown the fine structure of  RC, which includes two subclass stars, the main RC stars and the secondary RC stars (SRCs) .
 The main RC stars with low mass that we usually take as standard candles have almost the same luminosity and electronic-degenerate core; while the SRCs  whose luminosity are greatly changed contain  non-degenerate He-cores \citep{Girardi1999}. The largest sample of RC stars identified based on LAMOST survey DR3 \citep{Wan2015,Wan2017} shows  the ratio of the main and secondary RC stars is about 3:1.
To investigate  the effects of the  SRCs for this method, we test to use a double Gaussian function to fit  the distribution of RC stars of each strip.  No apparent secondary Gaussian component can be found. Hence, the effects of the SRCs can be neglected in this method.

\section{Compilation of A$_{\rm V}$, N$_{\rm H}$ and D }

 Drawing from the catalogs of \citet{Green2014a} and \citet{Ferrand2012}, we investigate each of 161 SNRs in the first Galactic quadrant. Among them, 47 SNRs have access to  the optical extinction or hydrogen column density data in the literatures. We collect their parameters on optical extinction A$_{\rm V}$, hydrogen column density N$_{\rm H}$ and the distance D, then discuss the methods for determing determine the three parameters. The three parameters and 
 the corresponding methods are listed in Tables ~\ref{tab1} and ~\ref{tab2} .

 \begin{table*}
\setlength{\tabcolsep}{3mm}
\begin{center}
\caption{ Optical extinction $\rm A_V$ and distances D}
\label{tab1}
\begin{tabular}{llllllll}
  \hline
  \hline
  \mc{1}{l}{Source}&  \mc{1}{c}{$\rm A_{V}$}& \mc{1}{l}{Method}& \mc{1}{l}{$\rm D_{known}$}&\mc{1}{l}{Method}& \mc{1}{l}{$\rm D_{this    paper}$}&\mc{1}{l}{Ref.}\\
  \mc{1}{l}{Name }&       \mc{1}{l}{(mag)}&   \mc{1}{l}{}& \mc{1}{l}{(kpc)}& \mc{1}{l}{}&  \mc{1}{l}{(kpc)}&   \mc{1}{l}{}\\                   
 \hline
G54.1+0.3& 7.3$\pm$0.4&associated stars&6.2&kinematic measurement&6.3$_{-0.7}^{+0.8}$&1, 2, 3 \\
G65.8-0.5&2.4$\pm$0.4&H$_{\alpha}$/H$_{\beta}$&-&-&2.4$_{-0.5}^{+0.3}$&4\\
G66.0-0.0&2.0$\pm$0.2&H$_{\alpha}$/H$_{\beta}$&-&-&2.3$\pm$0.3&4\\
G67.6+0.9&1.9$\pm$0.2&H$_{\alpha}$/H$_{\beta}$&-&-&2.0$\pm$0.2&4\\
G67.7+1.8 &1.7$\pm$0.7&H$_{\alpha}$/H$_{\beta}$&7-17&$\Sigma$-D+Extiction&$2.0_{-0.5}^{+3.7}$&5, 6\\
G69.0+2.7&2.5$\pm$0.3& H$_{\alpha}$/H$_{\beta}$&1.5, 3$\pm2$&kinematic measurement&4.6$\pm0.8$&7, 8, 9\\
G82.2+5.3& 2.8$\pm$0.2&H$_{\alpha}$/H$_{\beta}$&1.6, 2.0& $\Sigma$-D, HII distance& 3.2$\pm$0.4&10-12\\
 G89.0+4.7& 1.6$\pm$0.3&H$_{\alpha}$/H$_{\beta}$& 0.8-1.7&kinematic measurement&1.9$_{-0.2}^{+0.3}$&7, 13, 14\\
\hline
\end{tabular}
\end{center}
\begin{flushleft}

Reference. (1) \citet{Kim2013},
(2) \citet{Leahy2008}, 
(3) \citet{Lu2002},
(4) \citet{Sabin2013}, 
(5) \citet{Sezer2008},
(6) \citet{Hui2009}, 
(7) \citet{Zhu2017}, 
(8) \citet{Leahy2012},
(9) \citet{Verbiest2012},
(10) \citet{Mavromatakis2004b}, 
(11) \citet{Uyaniker2003},
(12) \citet{Rosado1981},
(13) \citet{Byun2006}, 
(14) \citet{Tatematsu1990}. 
\end{flushleft}
\end{table*}
\begin{table*}
\setlength{\tabcolsep}{3mm}
\begin{center}
\caption {Hydrogen column density $\rm N_H$ and distances D}
\label{tab2}
\begin{tabular}{lllllllll}
  \hline
  \hline
  \mc{1}{l}{Sourse}&  \mc{1}{l}{$\rm N_{H}$}& \mc{1}{l}{$\rm A_{V}$}& \mc{1}{l}{Model$^a$}&   \mc{1}{l}{$\rm D_{known}$}& \mc{1}{l}{Method}&
\mc{1}{l}{$\rm D_{this paper}$}&\mc{1}{l}{Ref.}\\
  \mc{1}{l}{Name }&       \mc{1}{l}{(10$^{21}Hcm^{-2}$)}& \mc{1}{l}{(mag)}& \mc{1}{l}{}&   \mc{1}{l}{(kpc)}&   \mc{1}{l}{}&\mc{1}{l}{(kpc)}&\mc{1}{l}{}&\\
  \hline
 G5.7-0.1&13.0$\pm$1.0&6.4$\pm0.5$&TP&3.1 or 13.7&kinematic measurement&2.9$\pm0.3$&1, 2\\
G11.0-0.0&$8.0\pm3.0$ &3.9$\pm1.5$&PL& 2.6&absorption column&2.4$\pm0.7$&3\\
G18.0-0.7&10.0$\pm$2.0&4.9$\pm1.0$&PL&3.9$\pm$0.4&pulsar distance&3.1$\pm0.2$&4, 5\\
G18.9-1.1&8.3$\pm0.5$&4.1$\pm0.2$&TP&2.0&kinematic measurement&$1.8\pm0.2$&6, 7\\
G34.7-0.4&13.0$\pm$2.0&6.4$\pm1.0$&TP+PL&2.6-3.2,2.5&kinematic measurement&2.1$\pm0.2$&8-10\\
G49.2-0.7&17.0&8.3$\pm$1.7&TP&4.3,5.6&kinematic measurement&5.7$_{-0.8}^{+0.7}$&11-13\\
 G85.4+0.7&8.3&4.1$\pm0.8$&TP&2.5-4.5&kinematic measurement&4.4$\pm0.8$&14, 15\\
\hline
\end{tabular}
\end{center}
\begin{flushleft}
$^a$Model abbreviations: TP: thermal plasma, PL: power law, BB: black body, TT:a two-component thermal model.

Reference:
(1) \citet{Jounert2016}, 
(2) \citet{Hewitt2009}, 
(3) \citet{Bamba2003}, 
(4) \citet{Gaensler2003},
(5) \citet{Cordes2002}, 
(6) \citet{Harrus2004}, 
(7) \citet{Aschenbach1991},  
(8) \citet{Uchida2012}, 
(9) \citet{Park2013}, 
(10) \citet{Frail2011}, 
(11) \citet{Hanabata2013}, 
(12) \citet{Tian2013}, 
(13) \citet{Koo1995}, 
(14) \citet{Jackson2008}, 
(15)\citet{Kothes2001}.
\end{flushleft}
\end{table*}

\subsection{Optical extinction A$_{\rm V}$}
\label{rv}
The interstellar extinction is the absorption and scattering of electromagnetic radiation by dust and gas.
 The most common method to obtain optical extinction A$_{\rm V}$ is  measuring the reddening via the intensity ratios between the two emission lines and converting the reddening into the colour excess $\rm E_{B-V}$. Then we gain the extinction values  $\rm A_V=R_V\times E_{B-V}$.  The total to selective extinction ratio  $\rm R_V$  is $\sim$ 3.1 for the diffuse interstellar medium in the Milky Way, which is widely 
 used \citep[e.g.][]{Fitzpatrick1999,Draine2003,Schlafly2011}. 
 \citet{Schlafly2016} measured the reddening  of  37,000 stars
  in the Galactic disk based on APOGEE, PS1, 2MASS and WISE data, 
  then determined the uncertainty of the $\rm R_V$, $\sim$ 0.18. 
  Although  $\rm R_V$ has significant variance in some regions, 
  there is a wide wedge of intermediate $\rm R_V$ in the first 
  quadrant (see the Figure 3 of the \citet{Schlafly2017}). 
  Hence it is robust to adopt $\rm R_V=3.1\pm 0.18$. T
  he uncertainties of $\rm A_V$ are  approximately estimated  
  as  $\rm  A_V\times \sqrt{(\frac{\sigma(R_V)}{R_V})^2+(\frac{\sigma(E(B-V))}{E(B-V)})^2}$. 
     
The frequently used line ratio we present in Table~\ref{tab1} is H$_{\alpha}$(6563\AA)/H$_{\beta}$(4861\AA) based on the Blamer decrements, which are strong enough to be resolved in optical band. Other line ratios involve
[S\scriptsize \uppercase\expandafter{\romannumeral2} \normalsize]($\sim$10320\AA)/[S\scriptsize \uppercase\expandafter{\romannumeral2} \normalsize]($\sim$4068\AA), [Fe\scriptsize \uppercase\expandafter{\romannumeral2} \normalsize]($\sim$1.6435 $\mu$m)/ [Fe\scriptsize\uppercase\expandafter{\romannumeral2} \normalsize]($\sim$4068$\mu$m).
The group lines from the transition with the same upper level weakly depend on the physical conditions
such as temperature and density of the gas. Hence, they are feasible to
estimate the extinction to the extended sources \citep[e.g.][SNRs and pulsar wind nebulae]{oliva1989}.
And another approach is to measure the extinction of the  individual stars with known distances that are associated with an SNR.

\subsection{N$_{\rm H}$}
Hydrogen column density N$_{\rm H}$ is usually used to approximately denote X-ray extinction which is caused by any element not fully ionized, especially the abundant heavy elements when energy is above 0.25 \keV. The dust grains of the same abundant heavy elements also contribute to the optical extinction $\rm A_V$ \citep{Tolga2009}. Both theoretical and observational studies for decades have indicated that there should be a reasonable correlation of  $\rm A_V$ and N$_{\rm H}$. We adopt the latest value  $\NH/\AV=\left(2.04\pm0.05\right)\times10^{21}\rmH\cm^{-2}\magni^{-1}$ for the first and fourth Galactic quadrants \citep{Zhu2017}.The conversion error is  approximately estimated  as  $\rm  \frac{N_H}{2.04}\times \sqrt{(\frac{\sigma(N_H)}{N_H})^2+(\frac{\sigma(\NH/\AV)}{\NH/\AV})^2}$.

More than half of SNRs have been detected in X-ray band \citep{Ferrand2012}. About 30\% of SNRs are associated with optical emission \citep{Green2014a}. Therefore, we can obtain more $\rm A_V$ values transformed from the N$_{\rm H}$.

 N$_{\rm H}$ is usually derived from the  best fitting of X-ray spectrum.
Here we only  collect the N$_{\rm H}$ derived from solar abundances \citep{AG89} to keep interstellar abundances consistent in the whole transition. If the uncertainty of  N$_{\rm H}$ has not been given in the literature,  we use the average errors of the $\rm N_H$, 20\%, which are derived from our sample with known uncertainties.
\subsection{Distance}

 The distance measurements of SNRs mainly
 include the $\Sigma$-D relation, the kinematic method, the proper motion measurements, extinction measurements, Sedov estimates, and associated objects with known distance.
Due to different distance measurements with varying uncertainties, we select distances of SNRs
 in the literature with the following priority: kinematic method, proper motion estimates, associated objects with known distance, Sedov estimates and $\Sigma$-D relation.
 
\section{Uncertainty analysis}
\label{sec:uncertain}
The uncertainties of the derived distances to SNRs are mainly attributed to the errors of  the SNRs' $A_V$  and the RC's distances. The errors of SNRs' $A_V$  are calculated by a standard  deviation formula. Here, we focus on the discussion on the errors caused by RC stars. The errors of RC's distances  mainly include the dispersion of absolute magnitude  and extinction traced by RC stars.

 The absolute magnitude M$_{\rm K}$ of RC stars is extensively studied from the perspective of observations, especially in the K band \citep[e.g.][]{Alves2000, van2007, Groenewegen2008, Laney2012, Yaz2013}. In this work, we assume that the mean value of M$_{\rm K}$ is -1.61 mag, which is consistent with M$_{\rm K}$=-1.61$\pm$0.03 from \citet{Alves2000} from a sample of 238 Hipparcos RC stars in the solar neighborhood,
 with M$_{\rm K}$=-1.61$\pm$0.04 mag derived by \citet{Grocholski2002} based on 14 open clusters, and also with the latest value M$_{\rm K}$=-1.61$\pm$0.01 mag determined by \citet{Hawkins2017}  based on the 2MASS, Gaia and WISE data. However, \citet{van2007} estimated a larger value of RC stars M$_{\rm K}$ =-1.57 $\pm0.05$ mag from 2MASS
data on 24 open clusters, which is in agreement with the value M$_{\rm K}$=-1.54$\pm$0.04 mag measured by \citet{Groenewegen2008} based on the revised Hipparcos parallaxes. Taking all of these studies into consideration, the variance of the absolute magnitude M$_{\rm K}$ leads to a systematic uncertainty of 0.1 mag that contributes about 5\% error in the mean distance calculated by equation (3).

The uncertainties of extinction are caused by the dispersion of intrinsic colour and the random errors of the Gaussian fitting. The absolute magnitudes of RC stars are more sensitive to $\rm [Fe/H]$ and age in the J band than the K band, which leads to the variation of the intrinsic colour $\rm (J-K)_0$ \citep{Tolga2010}. The values of intrinsic colour $\rm (J-K)_0$ for RC stars concentrate in the range from 0.5 to 0.75 mag \citep[e.g.][]{Yaz2013,Grocholski2002}. We adopt $\rm (J-K)_0$ as 0.63 mag, and assume that its dispersion is 0.1 mag which leads to about 3\% uncertainty in the distance, according to equations (2) and (3). Meanwhile we estimate the uncertainty of mean colour on the peak density location of RC stars :
$\rm \sigma_{J-K}=\sigma N_{RC}^{-1/2}$,
 where $\rm N_{RC}=A_{RC}\sigma \sqrt{2\pi}$ is the sum of  RC stars in each magnitude strip \citep{Durant2006}. It is a good estimate if the Gaussian fit is valid and the contamination is not significant. The typical error of the mean colour is about 0.05 mag, which brings $\sim$ 2\% uncertainties in distance estimation.
 
In summary, the systematic uncertainties of the distances traced by the RC stars are about 10\% in total.
 
\section{Results  and Discussion}
\begin{figure*}
\caption{The CMDs within 0.5 deg$^2$ in  each direction of SNRs, the grey colours denote stellar densities in the logarithmic scale. The red dot and  lines show the fitted location of the RC peak density and its extent with $1 \sigma$. The CMD  in the direction of G49.2-0.7 is built within 0.125 deg$^2$.}
\label{fig5}
\begin{tabular}{cc}
\includegraphics[width = 0.45\textwidth]{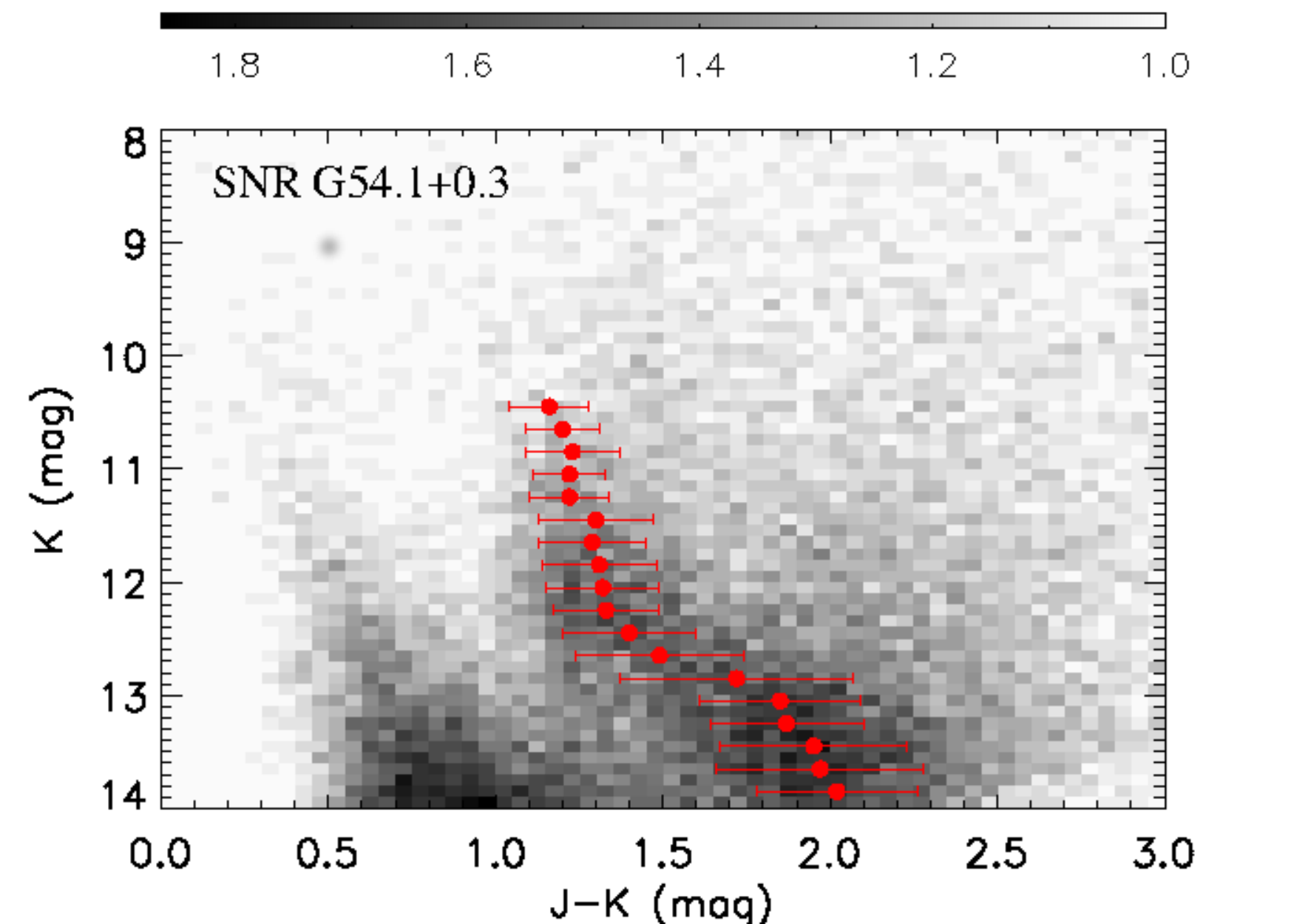}&
\includegraphics[width = 0.45\textwidth]{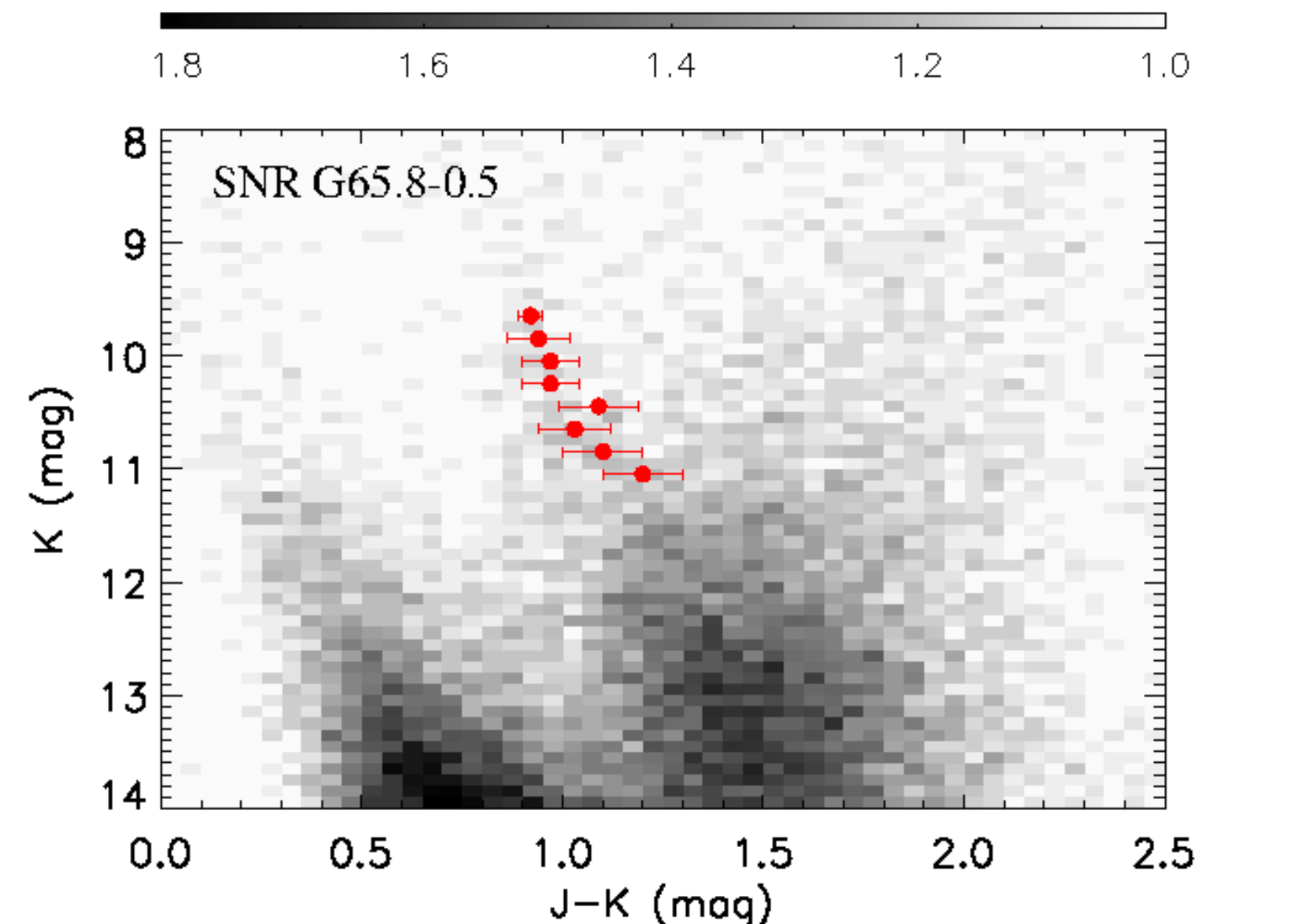}\\
\includegraphics[width = 0.45\textwidth]{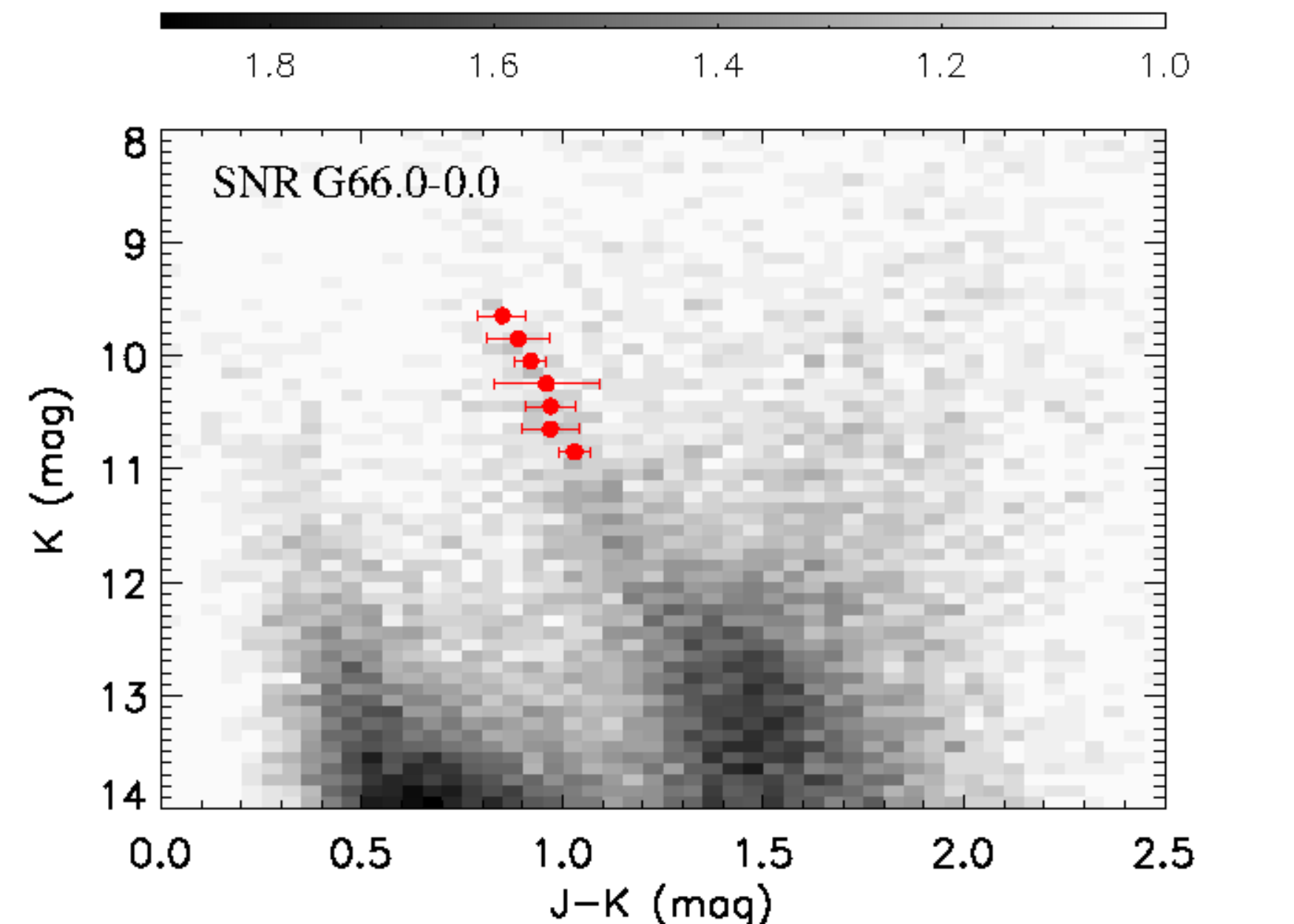}&
\includegraphics[width = 0.45\textwidth]{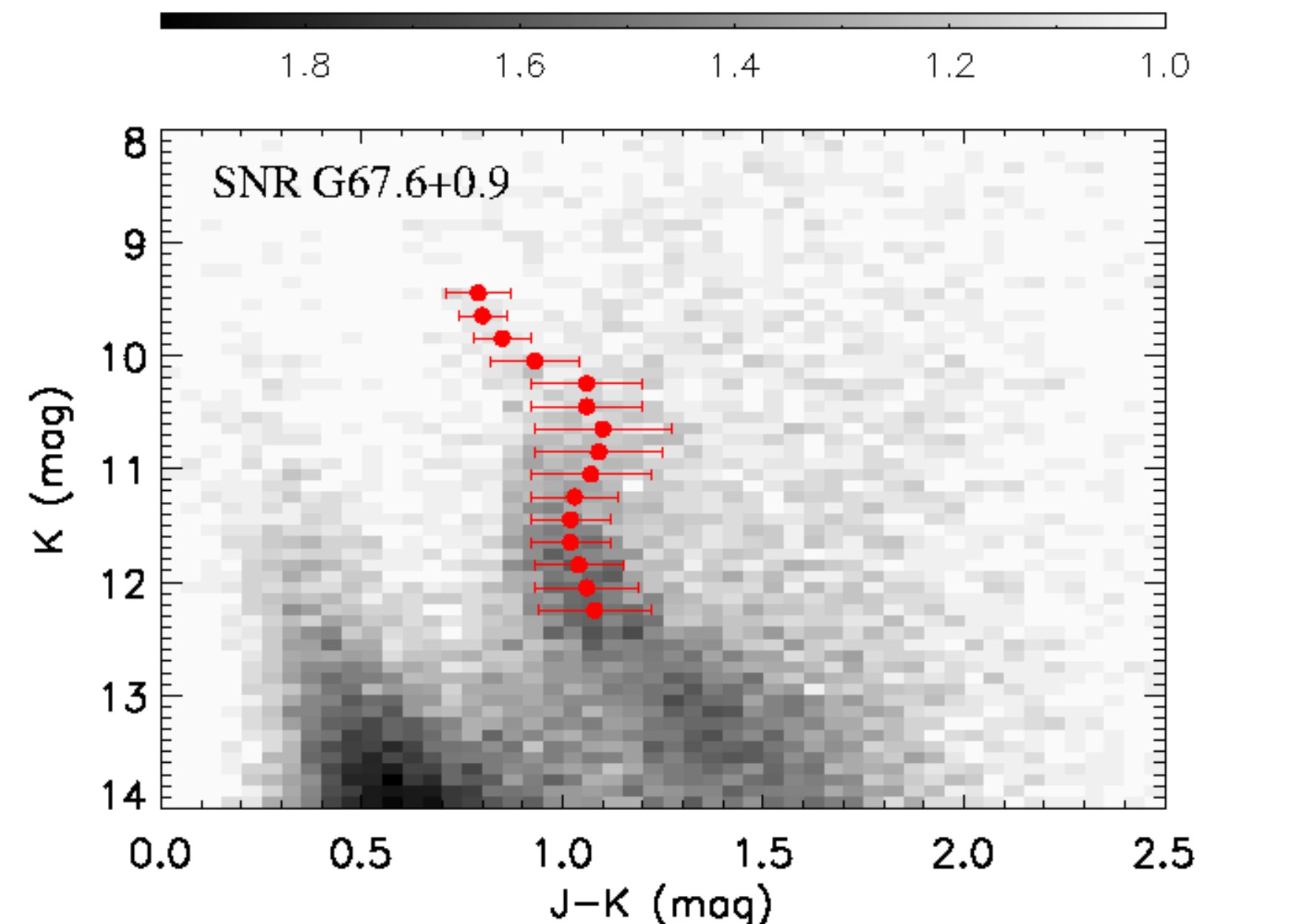}\\
\includegraphics[width = 0.45\textwidth]{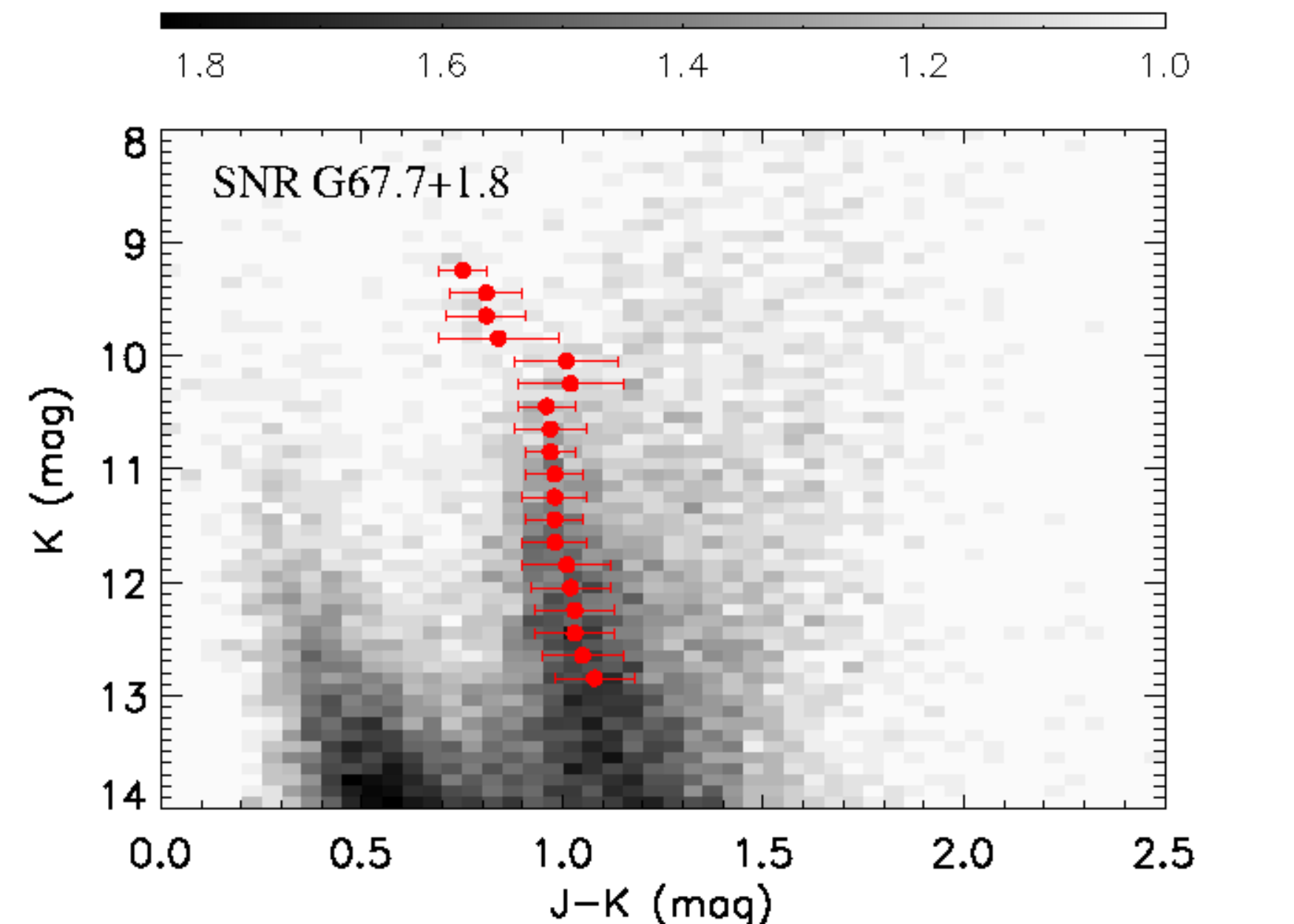}&
\includegraphics[width = 0.45\textwidth]{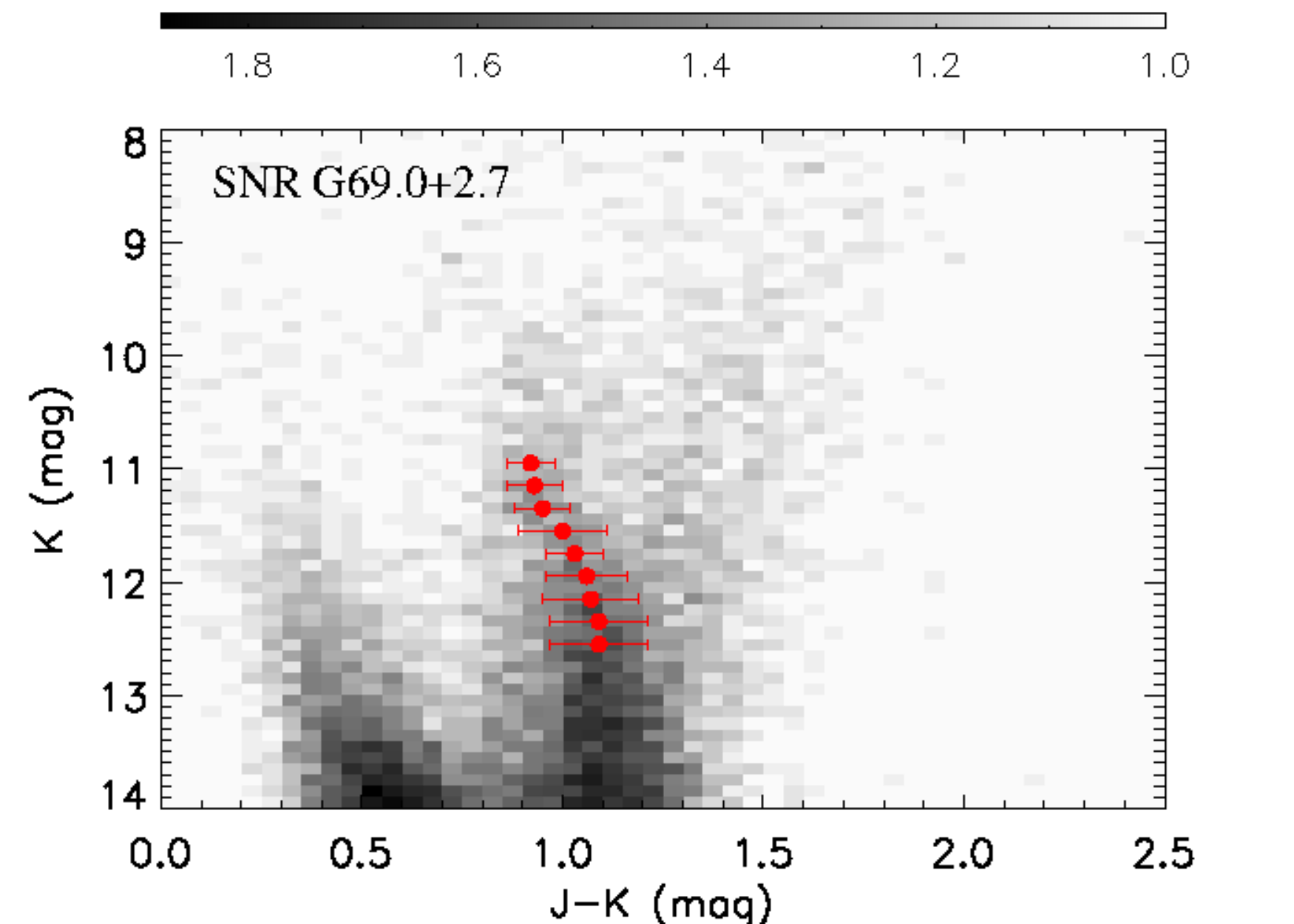}\\
\includegraphics[width = 0.45\textwidth]{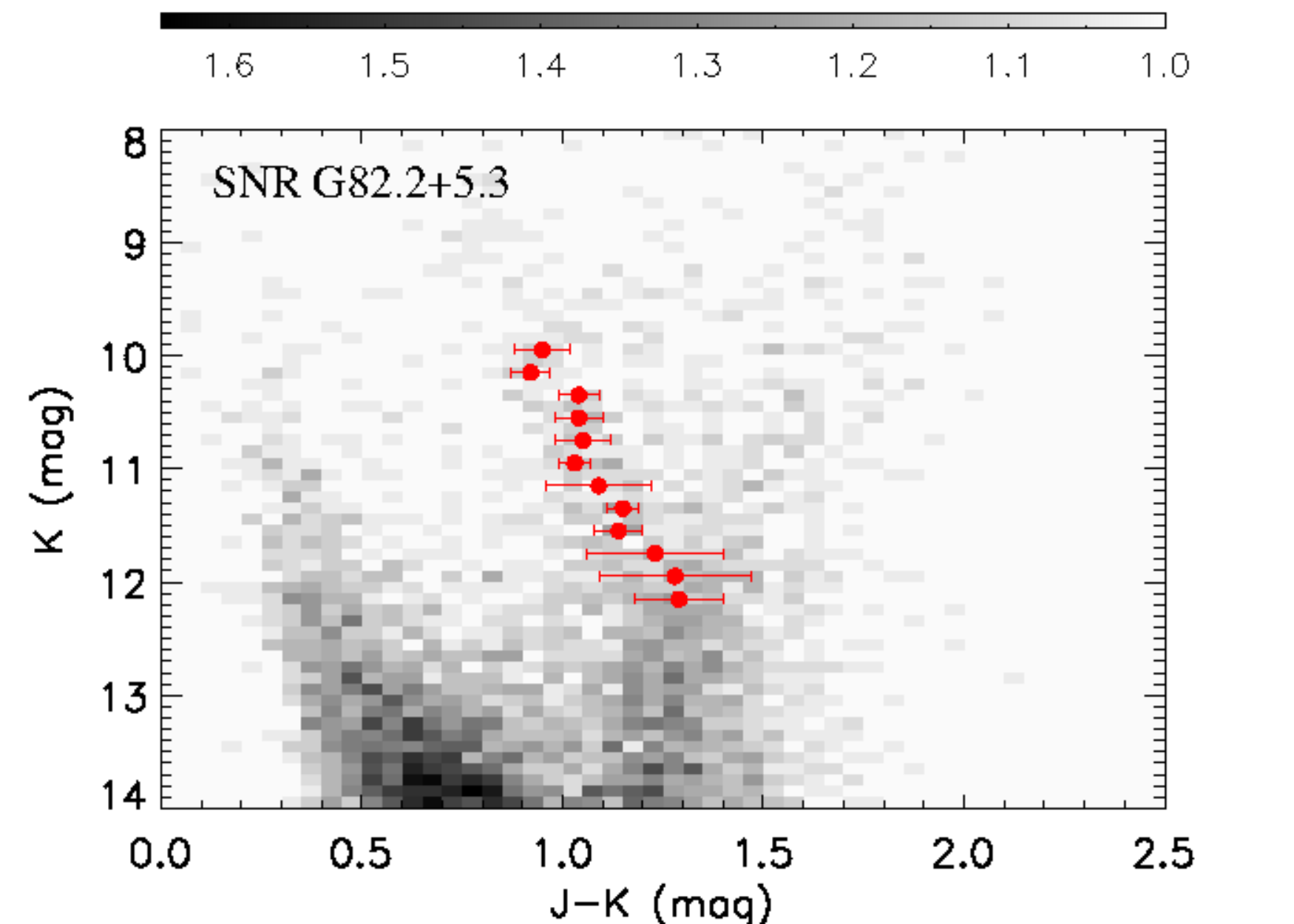}&
\includegraphics[width = 0.45\textwidth]{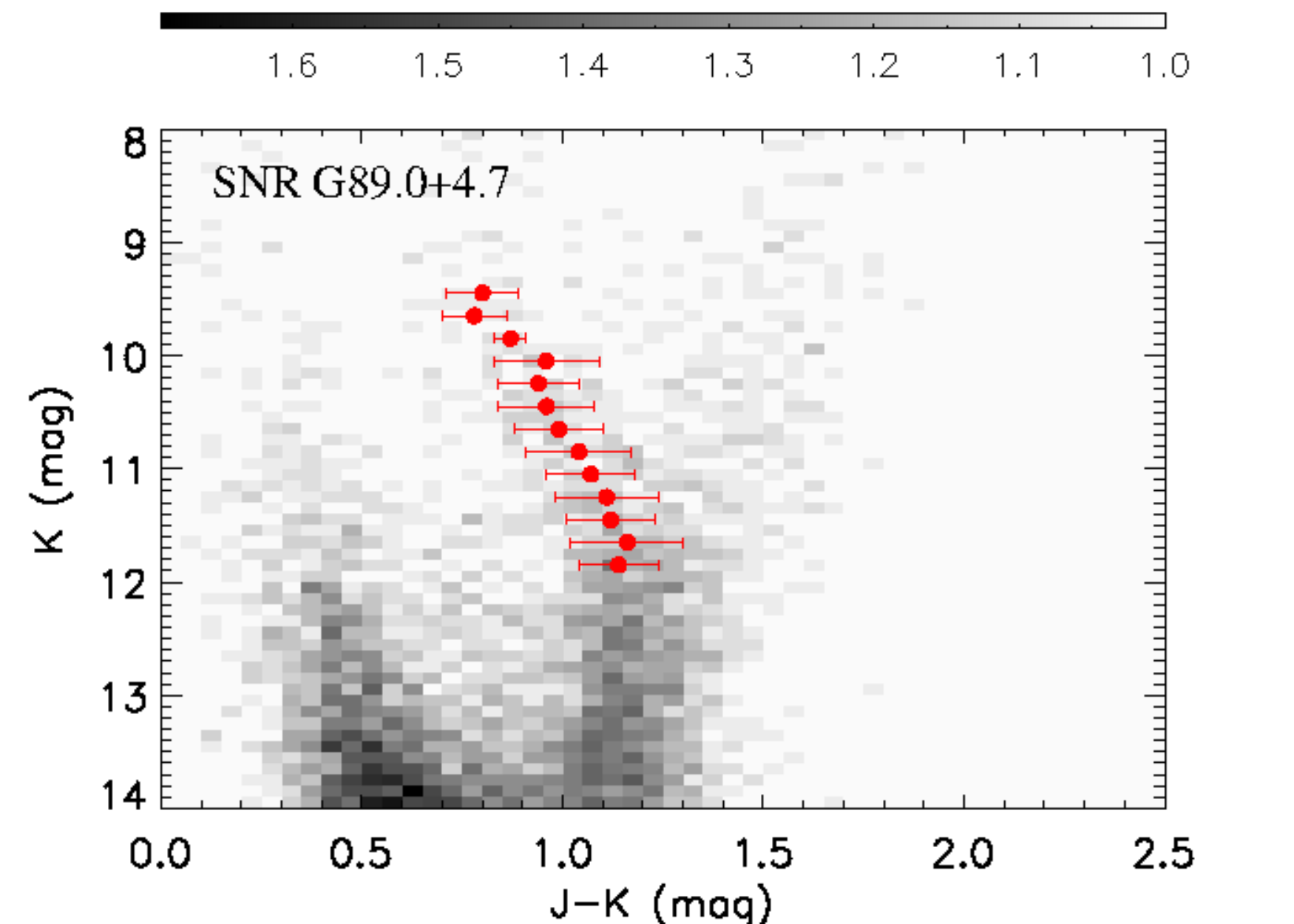}\\
\end{tabular}
\end{figure*}

\begin{figure*}
\begin{tabular}{cc}
\includegraphics[width = 0.45\textwidth]{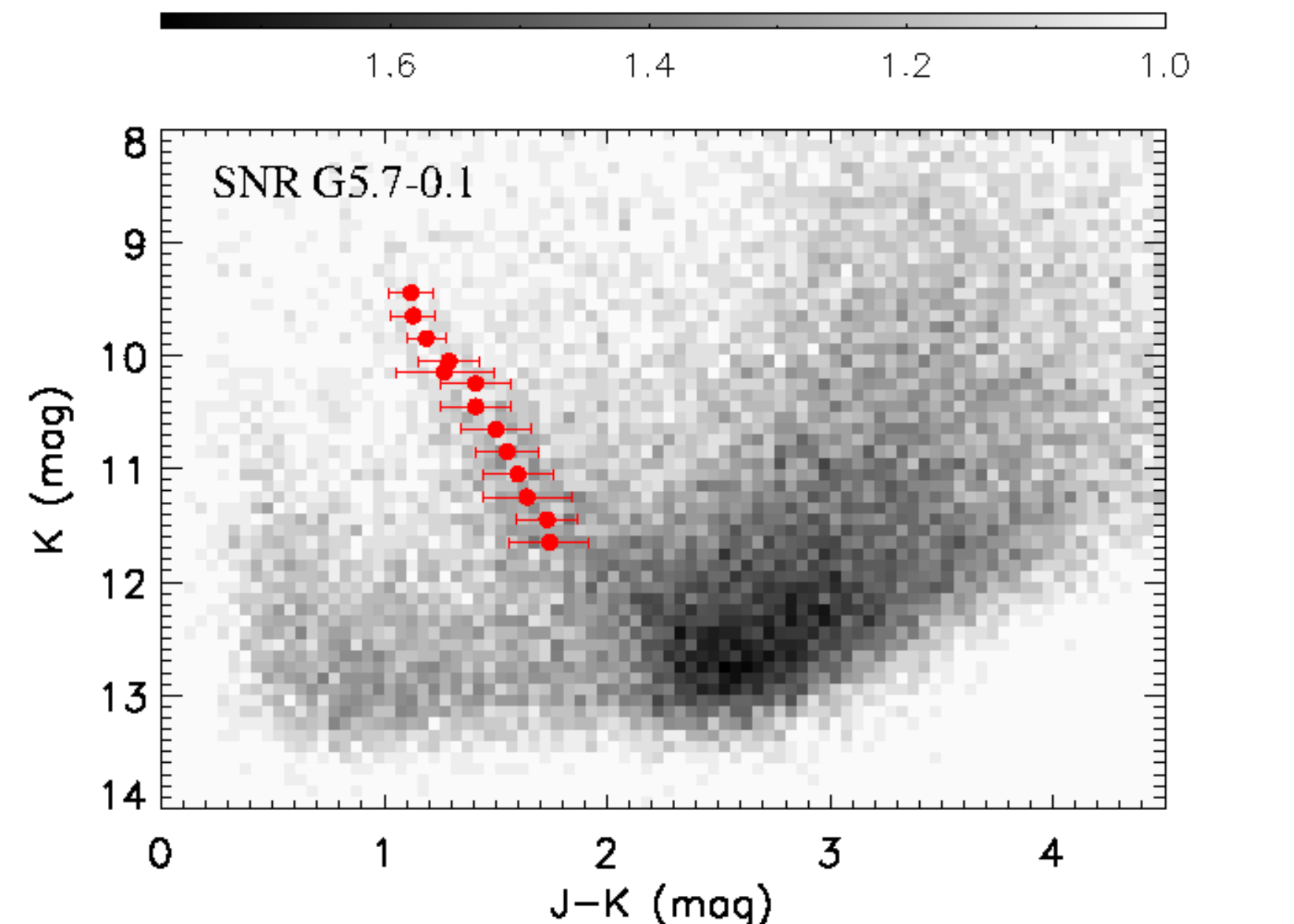}&
\includegraphics[width = 0.45\textwidth]{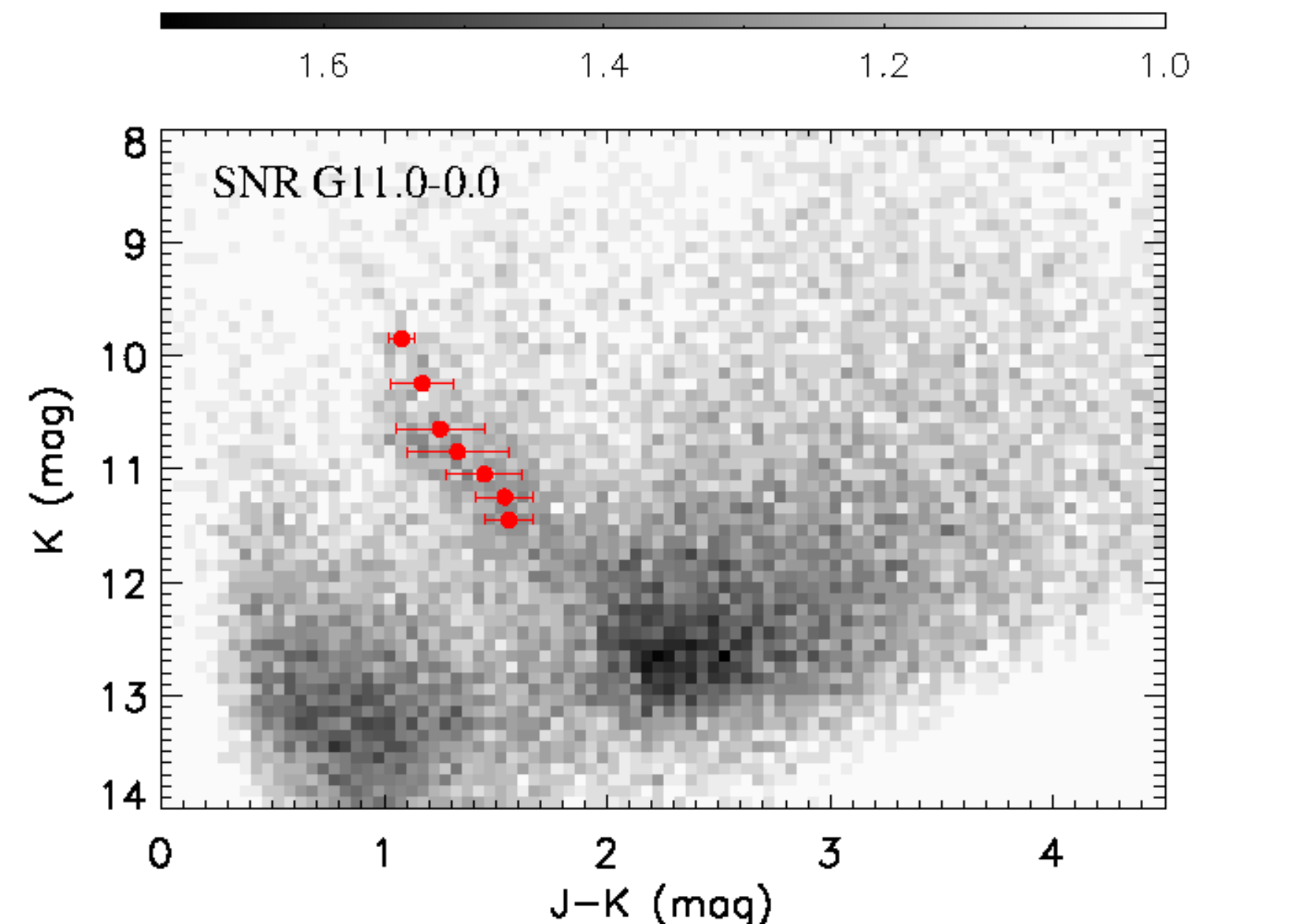}\\
\includegraphics[width = 0.45\textwidth]{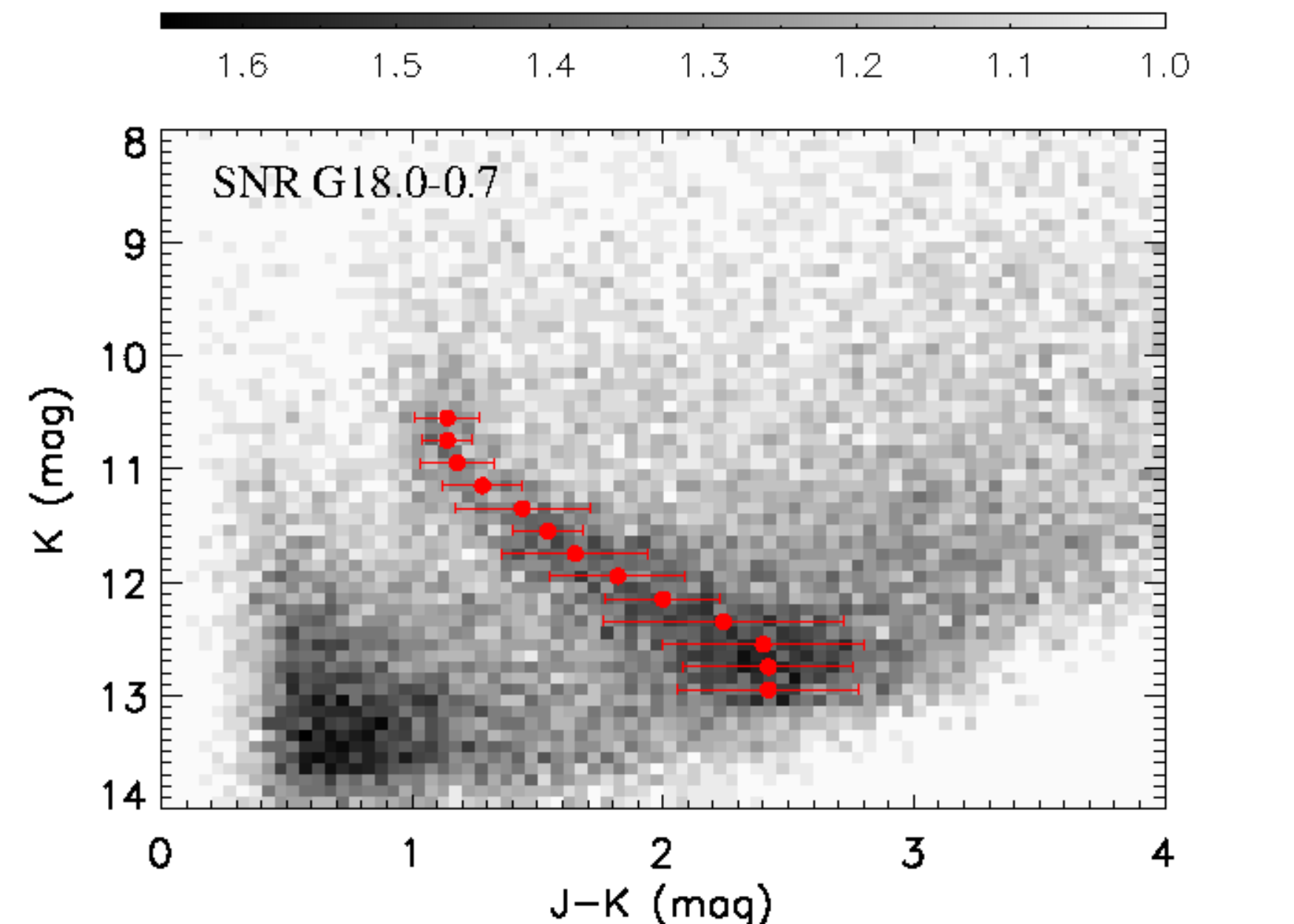}&
\includegraphics[width = 0.45\textwidth]{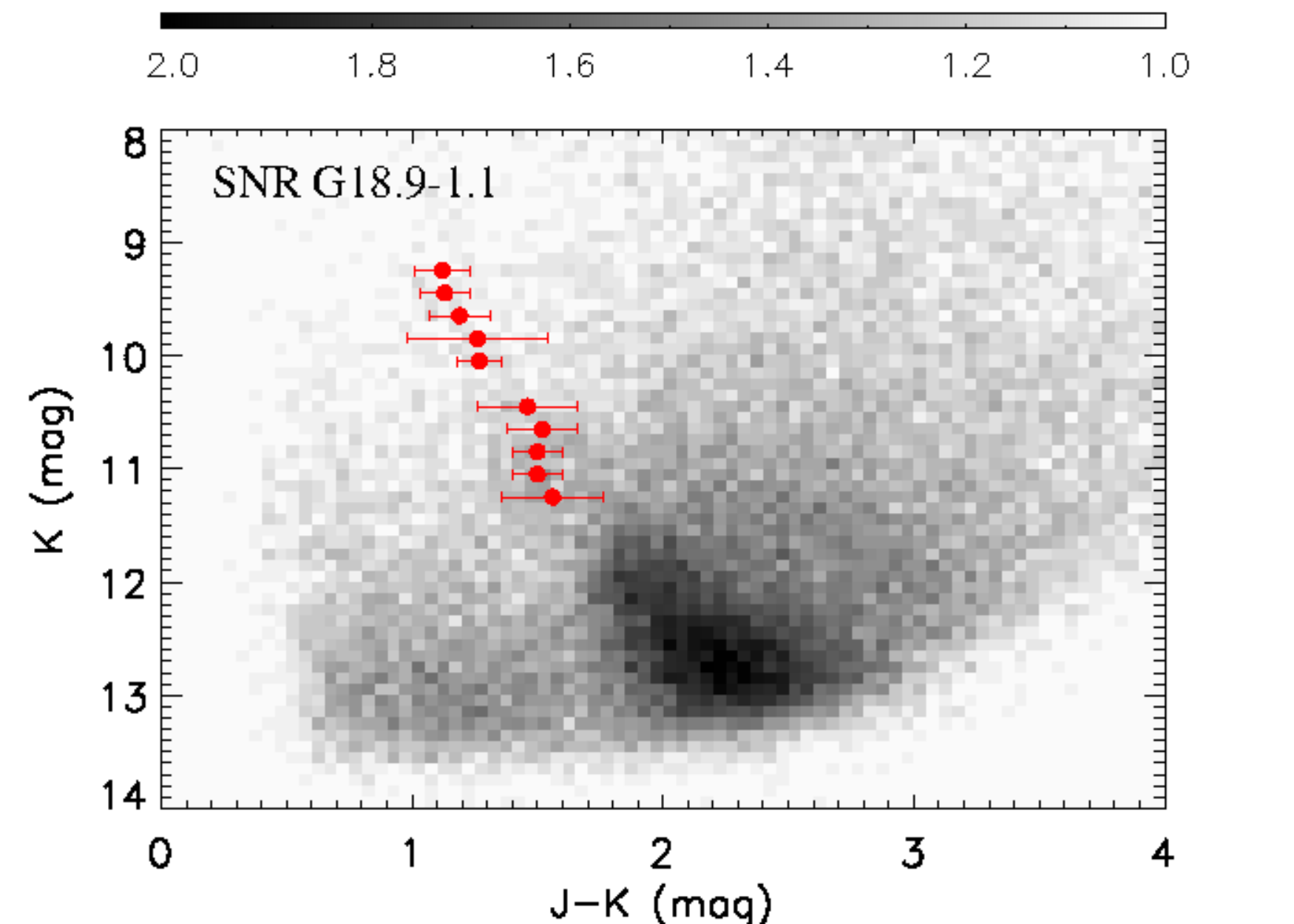}\\
\includegraphics[width = 0.45\textwidth]{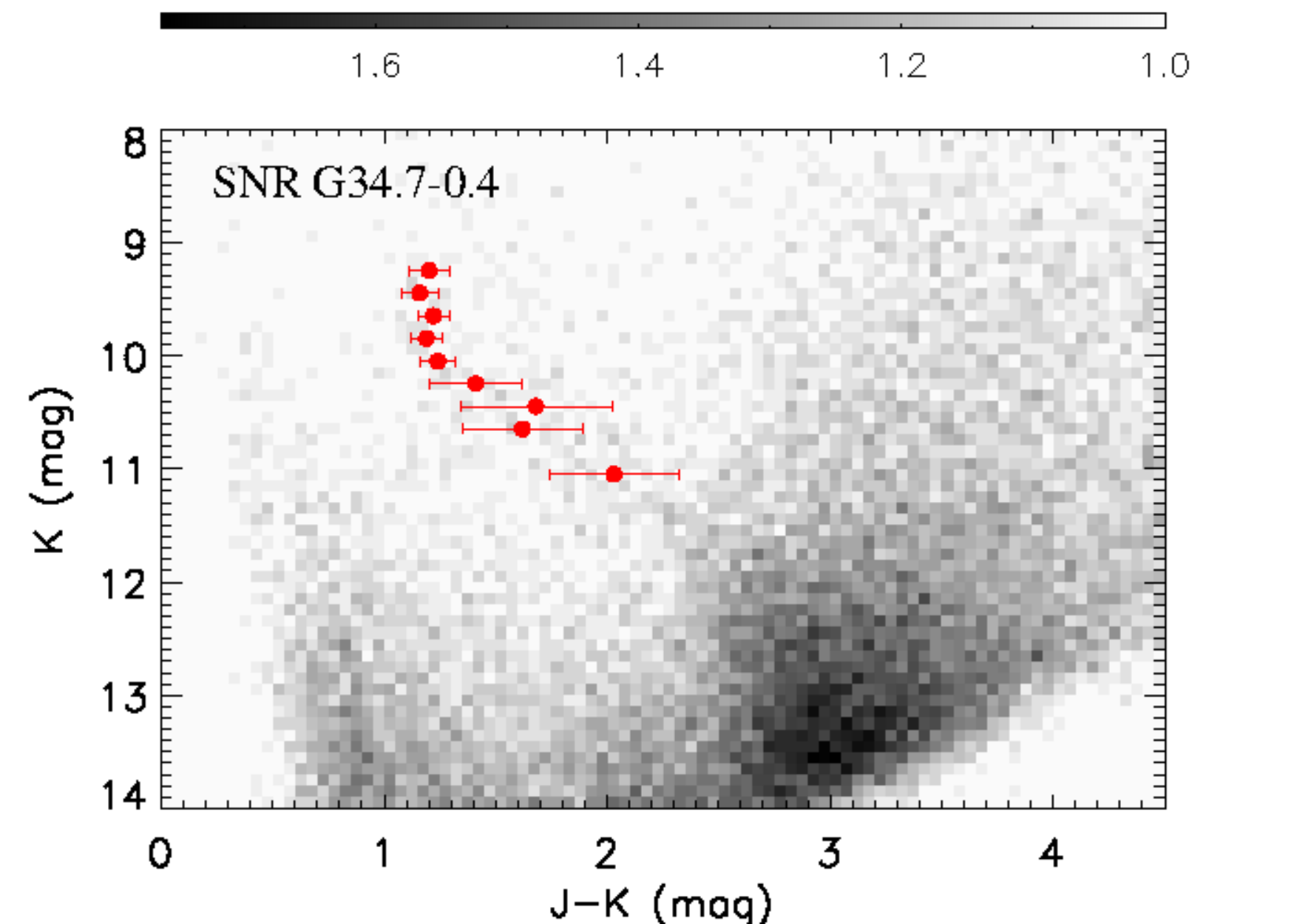}&
\includegraphics[width = 0.45\textwidth]{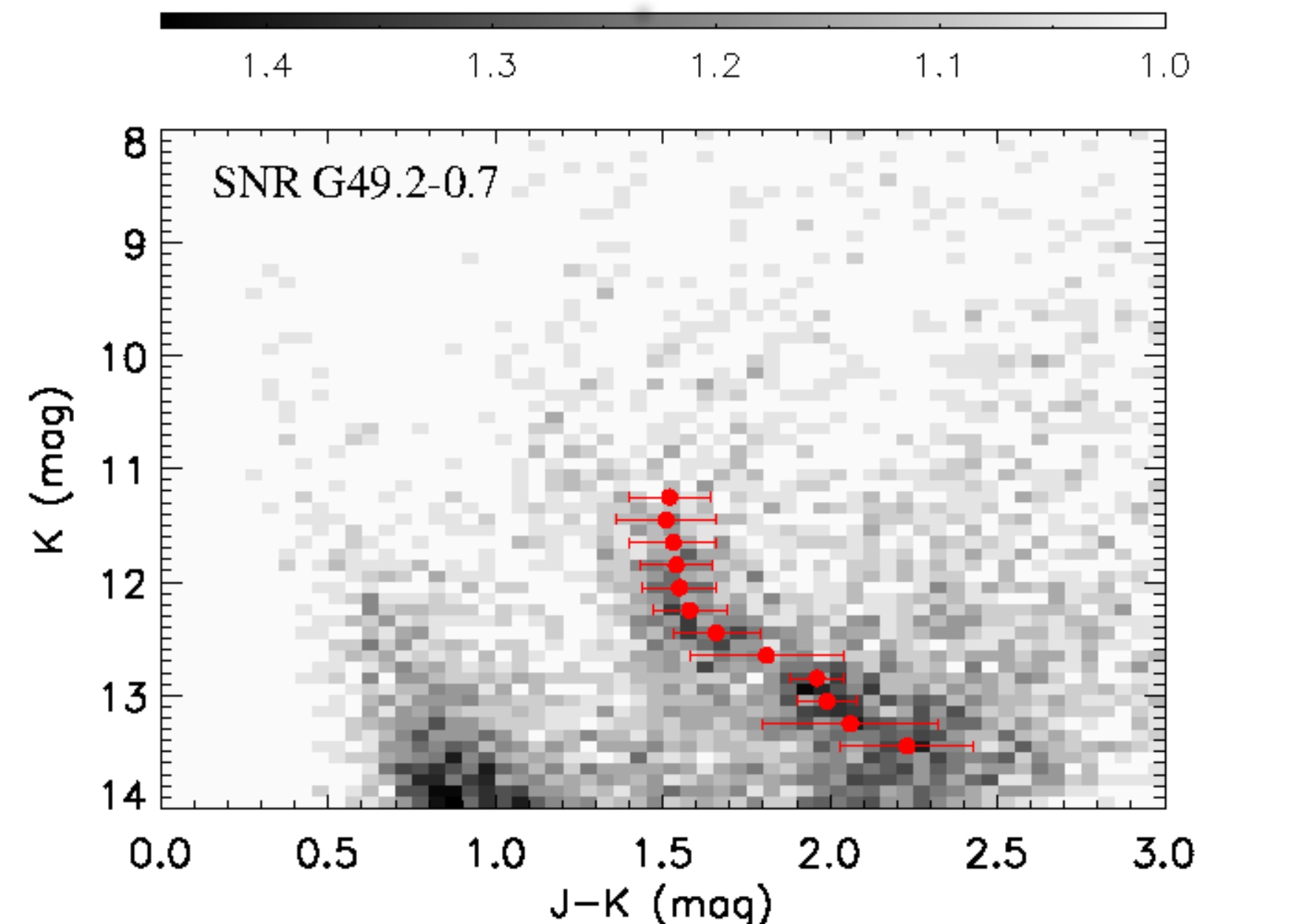}\\
\includegraphics[width = 0.45\textwidth]{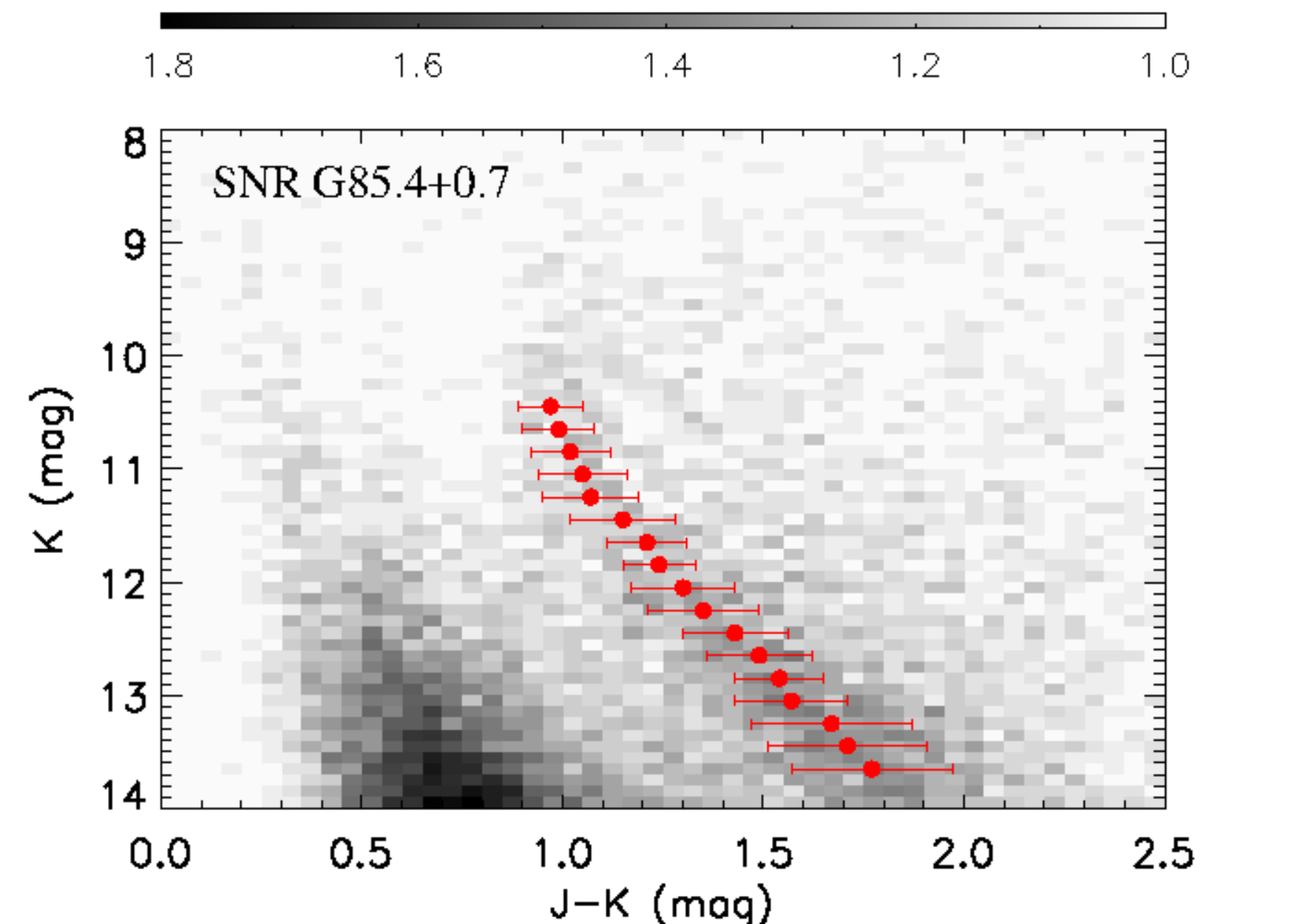}&
\end{tabular}
\end{figure*}

\begin{figure*}
\caption{Left column: The  $\rm A_V$-D relation  traced by RC stars along the direction of each SNR. 
The dashed line is $\rm A_v$ value of each SNR.The dotted lines are the uncertainties of $\rm A_v$. Right column: Probability distribution over distance to the SNRs and the best-fit Gaussian model with the cutoffs.}
\label{fig6}
\begin{tabular}{cc}
\includegraphics[width = 0.45\textwidth]{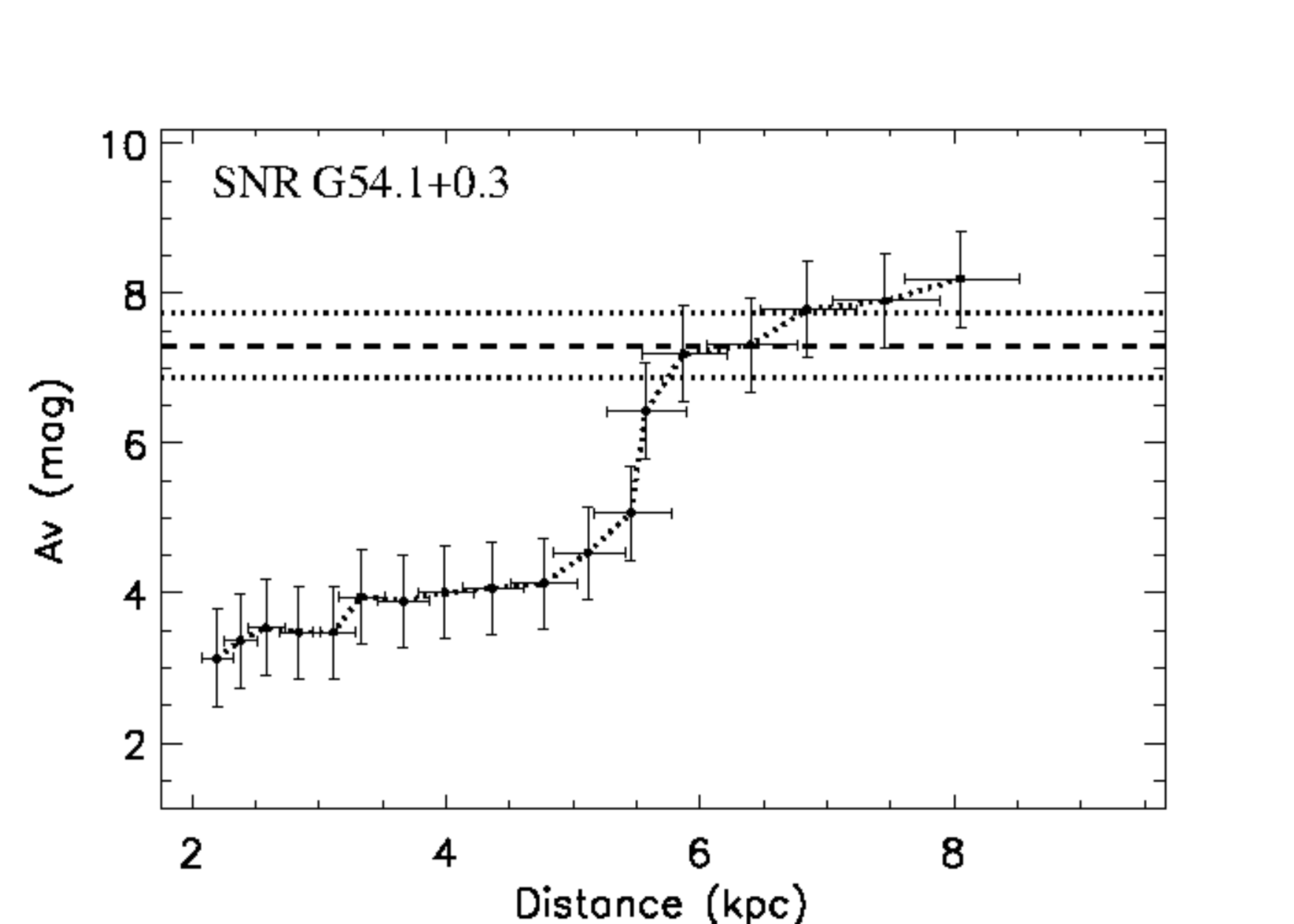}&
\includegraphics[width = 0.45\textwidth]{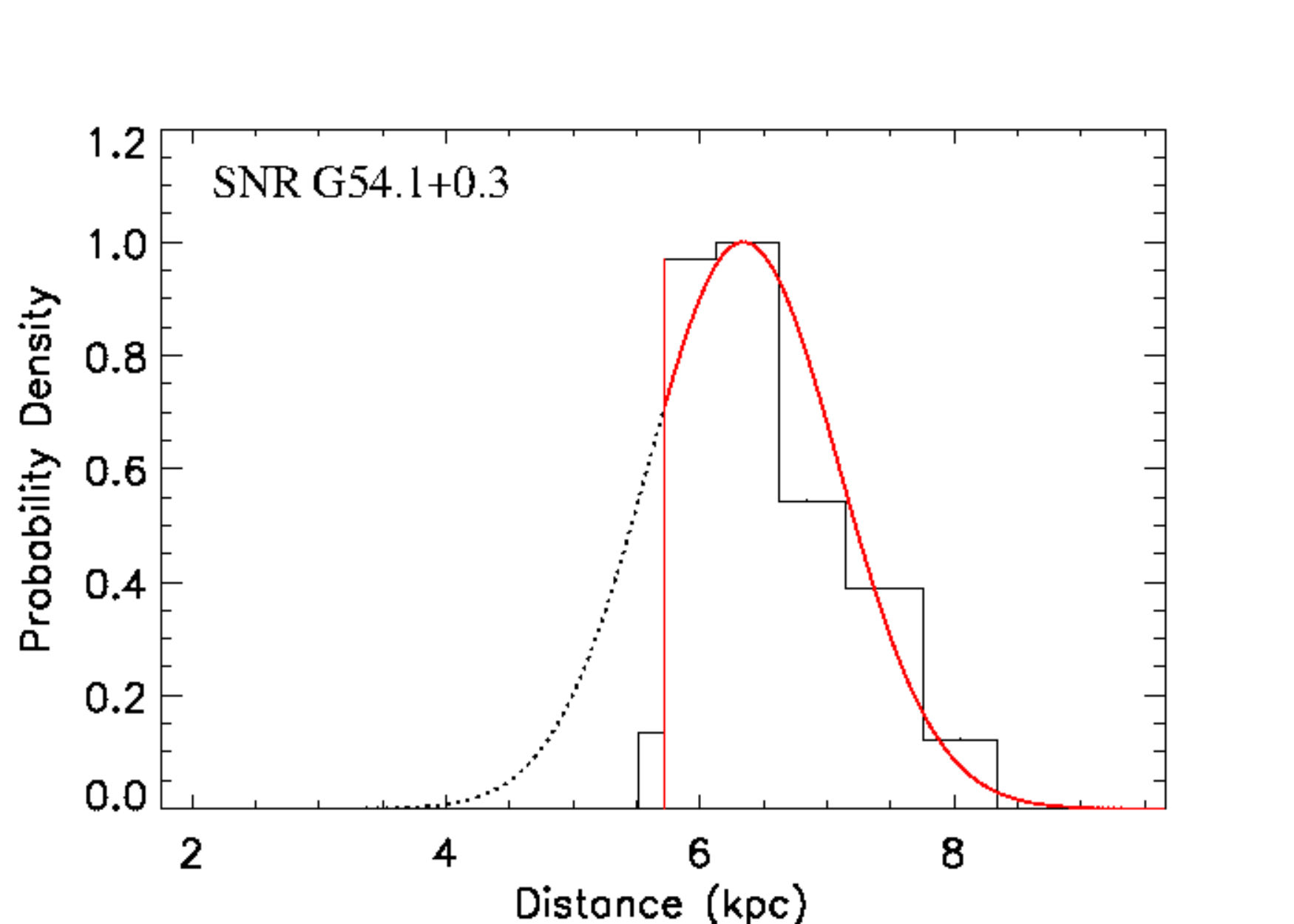}\\
\includegraphics[width = 0.45\textwidth]{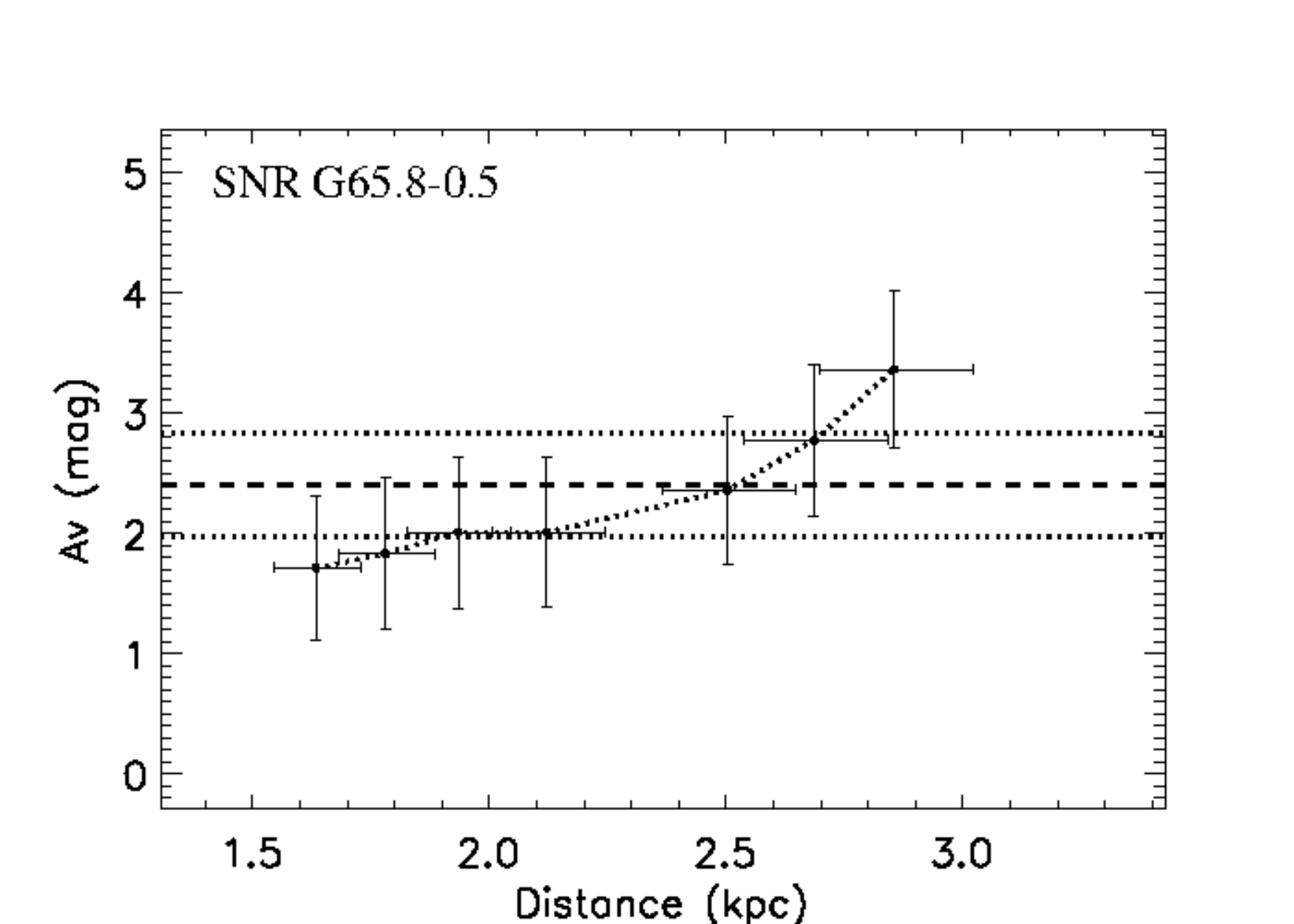}&
\includegraphics[width = 0.45\textwidth]{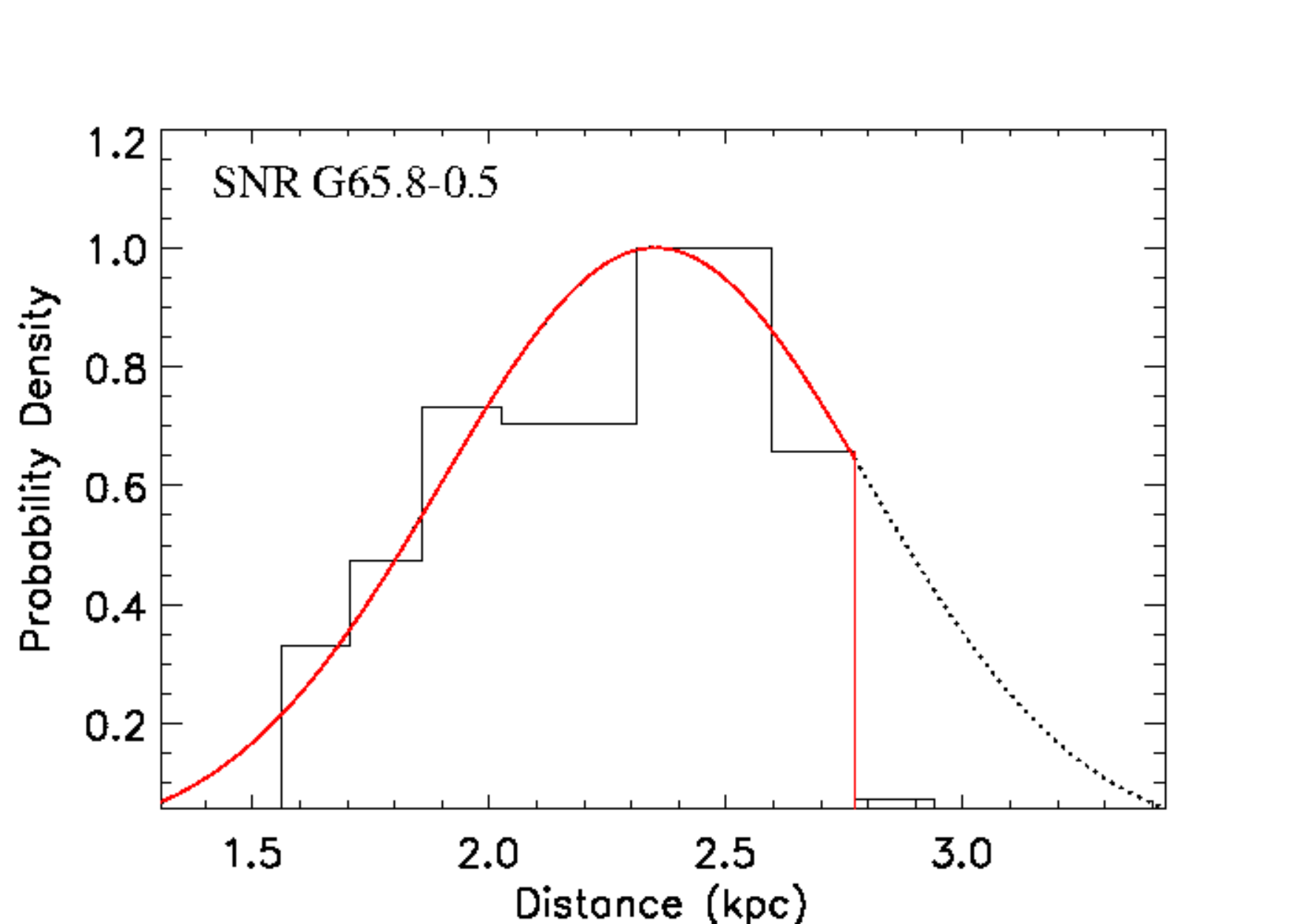}\\
\includegraphics[width = 0.45\textwidth]{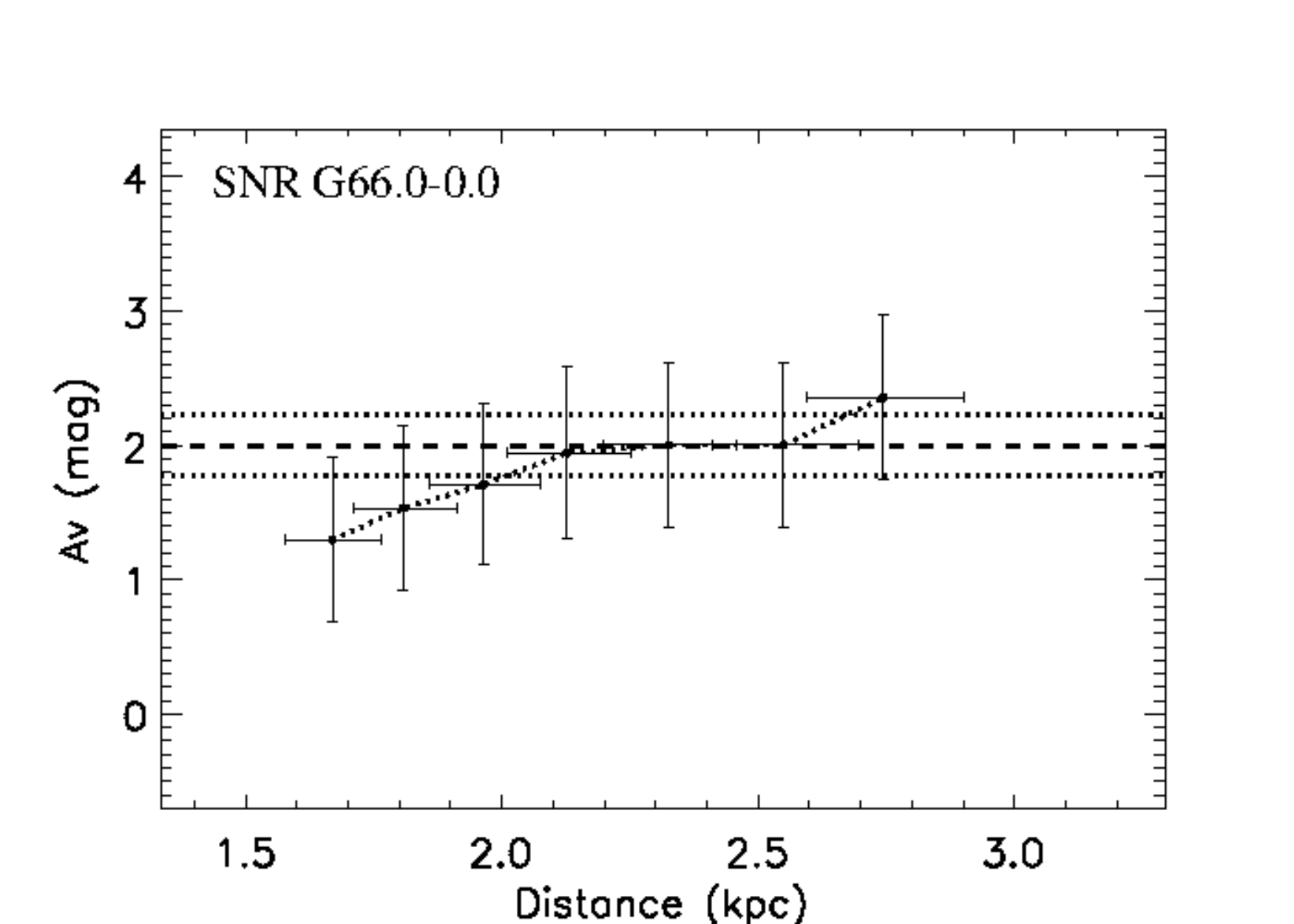}&
\includegraphics[width = 0.45\textwidth]{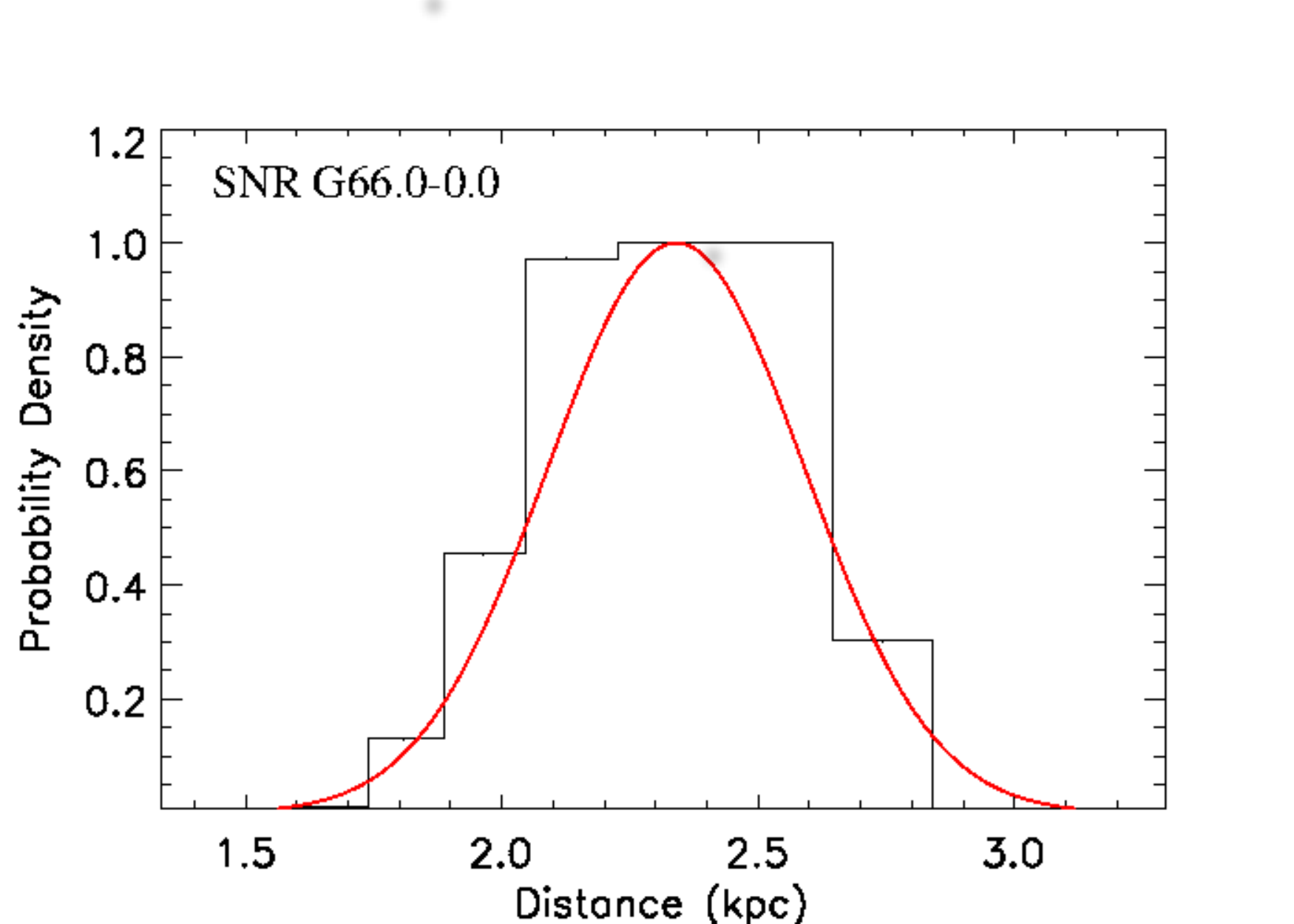}\\
\includegraphics[width = 0.45\textwidth]{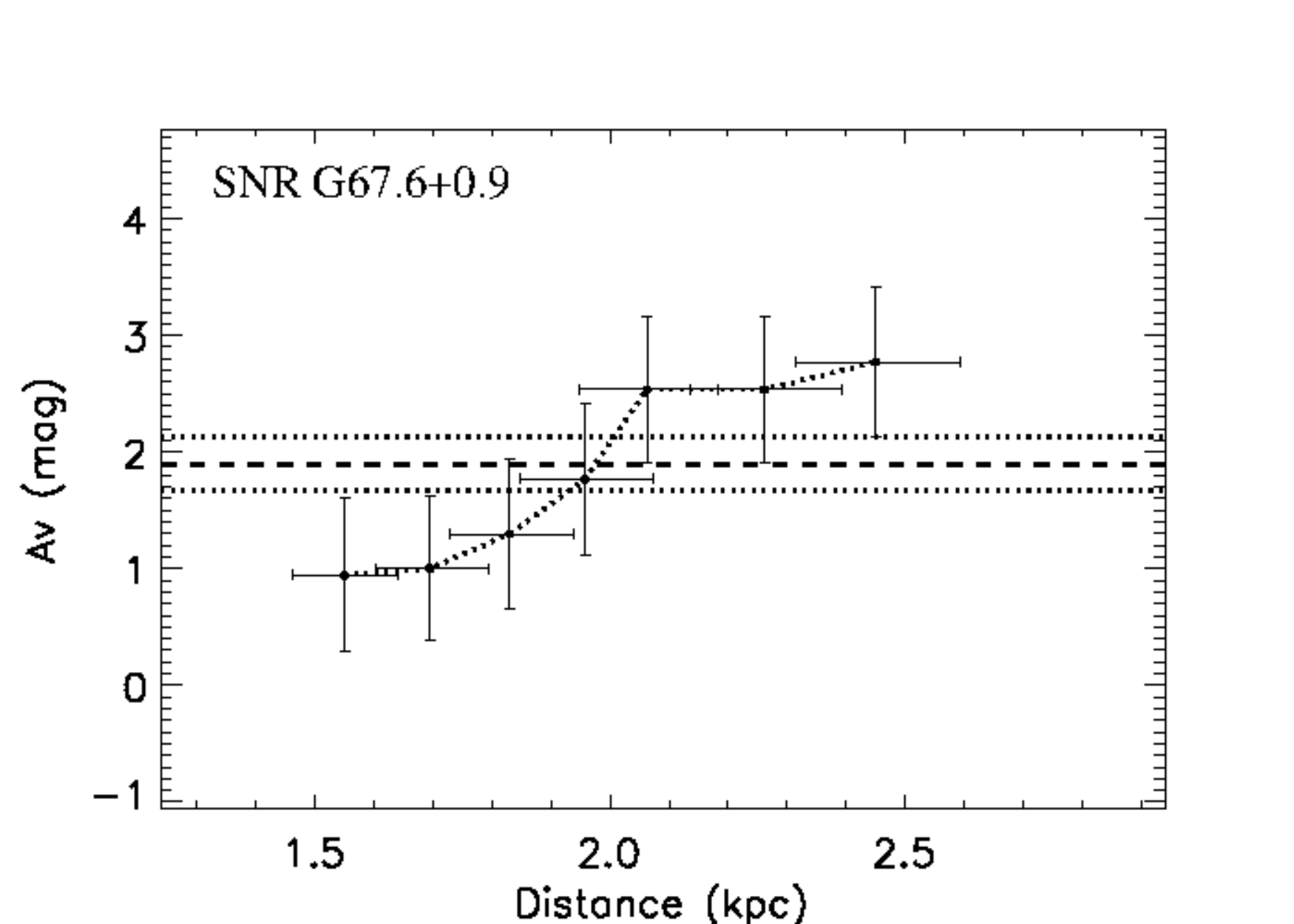}&
\includegraphics[width = 0.45\textwidth]{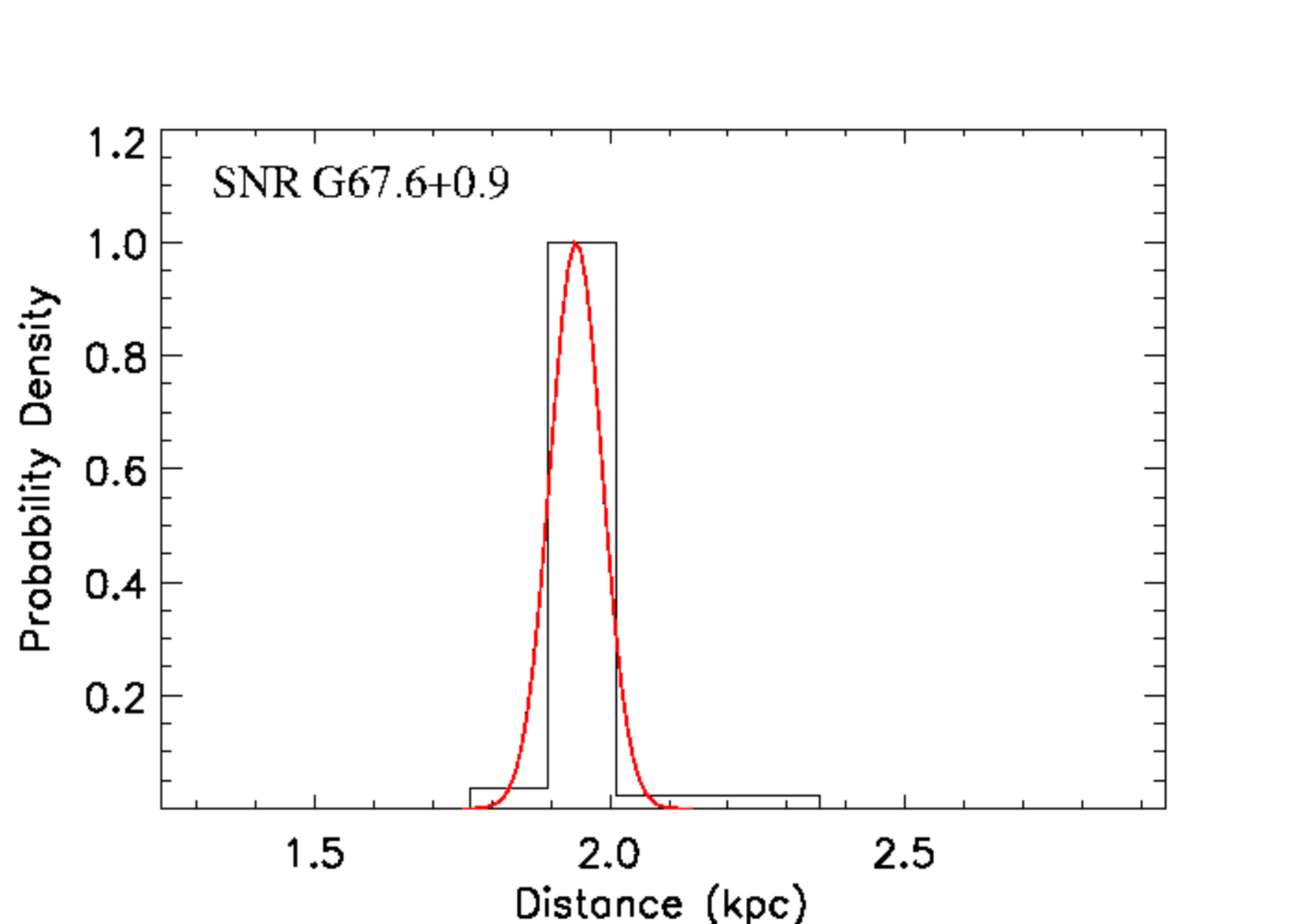}\\
\end{tabular}
\end{figure*}
\begin{figure*}
\begin{tabular}{cc}
\includegraphics[width = 0.45\textwidth]{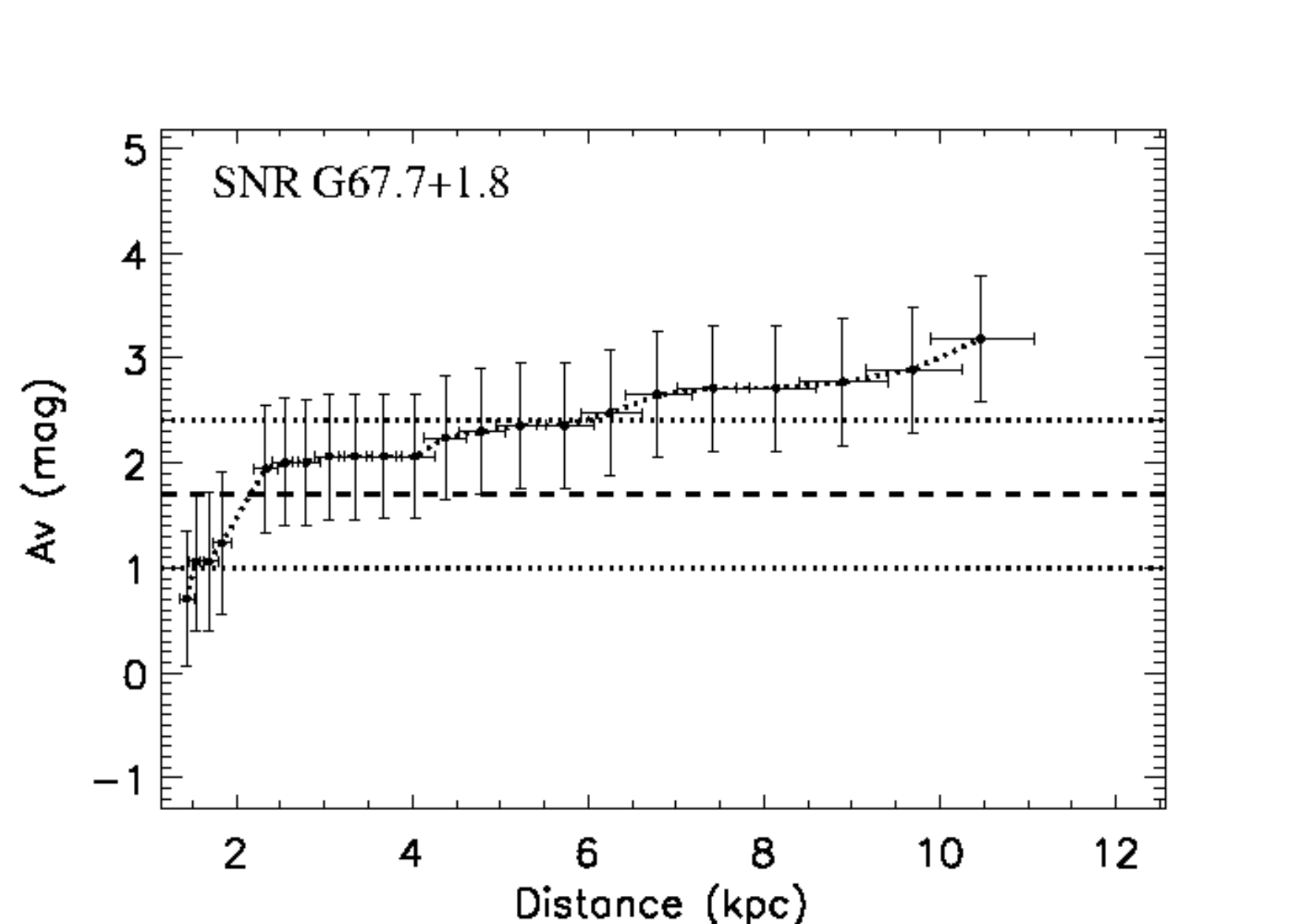}&
\includegraphics[width = 0.45\textwidth]{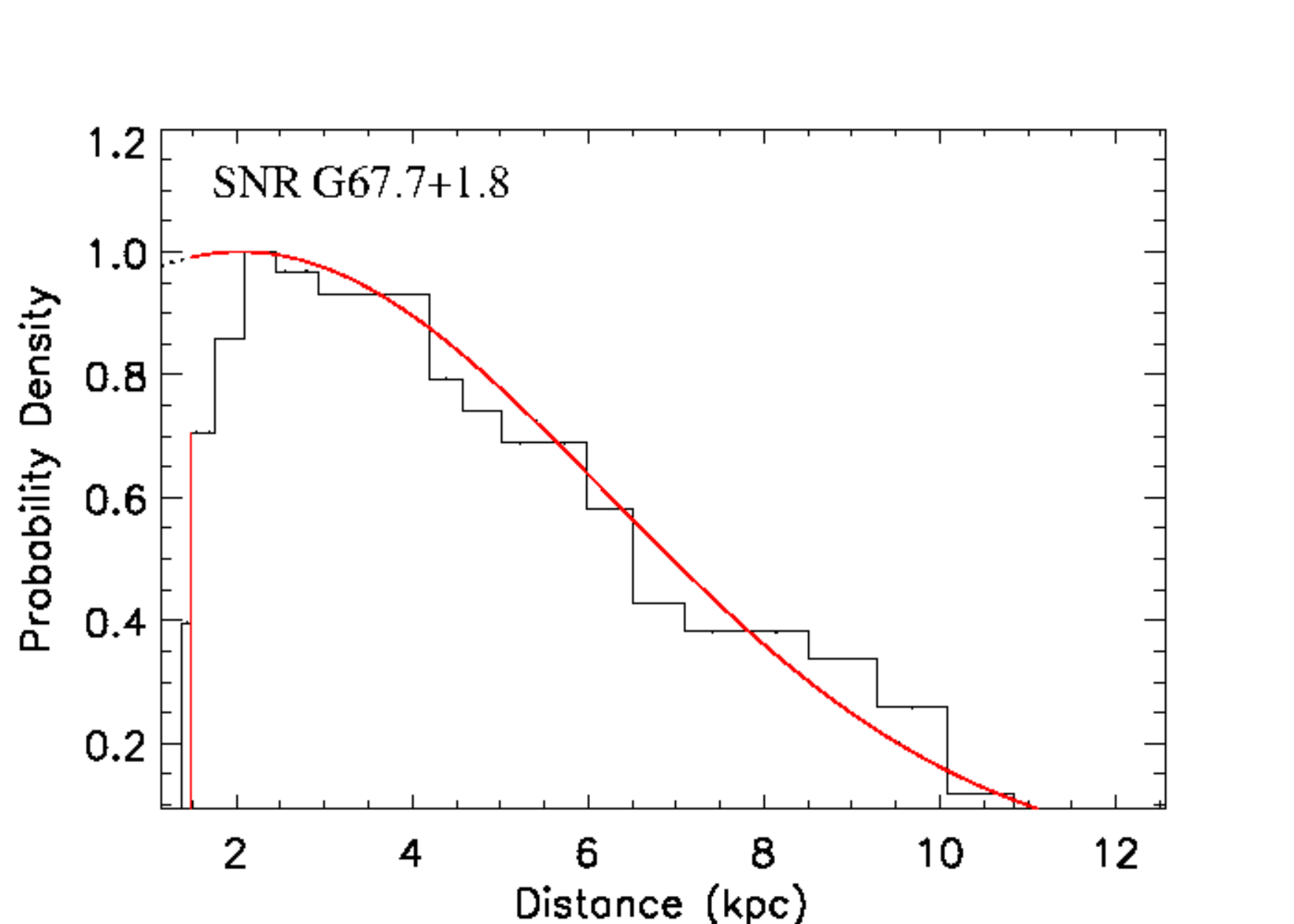}\\
\includegraphics[width = 0.45\textwidth]{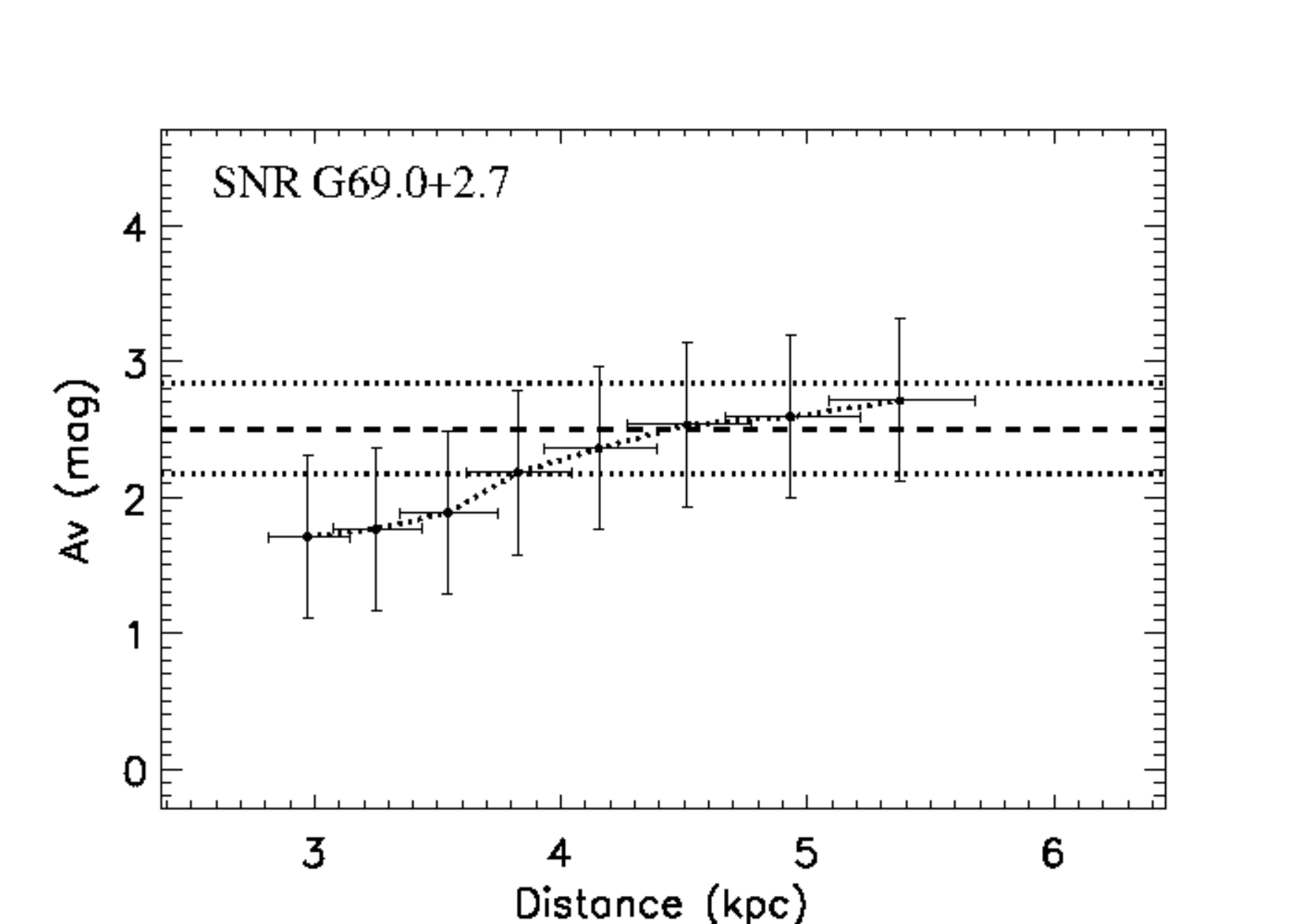}&
\includegraphics[width = 0.45\textwidth]{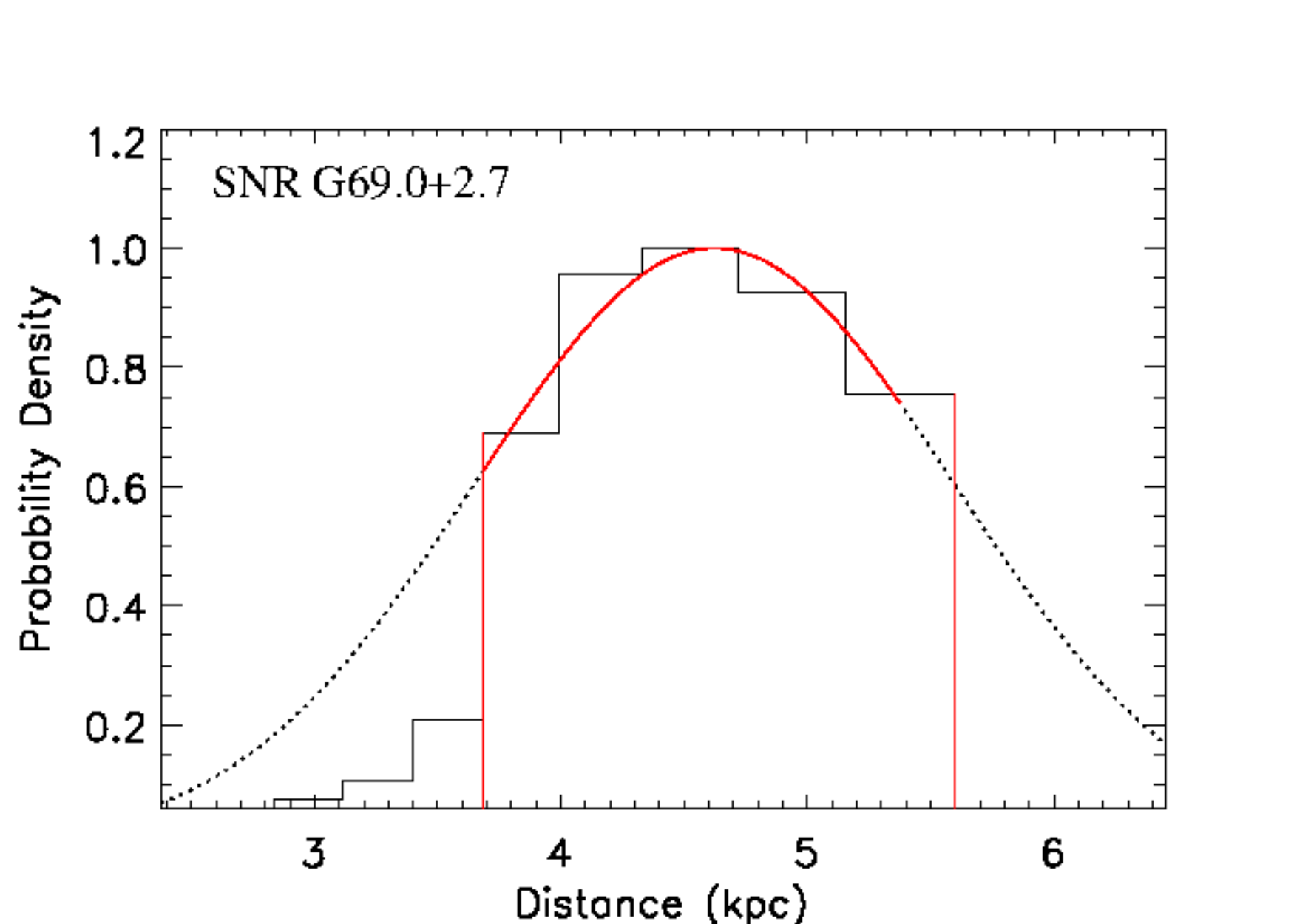}\\
\includegraphics[width = 0.45\textwidth]{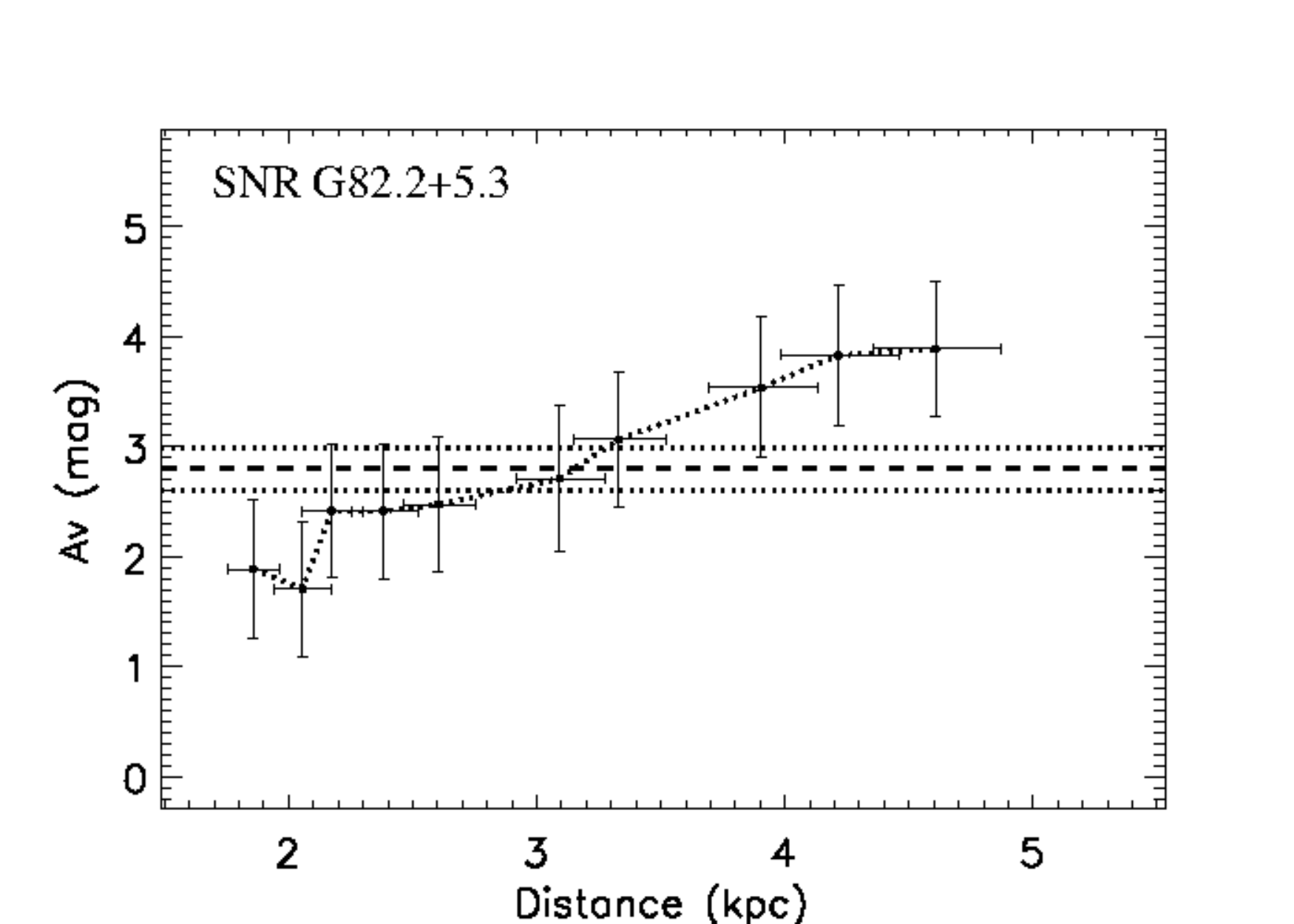}&
\includegraphics[width = 0.45\textwidth]{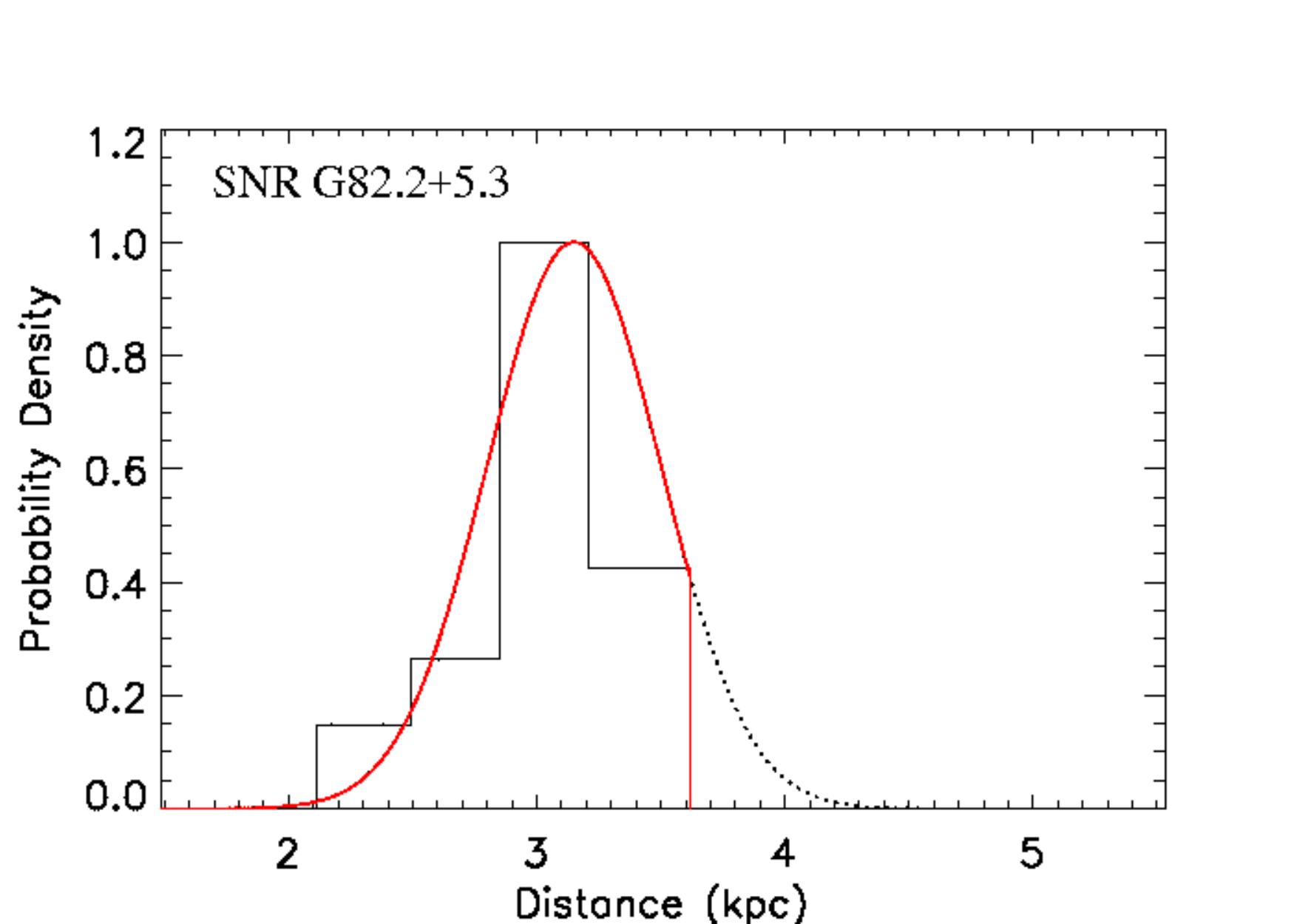}\\
\includegraphics[width = 0.45\textwidth]{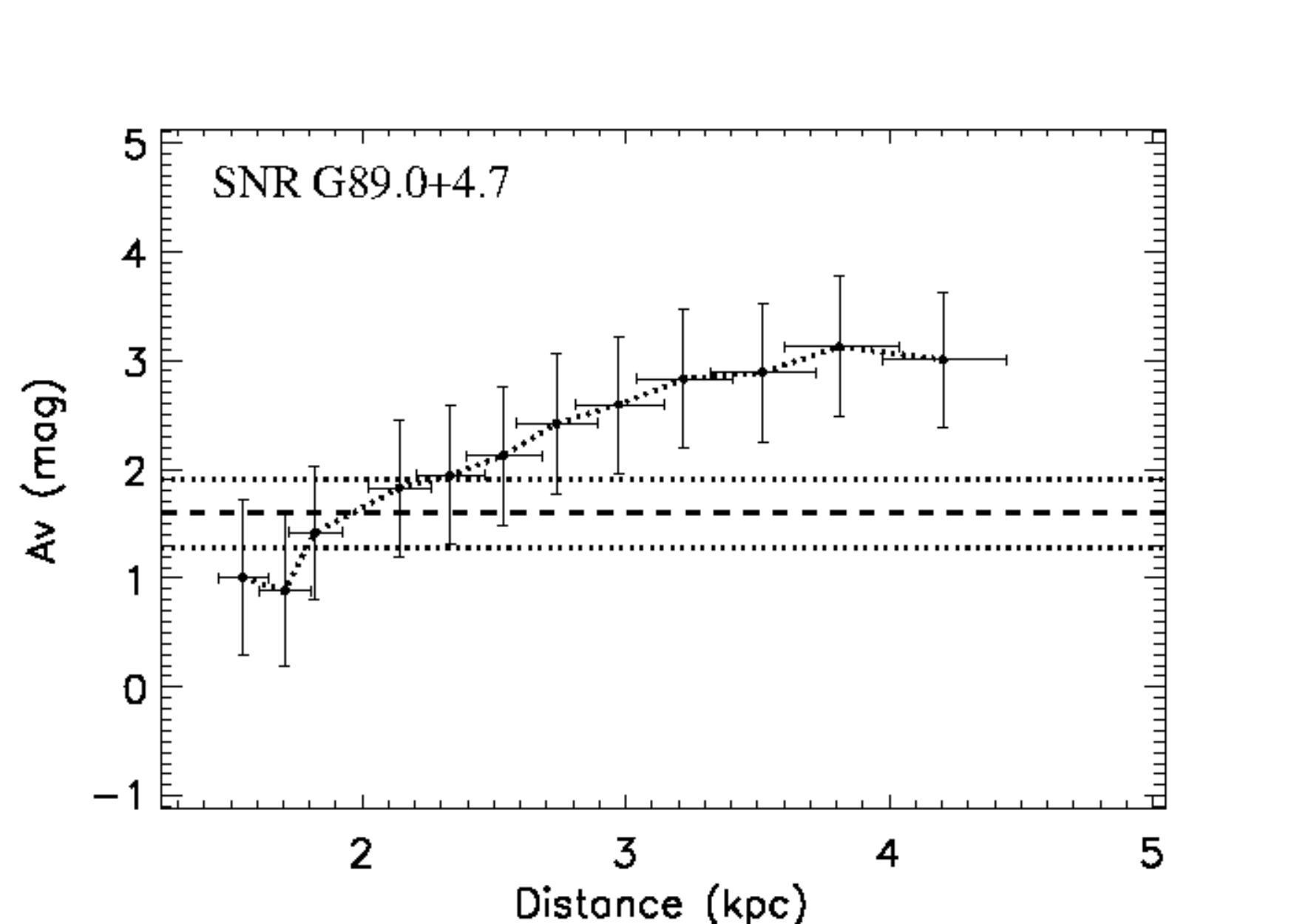}&
\includegraphics[width = 0.45\textwidth]{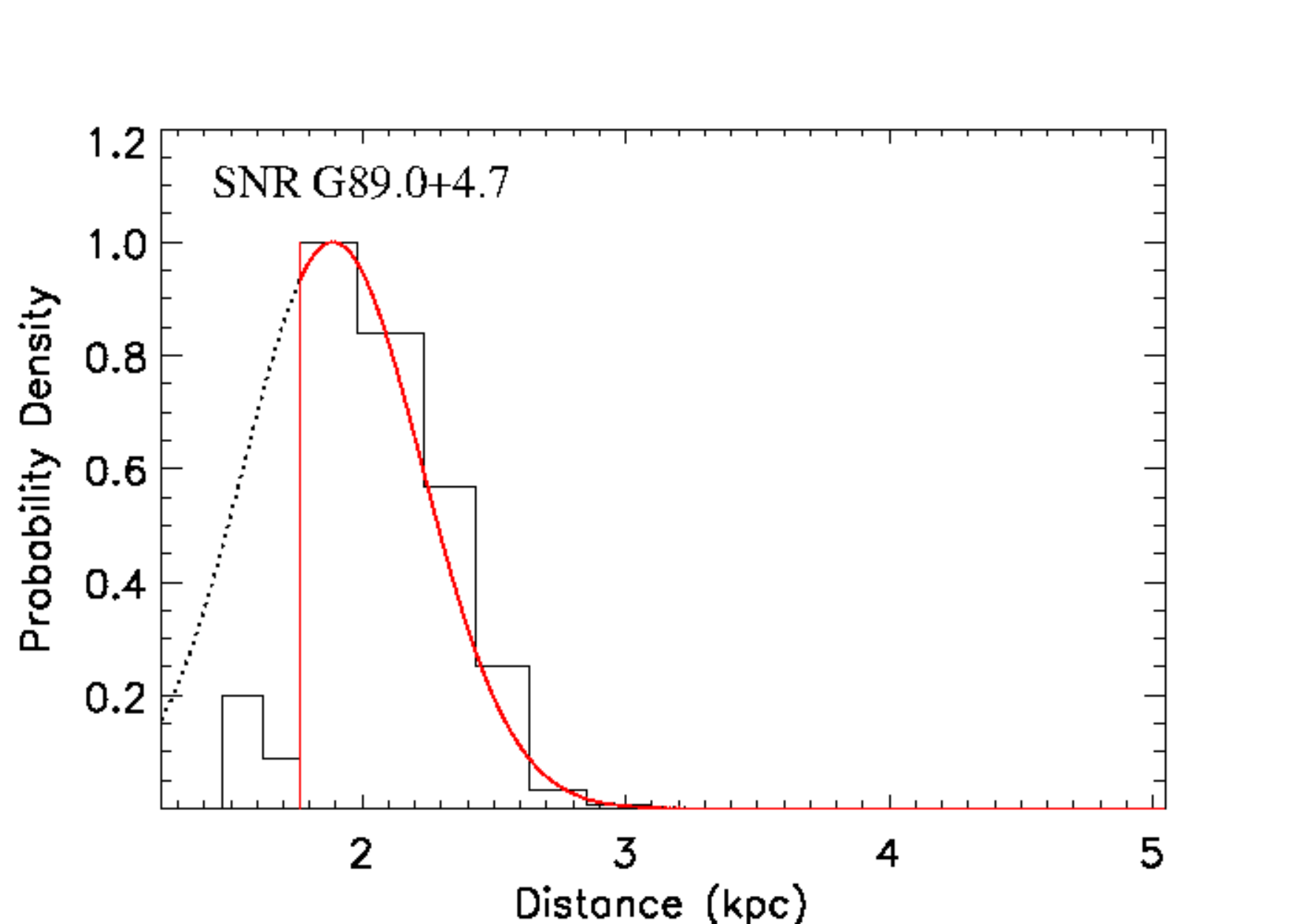}\\
\end{tabular}
\end{figure*}
\begin{figure*}
\begin{tabular}{cc}
\includegraphics[width = 0.45\textwidth]{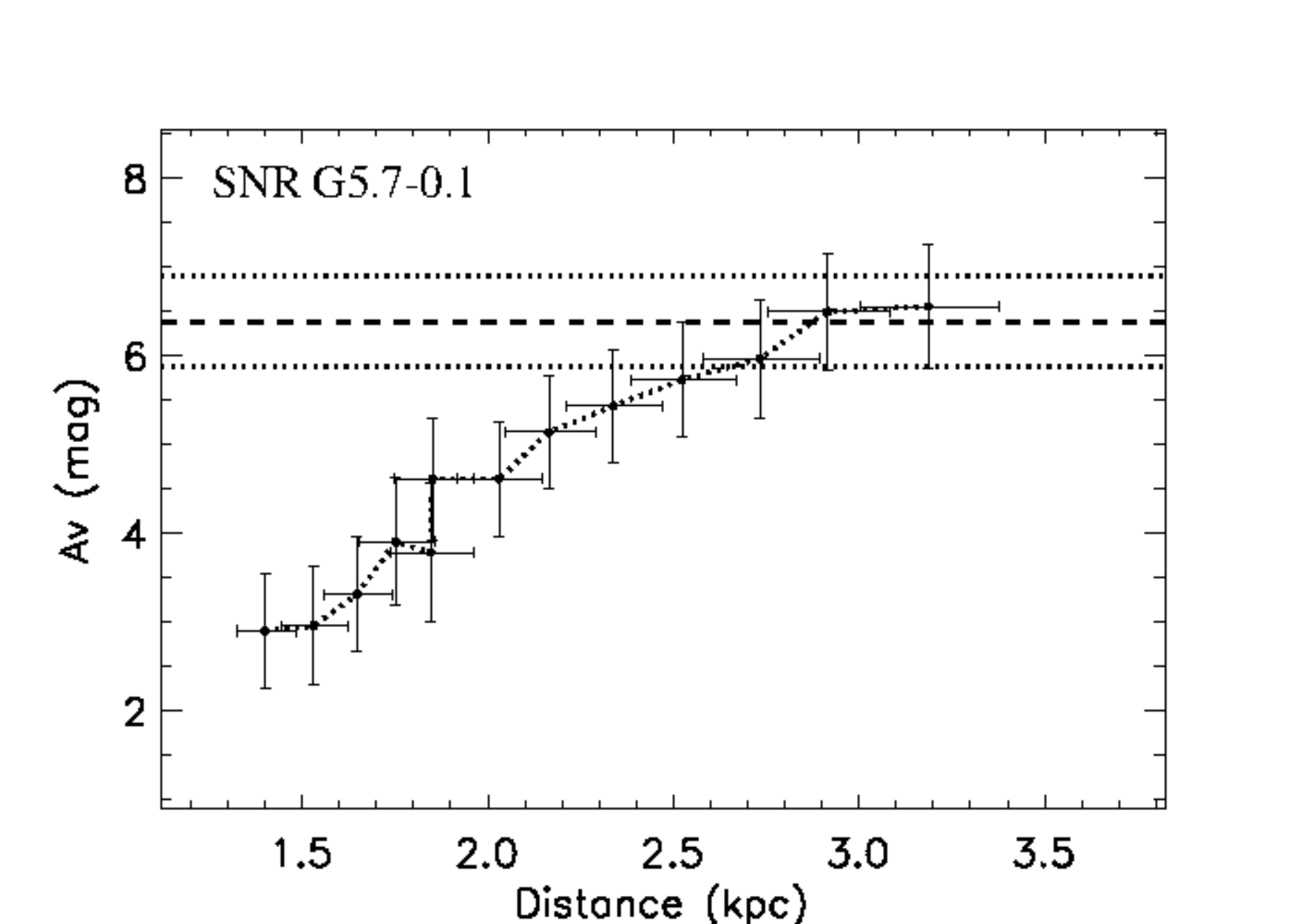}&
\includegraphics[width = 0.45\textwidth]{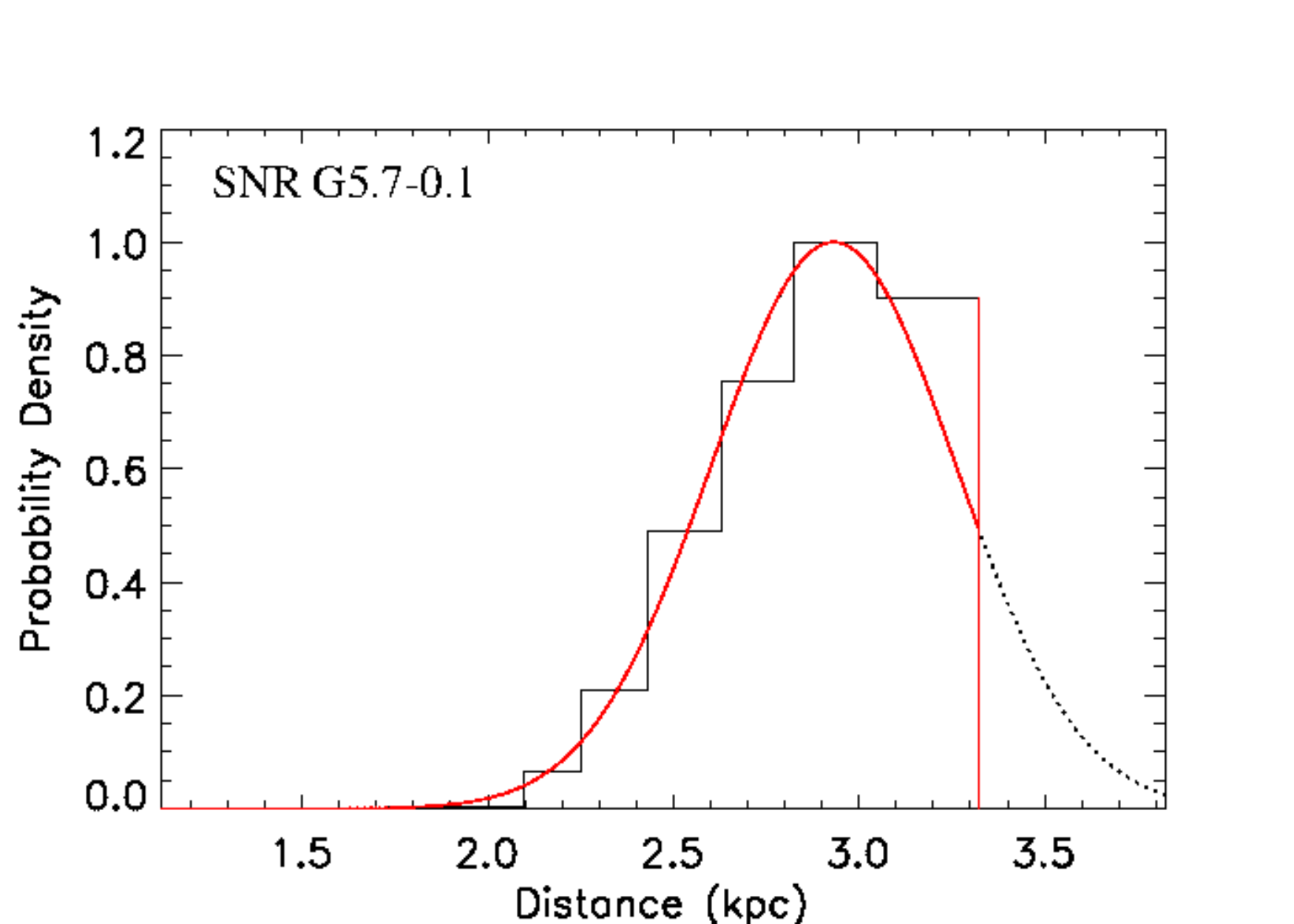}\\
\includegraphics[width = 0.45\textwidth]{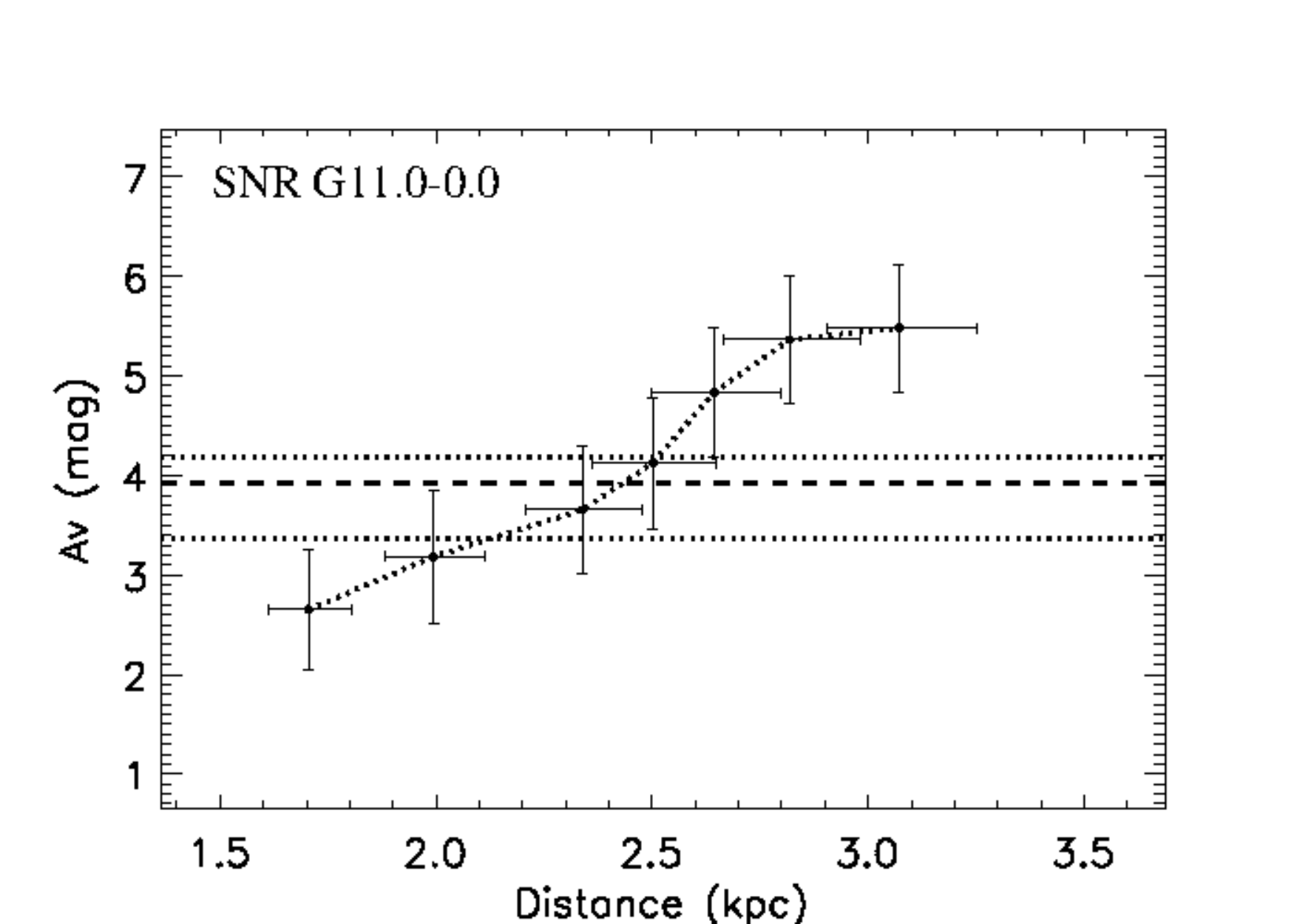}&
\includegraphics[width = 0.45\textwidth]{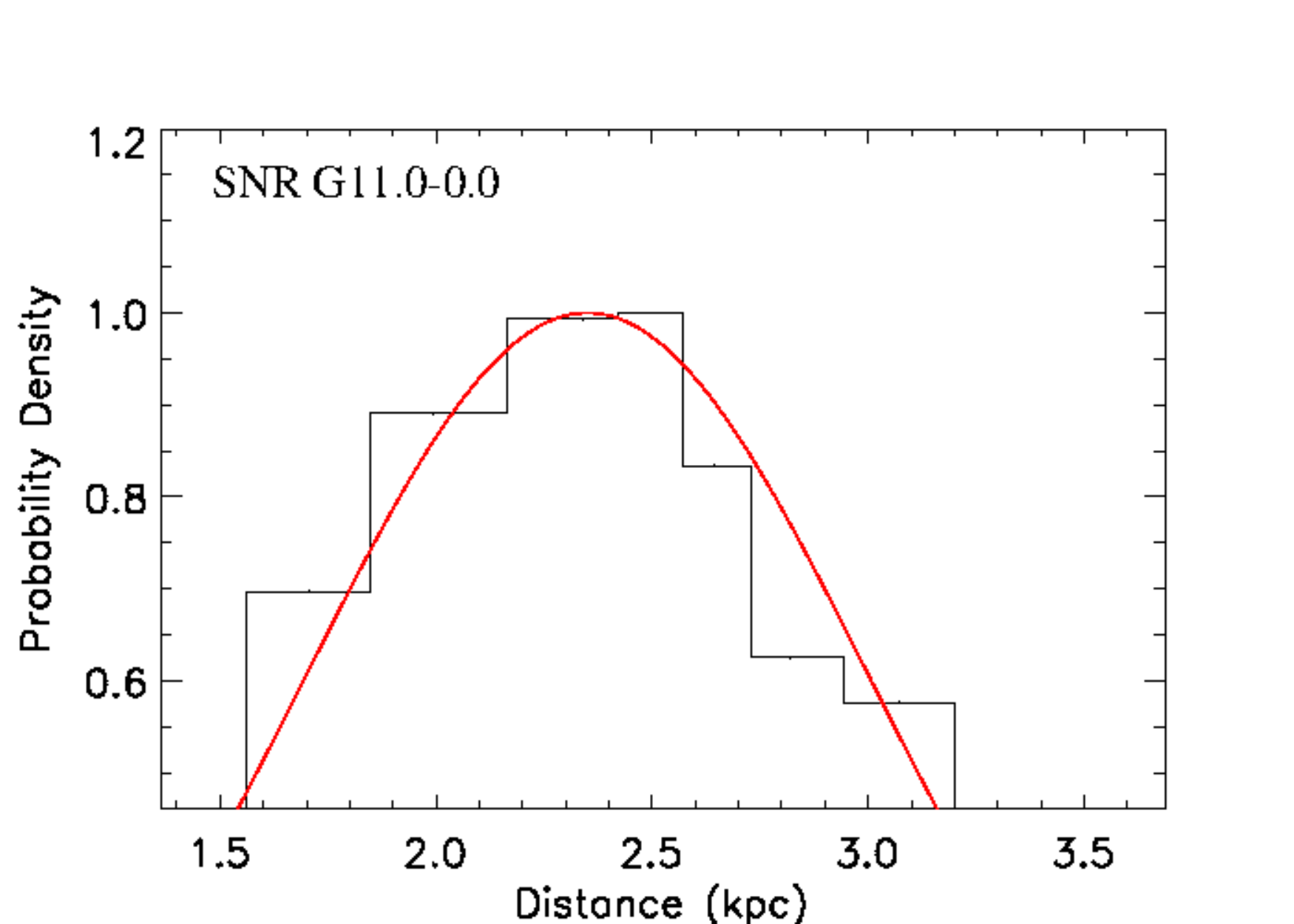}\\
\includegraphics[width = 0.45\textwidth]{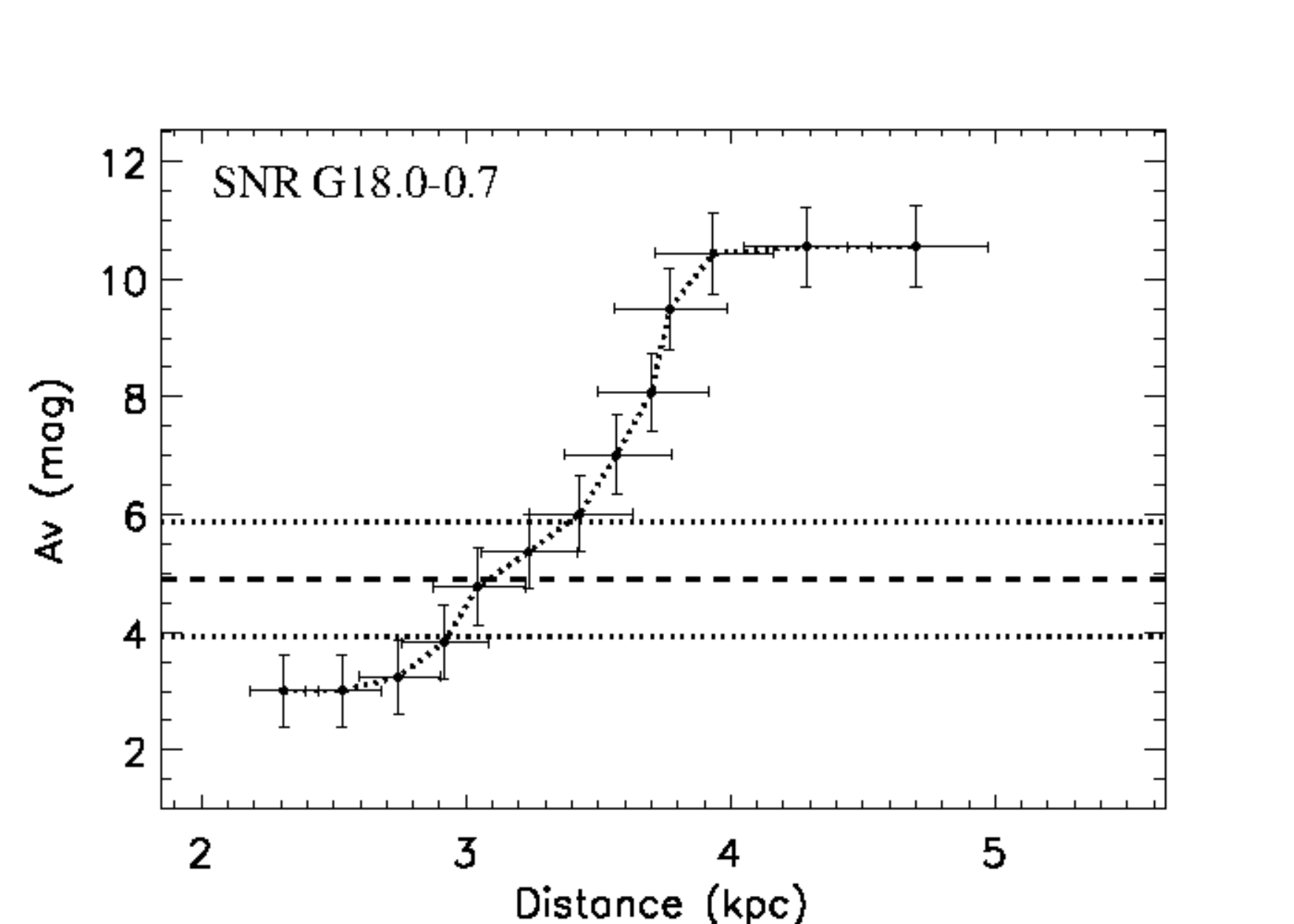}&
\includegraphics[width = 0.45\textwidth]{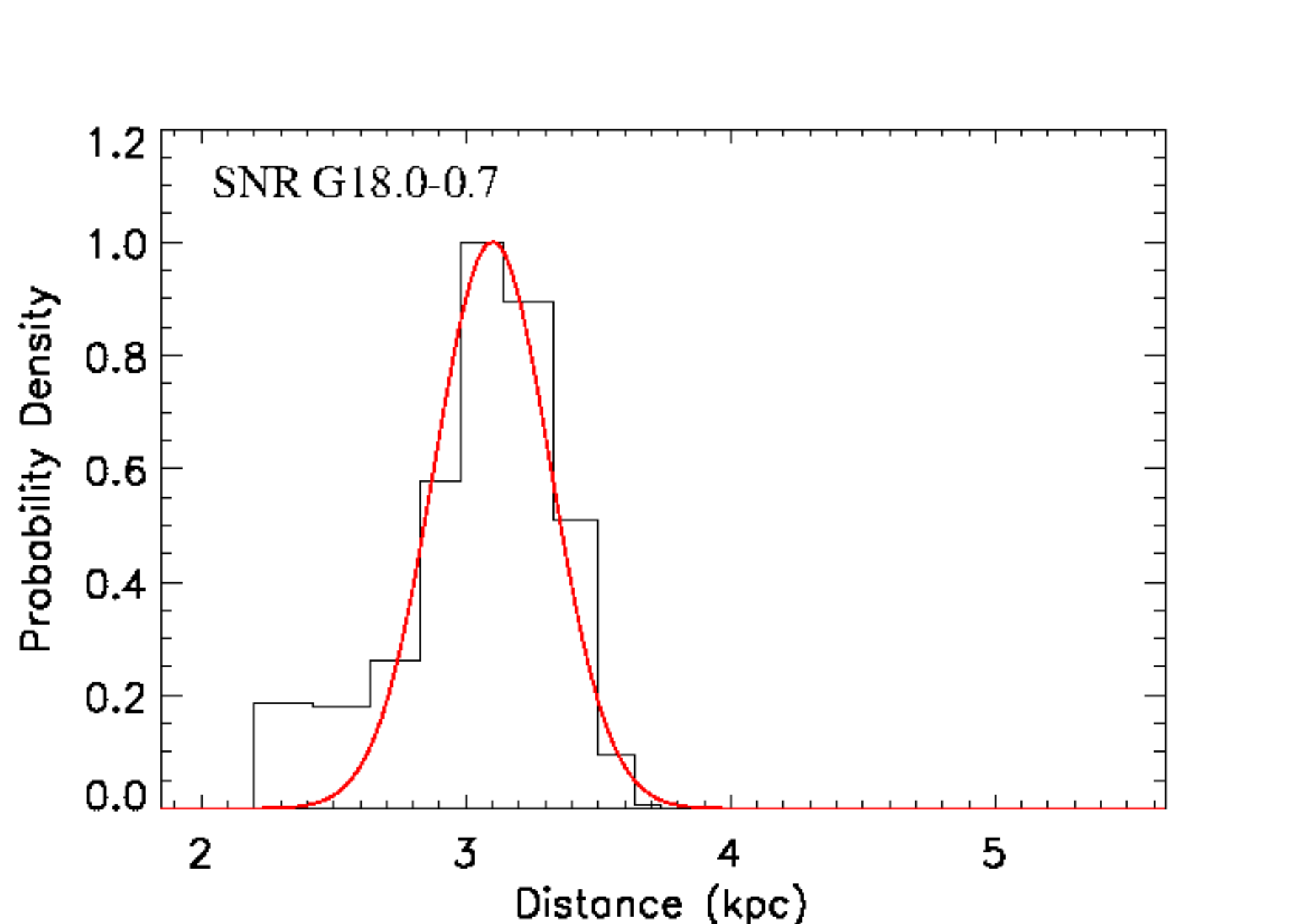}\\
\includegraphics[width = 0.45\textwidth]{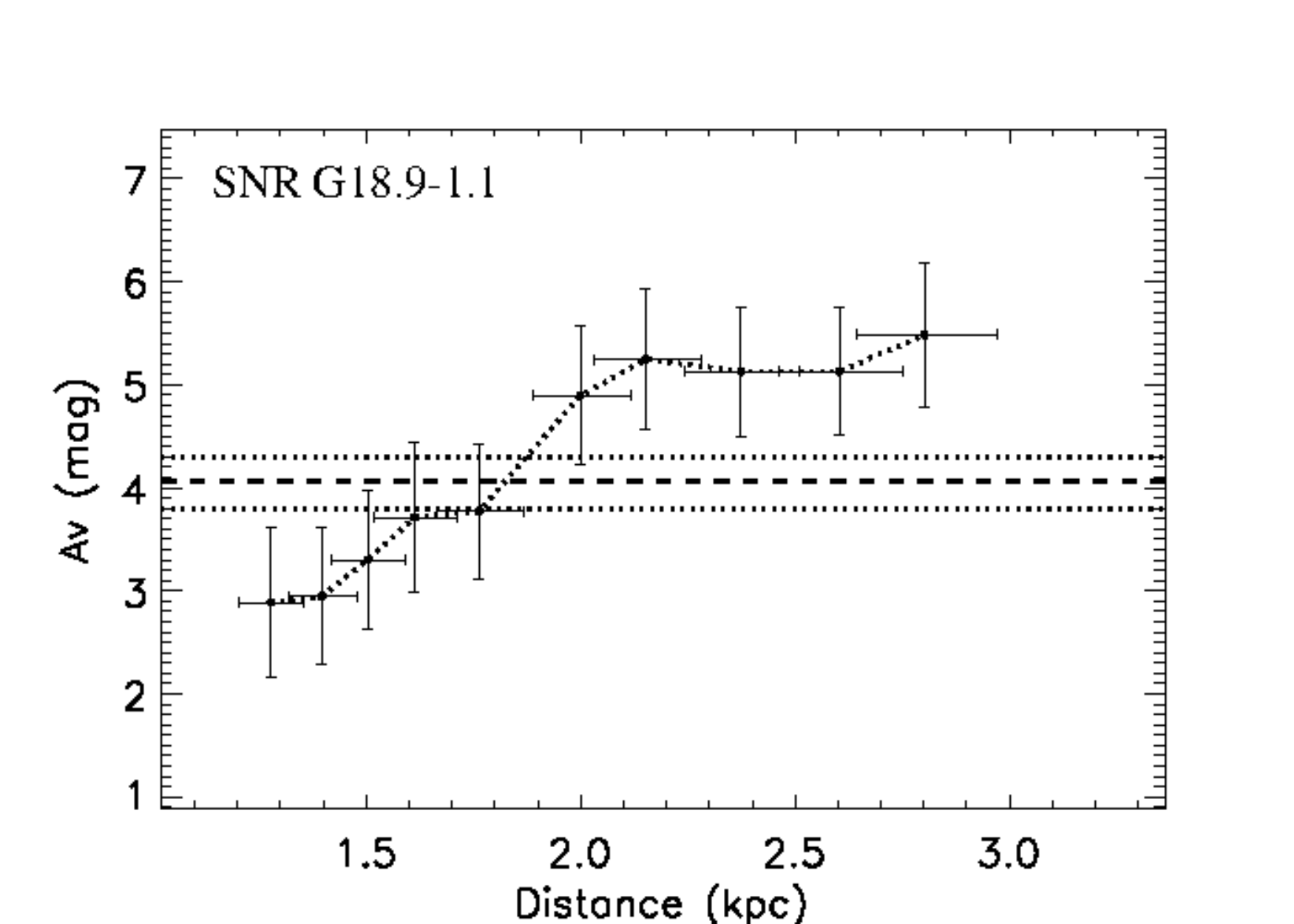}&
\includegraphics[width = 0.45\textwidth]{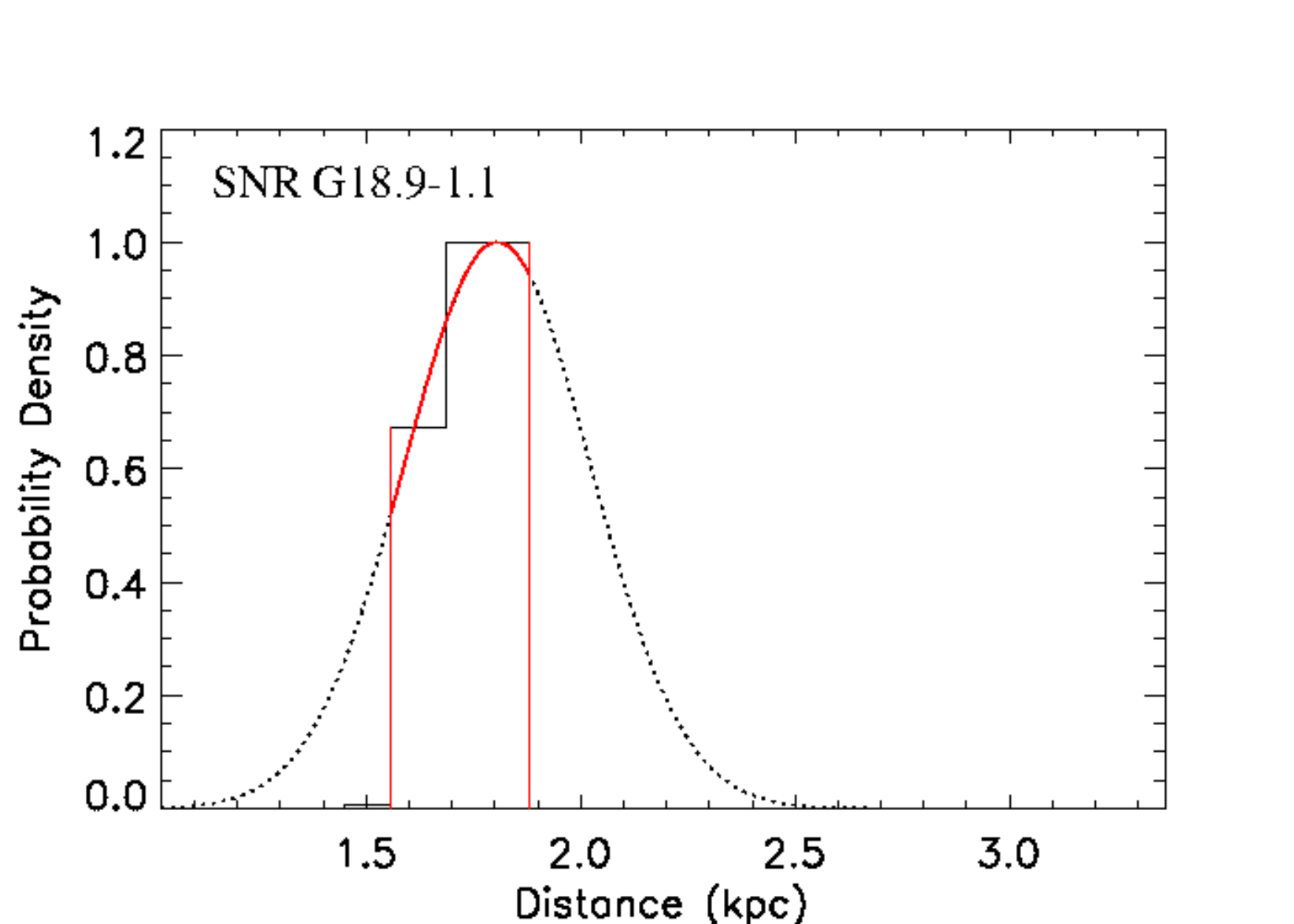}\\
\end{tabular}
\end{figure*}
\begin{figure*}
\begin{tabular}{cc}
\includegraphics[width = 0.45\textwidth]{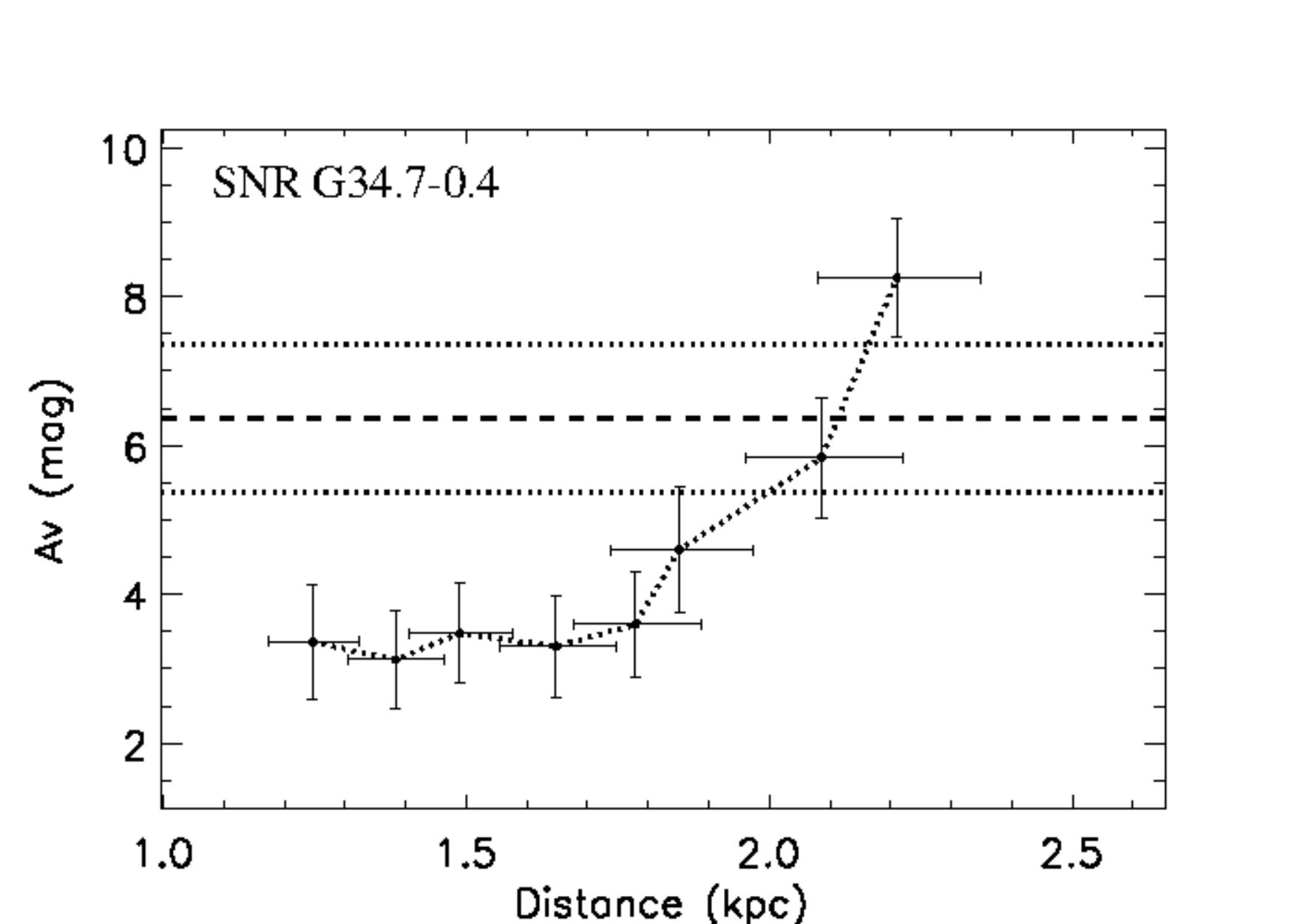}&
\includegraphics[width = 0.45\textwidth]{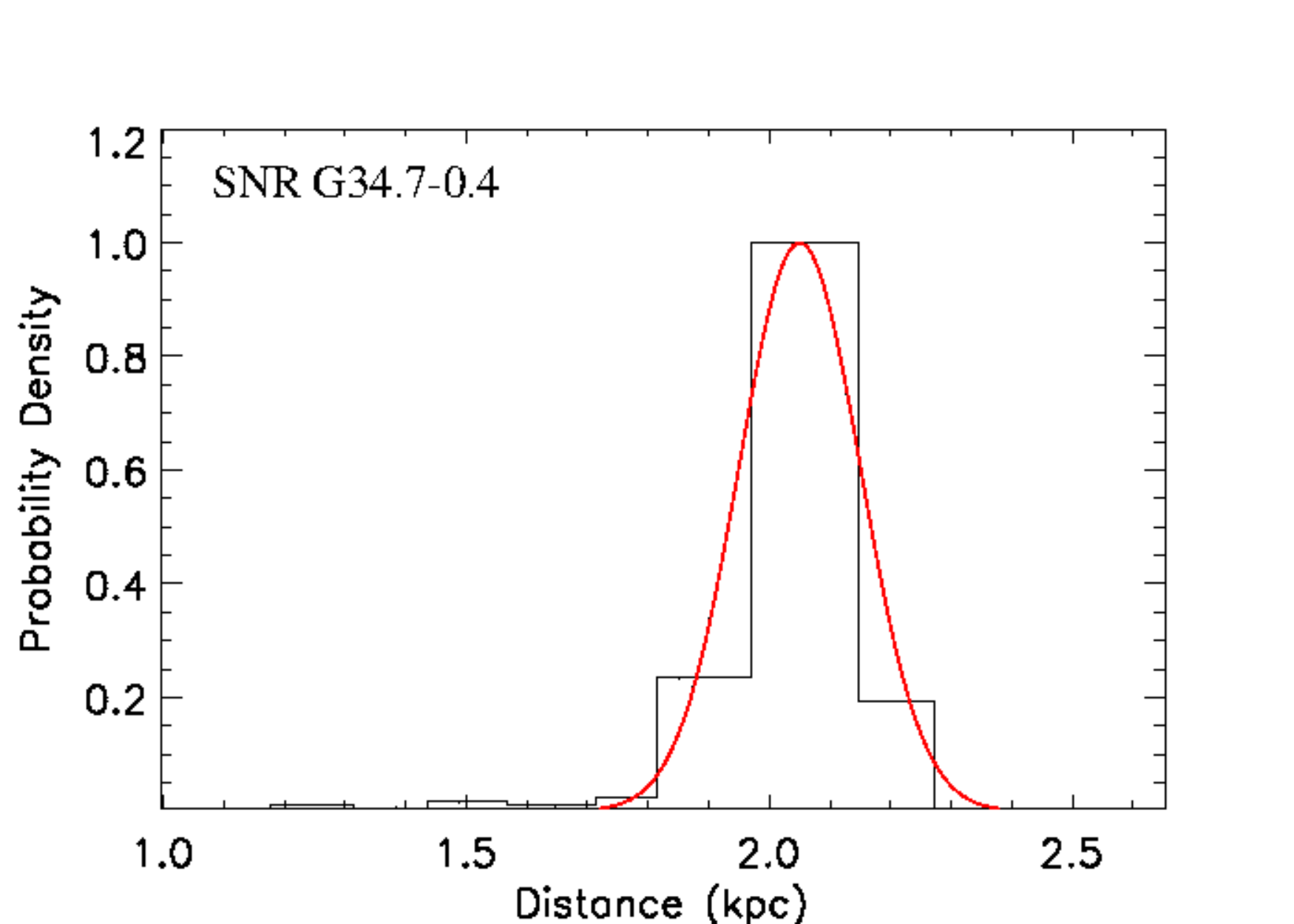}\\
\includegraphics[width = 0.45\textwidth]{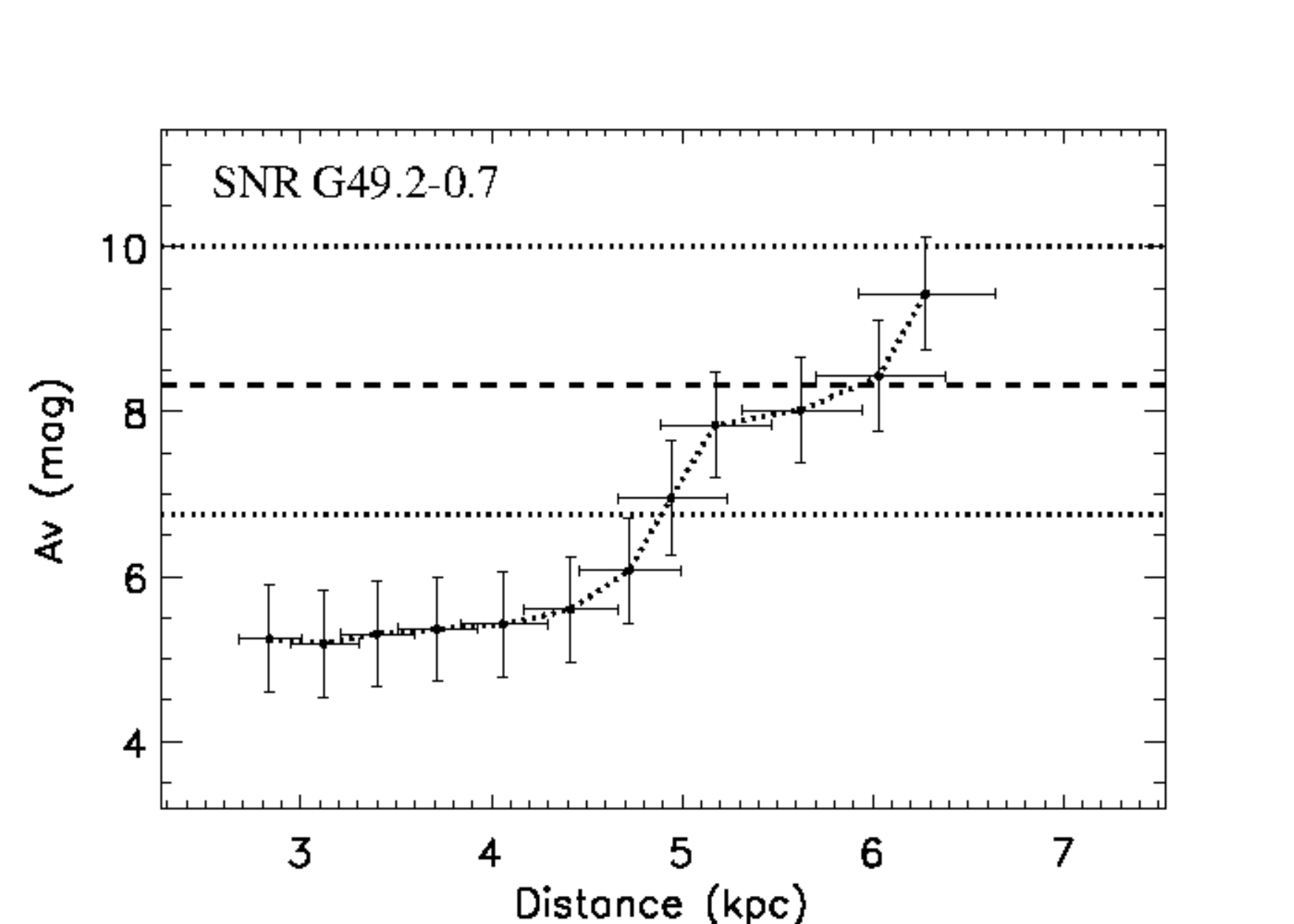}&
\includegraphics[width = 0.45\textwidth]{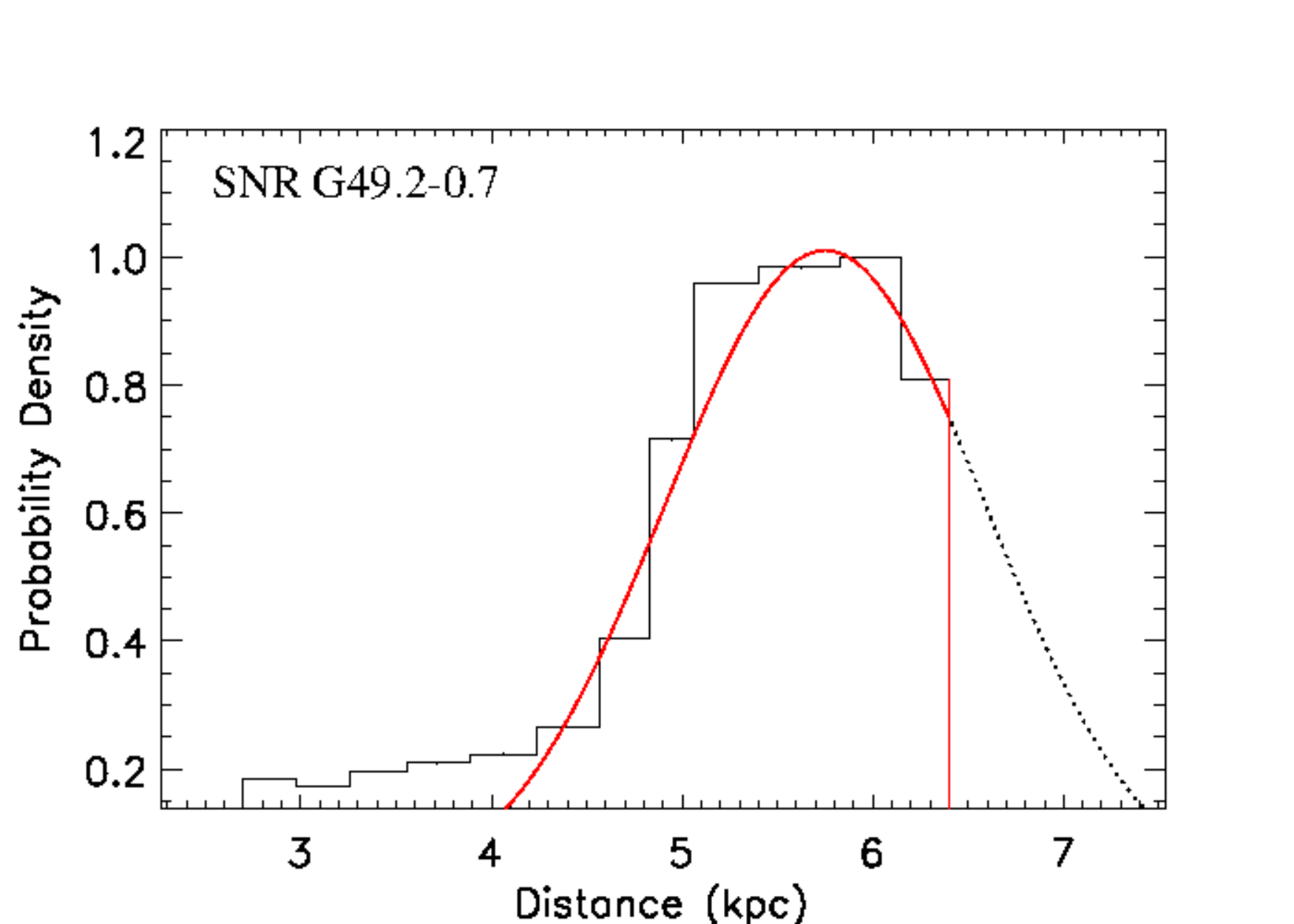}\\
\includegraphics[width = 0.45\textwidth]{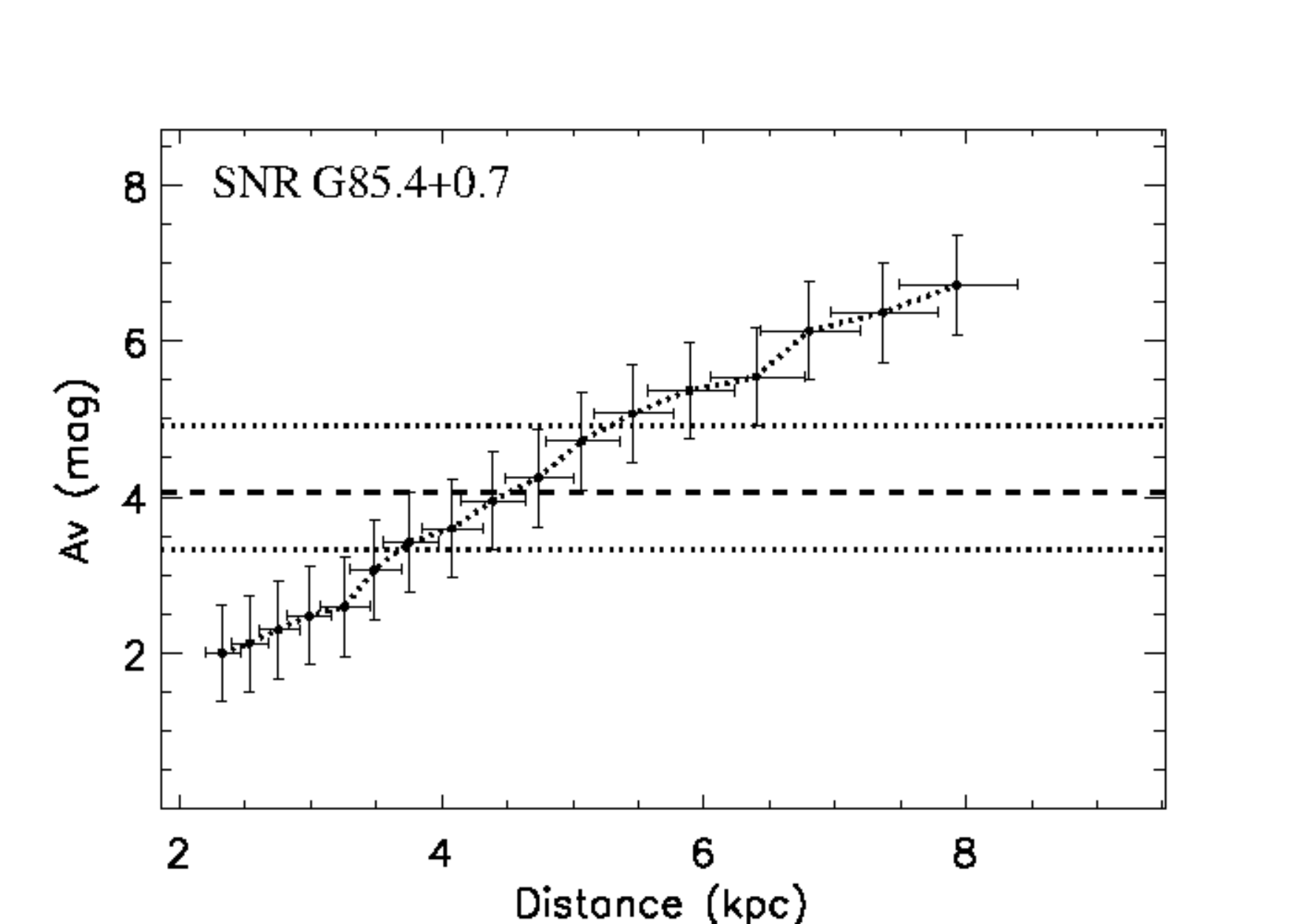}&
\includegraphics[width = 0.45\textwidth]{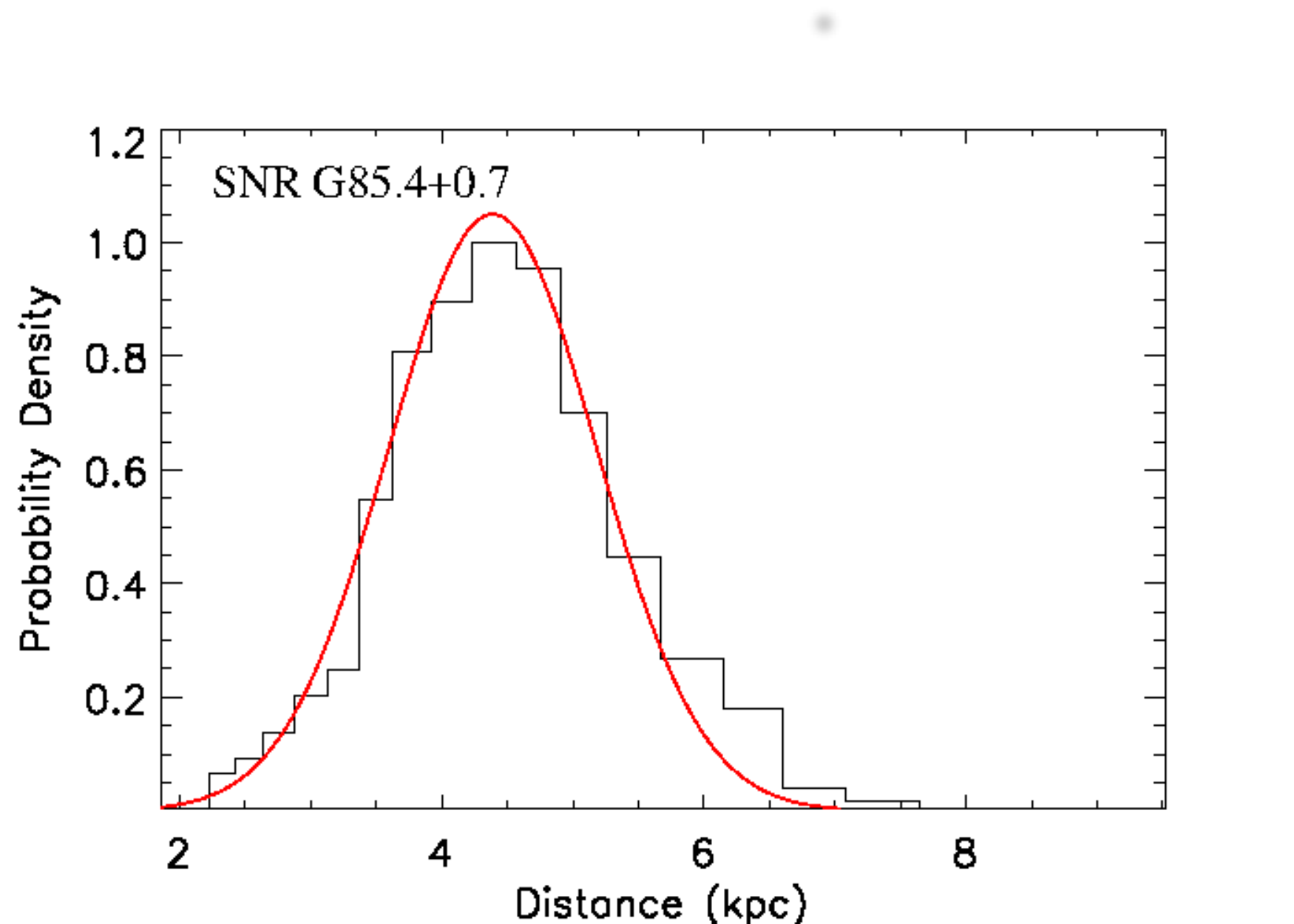}\\
\end{tabular}
\end{figure*}

We measure the run of reddening along distance using the RC method in each line of sight of 47 SNRs in the first Galactic quadrant. Among them, 32 SNRs' extinctions  are beyond the range of $\rm A_V$ traced by the RC method, hence the upper/lower limits of distances are obtained. Fortunately, there are 15 SNRs' extinction bands  overlapping
with the extinction measured by the RC stars, which provides an opportunity  to estimate the distance  accurately and with precision. Figure~\ref{fig5} presents the CMDs with the locations of RC's peak density for each of the 15 SNRs. Figure~\ref{fig6} shows the corresponding A$_{\rm V}$-D relations and the probability distribution over distance to the SNRs.
\subsection{Derive the Distances of SNR}
  To determine the distances of SNRs,  we calculate the probability distribution of distance using the product of  two probability distributions and marginalizing over the extinction \citep{{Tolga2010}}:
\begin{eqnarray}
\label{pp}
P(D)= \int P_{SNR}(A_{K})P_{RC}(D|A_{V})dA_{V}.
\end{eqnarray}

where $\rm P_{SNR}(A_{V})$ presents the probability distribution of an SNR's extinction.  We assume $\rm P_{RC}(D|A_{V})=P_{RC}(A_{V}|D)$.  $\rm P_{RC}(A_{V}|D)$ presents the distribution of the extinction traced by RC  at each distance bin. Both distributions are denoted as Gaussian functions. 

In  Figure~\ref{fig6} (right column), the panels show the probability distributions over distance calculated by equation \ref{pp}. Then, we fit these distributions with a Gaussian function, yielding the distance with the highest probability. For the objects with good Gaussian fitting, the uncertainty of the distance is equal to the standard deviation of the Gaussian. However, for some objects there are apparent and sudden decreases in the distance probability. The red lines mark such decreases (see Figure~\ref{fig6}). In this case, the uncertainty of distance reflects the cut-off distance. The results are listed in Tables~\ref{tab1} and~\ref{tab2}.

\subsection{Summarize the results }
  We obtain 15 new distances, three of which are given for the first time. G65.8-0.5\footnote{\citet{Anderson2017} suggested
 that G65.8-0.5 is likely an HII region.}, G66.0+0.0 and G67.6+0.9 are identified as SNRs by \citet{Sabin2013}. We estimated their distances
 as 2.4 kpc, 2.3 kpc, 2.0 kpc, respectively. Note that G66.0+0.0 is not detected in the most sensitive  Galactic Plane surveys \citep{Anderson2017}. Maybe more observations are needed for its classification.

We have also given 20 lower distance limits and 12 upper distance limits. Among them,
the distances of 4 SNRs have been further constrained by combining the lower or upper limits inferred by the RC method and the previous results.

  The distance of G5.7-0.1 is ambiguous, at either 3.1 or 13.7 kpc,  as inferred by the OH maser velocity \citep{Hewitt2009}.  The RC method's distance is about 2.9 kpc. Therefore, we predict  its distance is around 3 kpc.

 \citet{Boumis2008} suggested a lower limit distance of 2.2 kpc for G15.1-1.6. The RC method gives an upper limit distance of  2.1 kpc. Hence, we  conclude that the distance of G15.1-1.6 is around 2.2 kpc.

  The distance to G28.8+1.5 is estimated to be less than 3.9 kpc by \citet{Schwentker1994}. We obtain its lower limit as 2.8 kpc, so we suggest a distance of G28.8+1.5: 3.4$\pm0.6$ kpc.

    The distance to G78.2+2.1 is around 1.7-2.6 kpc by HI absorption \citep{Schwentker1994}. Our result is less than 2 kpc. Therefore, the distance of G78.2+2.1 is 1.9$\pm0.2$ kpc.

\subsection{Discussion}
We have estimated distances for 15 SNRs and upper or lower limits for 32 SNRs using the RC method.

\begin{figure*}
\begin{tabular}{cc}
\includegraphics[width = 0.45\textwidth]{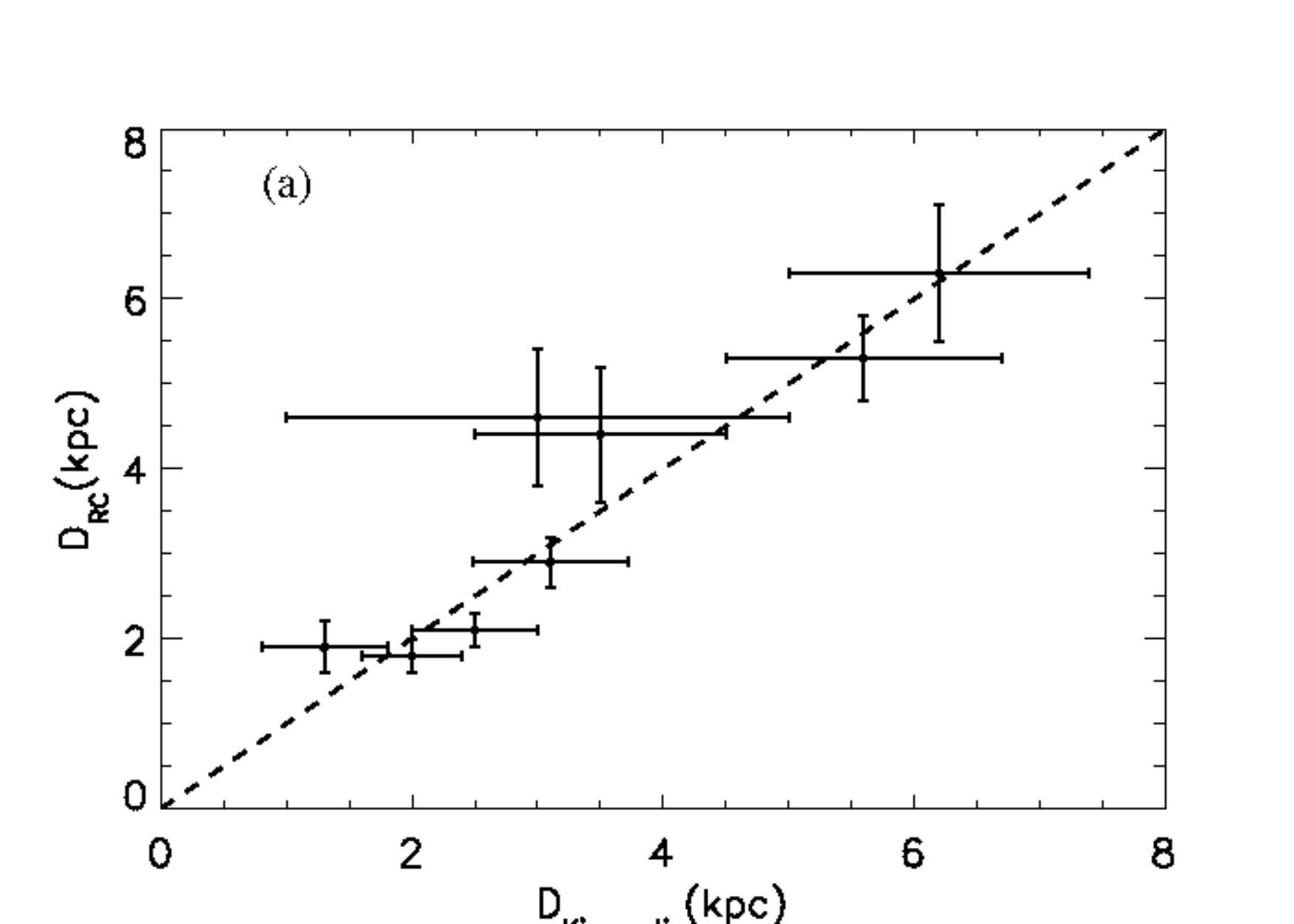}&
\includegraphics[width = 0.45\textwidth]{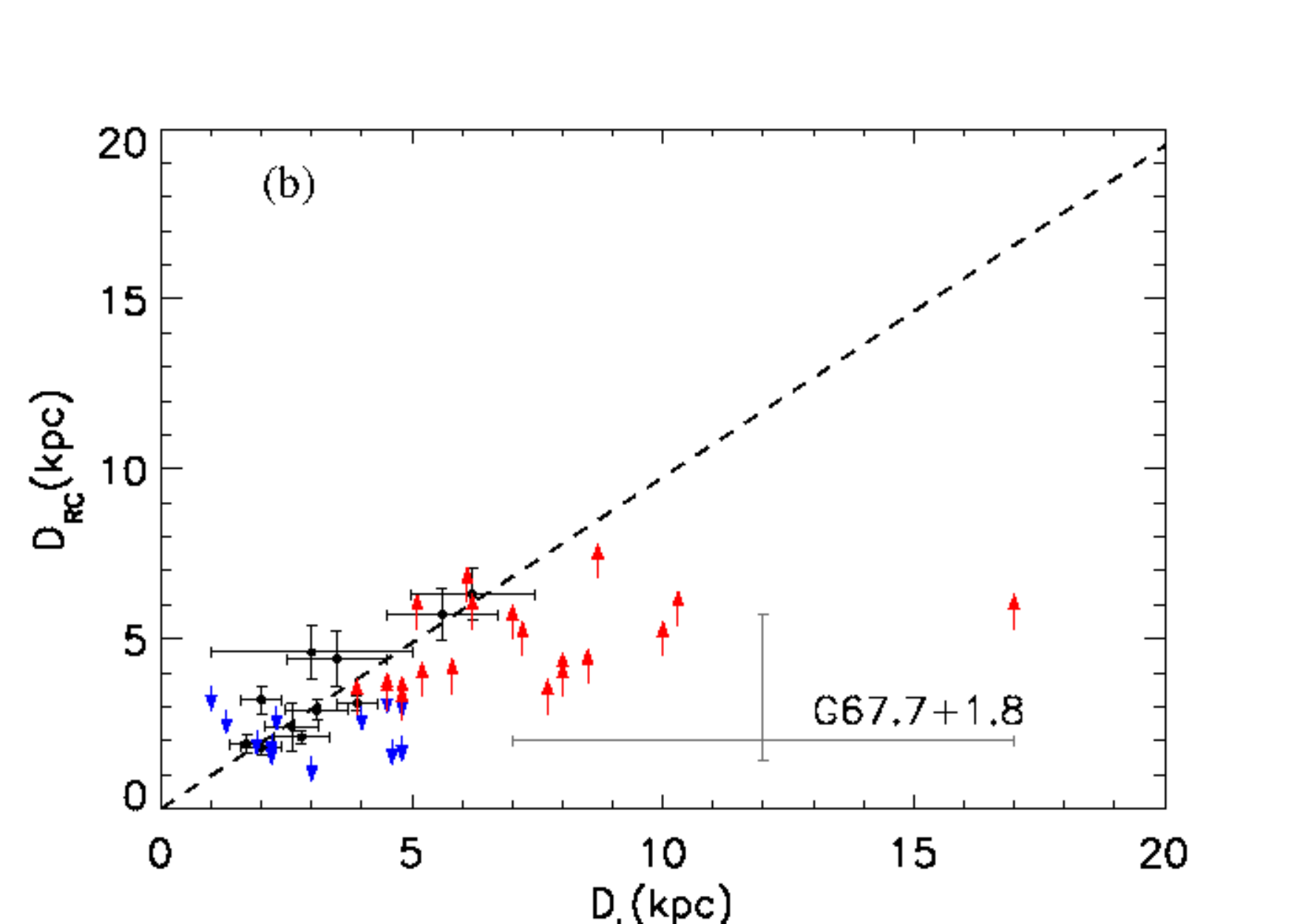}\\
\end{tabular}
\caption{(a) Correlation betweem the RC distances and kinematic distances of SNRs. (b) Comparison of the distances determined by the RC method with the distances determined by other measurements. The red arrows present lower limits and the blue arrows represent upper limits. The dashed line is fitted by black data points associated with uncertainties.}
\label{fig7}
\end{figure*}

 To further understand the precision of the distances indicated from the RC method,  we compare these results with the previous studies by two steps. 
Firstly, we compare 8 new SNRs'  RC distances with their kinematic distances.
The kinematic distances are denoted as $\rm D_{Kinematic}$. When the corresponding
 uncertainties are not given by the literature, we will empirically assume 20\% of their distances as  uncertainties.
 The distances measured by the RC method are denoted as $\rm D_{RC}$.
Assuming that each measurement with no errors for the same source, then regression function will be $\rm D_{RC}=D_{Kinematic}$. The MPFITXY routine based on the MPFIT package is used to fit a straight line via data with errors in both coordinates  \citep{Markwardt2009, Williams2010}. In Figure~\ref{fig7}(a), the fitted regression equation, $\rm D_{RC}=(1.0\pm0.1)\times D_{Kinematic}$,
means the RC distances of SNRs are highly consistent with their kinematic method within the range of uncertainty. Secondly,  we compare 44  distances constrained by the RC method with the corresponding distances measured by other methods. As  Figure~\ref{fig7} (b) shows, all distances estimated by the RC method are all in the range of 1.5-8 kpc, which is consistent with our expectations. All  20 lower limits of SNRs' distances coincide  with the trend of the fitting lines, while
 6 of the  12 upper limits are in agreement with other distance measurements. 
We conclude that  most of RC distances are in agreement with the previous measurements, and the lower limit distances are more reliable. Therefore, SNRs' distances can be independently constrained by the RC method. 

We analyse the reasons why the RC method only can trace either upper  or lower limits.
For 20 relatively distant SNRs, we only draw their lower limits for two reasons. One restricted condition is the 2MASS Survey completeness limit as J=15.8, $\rm K_S$=14.3 mag. The other is that  the  RC stars are likely mixed with the highly reddening main-sequence stars in CMDs when the apparent magnitudes begin to be fainter than 13 mag (see Figure~\ref{fig5}).
For 12 relatively closer SNRs, we only obtain their upper limits because the sample of RC stars is not enough for statistics when their apparent magnitudes are brighter than 9 mag in most cases. Hence, the RC method based on 2MASS data can be effective in the range from 1.5-8 kpc and the specific range of distance it can trace depends on the RC stars sample in a given direction.

We next check the seven discrepant measurements. First,  the differences between old and new distances for  G32.8-0.1, G39.7-2.0, G73.9+0.9 are less than 30\%, which may be caused  by the uncertainties  of the two different methods. Then, the key investigation is conducted on the other  four SNRs. The optical extinction values towards SNRs  G13.3-1.3,  G85.9-0.6 are smaller than 1 mag.  We expect that the  extinction values are not sensitive to the distance when extinction is extremely slight in the line of sight. For G32.1-0.9, the difference between RC and kinematic distance is greater than 50\% likely due to  the large error of $\rm N_H$ that is up to 40\%. Note the distance of G67.7+1.8 listed in Table~\ref{tab1} is not reliable since its probability distribution  over distance is not well fitted by a Gaussian function. It might be product of the broad range of the SNR's $\rm A_V$ and the slight extinction which is much lower than the average  magnitudes of extinction 
per kiloparsec ($\rm c_V \sim0.7 mag$ $\rm kpc^{-1}$ ) along the line
of sight \citep{Indebetouw2005}. These discrepant results suggest that the slight extinction or large uncertainties of extinction might significantly affect the accuracy of the RC method.

\section{SUMMARY}
We have taken advantage of the RC stars from 2MASS data to construct the A$_{\rm V}$-D relations along the directions of 47 SNRs in the first Galactic quadrant. In total,  15 distances and 32 upper or lower limit distances of SNRs have been obtained by overlapping their extinction values on the A$_{\rm V}$-D relations.  Among them, the distances of SNRs G65.8-0.5, G66.0-0.0 and G67.6+0.9
are estimated as 2.4 kpc, 2.3 kpc, and 2.0 kpc for the first time. Distances of SNRs G5.7-0.1, G15.1-1.6, G28.8+1.5 and G78.2+2.1 have been better constrained as about  3 kpc, 2.2 kpc, 3.4 kpc, and 1.9 kpc, respectively.

 By comparison, distances estimated by the RC method are consistent with other measurements within the range of the allowed errors. In addition, the RC method tends to give a reliable lower limit distance. In addition, we analyze the possible reasons why six upper limits are incompatible with the previous results. Finally, we highlight the RC method 
 can independently constrain distances of SNRs in the range of 1.5 kpc to 8 kpc and the distances  can be well determined by this method when the  samples of RC stars are relatively abundant  along the line of sight.

\acknowledgments
We all acknowledge supports from NSFC program (11473038, 11603039, U1831128). We appreciate it that the referee provided constructive comments and useful suggestions. We also thank Dr. Liu Chao, Dr. Xiang Mao-Sheng, Dr. Chen Bing-Qiu and Dr. Wan Jun-Chen for helpful discussion.
This publication makes use of data products from the Two Micron All Sky Survey and \software{MPFIT \citep{Markwardt2009}}

\bibliographystyle{aasjournal}
\bibliography{ref}

\appendix
\section{Supplementary data}

\begin{table*}
\setlength{\tabcolsep}{3mm}
\begin{center}
\caption{ Optical extinction $\rm A_V$ and limits of distances}
\label{tab3}
\begin{tabular}{llllllll}
  \hline
  \hline
  \mc{1}{l}{Source}&  \mc{1}{c}{$\rm A_{V}$}& \mc{1}{l}{Method}& \mc{1}{l}{$\rm D_{known}$}&\mc{1}{l}{Method}& \mc{1}{l}{$\rm D_{this    paper}$}&\mc{1}{l}{Ref.}\\
  \mc{1}{l}{Name }&       \mc{1}{l}{(mag)}&   \mc{1}{l}{}& \mc{1}{l}{kpc}& \mc{1}{l}{}&  \mc{1}{l}{kpc}&   \mc{1}{l}{}\\                   
 \hline
 G6.4-0.1 & 3.3$\pm$0.4& H$_{\alpha}$/H$_{\beta}$& 1.9$\pm$0.3& kinematic measurement&  $\leq2.3$&1, 2 \\ 
 G11.2-0.3& 13.0$\pm$1.9& Fe\scriptsize\uppercase\expandafter{\romannumeral2}  \normalsize  ratio&7.2, 5 &kinematic measurement, pulsar distance& $>4.5$&3, 4, 15\\
  G13.3-1.3& 0.5$\pm0.1$&H$_{\alpha}$/H$_{\beta}$& 2-4&CO absorbtion& $<1.5$&6\\
 G15.1-1.6&3.1$\pm$0.6&H$_{\alpha}$/H$_{\beta}$&$>2.2$& Blast wave energy& $<2.1$&7\\
G39.2-0.3&19$\pm$2.3&Fe\scriptsize \uppercase\expandafter{\romannumeral2}  \normalsize  ratio&6.2&kinematic measurement&$>5.3$&8, 9\\
 G39.7-2.0 & 2.3$\pm$0.3&H$_{\alpha}$/H$_{\beta}$&4.5$\pm$0.2, 5.5-6.5 &proper motion, kinematic measurement& $<3.5$&10, 11, 12\\
G53.6-2.2 & 3.4$\pm$0.5&S\scriptsize \uppercase\expandafter{\romannumeral2} \normalsize ratio& 3.8-6.3, 2.3$\pm0.8$&$\Sigma$-D, kinematic measurement& $>5.3$&13, 14\\
 G59.5+0.1& 3.1$\pm$0.7& H$_{\alpha}$/H$_{\beta}$&11, 2.3&$\Sigma$-D, kinematic measurement&$<3.0$&15, 16, 17\\
 G73.9+0.9&1.7$\pm$0.4&H$_{\alpha}$/H$_{\beta}$&4,4.3-4.5&$\Sigma$-D, kinematic measurement&$<3.0$&18, 19\\
 G78.2+2.1& 3.4$\pm$0.6&H$_{\alpha}$/H$_{\beta}$&1.7-2.6 &kinematic measurement&$<2.0$&1, 20\\
 G85.9-0.6&0.7$\pm$0.1&H$_{\alpha}$/H$_{\beta}$&4.8$\pm$1.6&kinematic measurement&$<2.1$&21, 22\\
\hline
\end{tabular}
\end{center}
\begin{flushleft}
Reference: 
(1) \citet{Zhu2017}; 
(2) \citet{Velazquez2002}; 
(3) \citet{Koo2007}; 
(4) \citet{Kilpatrick2016};
(5) \citet{Green1988};
(6) \citet{Seward1995}; 
(7) \citet{Boumis2008}; 
(8) \citet{Lee2009}; 
(9) \citet{Su2011}; 
(10) \citet{Boumis2007}; 
(11) \citet{Marshall2013};
(12) \citet{Lockman2007} 
(13) \citet{Long1991};
(14) \citet{Giacani1998};
(15) \citet{Sezer2008};
(16) \citet{Xu2012};
(17) \citet{Guseinov2003};
(18) \citet{Mavromatakis2003};
(19) \citet{Zdziarski2016};
(20) \citet{Leahy2013};
(21) \citet{Gok2009};
(22) \citet{Jackson2008}.
\end{flushleft}
\end{table*}

\begin{table*}
\setlength{\tabcolsep}{3mm}
\begin{center}
\caption {Hydrogen column density $\rm N_H$ and limits of distances}
\label{tab4}
\begin{tabular}{llllllll}
  \hline
  \hline
  \mc{1}{l}{Sourse}&  \mc{1}{l}{$\rm N_{H}$}&  \mc{1}{l}{Model$^1$}&   \mc{1}{l}{$\rm D_{known}$}& \mc{1}{l}{Method}&
\mc{1}{l}{$\rm D_{this paper}$}&\mc{1}{l}{Ref.}\\
  \mc{1}{l}{Name }&       \mc{1}{l}{(10$^{21}Hcm^{-2}$)}& \mc{1}{l}{}&   \mc{1}{l}{kpc}&   \mc{1}{l}{}&\mc{1}{l}{kpc}&\mc{1}{l}{}&\\
  \hline
 G1.0-0.1& 75.0$\pm$15.0& TP&8.0&proper motion& $>3.3$&1, 2\\
 G5.4-1.2&$35.0_{-10.0}^{+7.6}$&TP&5.2$\pm0.5$&pulsar distance&$>3.3$&3, 4\\
 G8.7-0.1&12.0&TP&4.5, 4.4&kinematic measurement, pulsar distance &$\ge 2.9$&3, 5\\
 G12.8-0.0&100.0$\pm$20.0&PL&4.8&kinematic measurement&$>2.6$&6, 7\\
 G15.9+0.2&39.0$\pm$2.0&TP&8.5&kinematic measurement&$>3.7$&8, 7\\
G20.0-0.2&41.0$_{-13.0}^{+24.0}$&PL&4.5&kinematic measurement&$>3.0$&9, 10\\
 G21.5-0.9&22.4$\pm$0.3&PL&4.8&kinematic measurement&$>2.9$&11, 12\\
G26.6-0.1 &4.9$\pm$1.7&TP&1.3&absorption column&$<2.9$&13\\
 G27.4+0.0&26.0$_{-3.0}^{+4.0}$&TT&8.7$\pm1.2$&kinematic measurement&$>6.8$&14, 15\\
 G28.6-0.1&37.0&TP&7.0&absorption column&$>5.0$&13\\
 G28.8+1.5&20.0&PL&$<3.9$&Sedov estimates&$>2.8$&16, 17\\
 G29.7-0.3&29.0&TP&6.3$\pm$1.2,5.8$^{+0.5}_{-0.4}$,10.6&kinematic measurement&$\geq3.4$&15, 18-20\\
G32.1-0.9&2.3$_{-0.8}^{+1.1}$&TP&4.6&Sedov estimates&$<2.0$&21\\
 G32.4+0.1&52.0$\pm$13.0&PL&17.0&absorption column&$>5.3$&22\\
 G32.8-0.1&8.1$\pm$0.7&TP&4.8&kinematic
  measurement&$<3.4$&18,23\\
G41.1-0.3&31.0$_{-3.0}^{+2.0}$&TP&10.3&kinematic measurement&$>5.4$&24, 25\\
 G42.8+0.6&23.0$\pm$10&PL+BB&7.7&PSR distance&$>2.8$&26, 27\\
 G43.3-0.2&51.8$\pm$0.5&TP&10.0&kinematic measurement&$>4.5$&28-30\\
 G65.7+1.2&2.6$_{-0.4}^{+0.5}$&BB&1$\pm0.4$&kinematic measurement&$<3.6$&31, 32\\
 G74.9+1.2&13.8$\pm$1.7&PL&6.1$\pm$0.9&extinction measurement&$>6.1$&33, 34\\
 G76.9+1.0&17.0$\pm$3.0&PL&8.0, 10.0&pulsar distance&$\geq3.6$&35, 36\\
 \hline
\end{tabular}
\end{center}
\begin{flushleft}
$^1$Model abbreviations: TP: thermal plasma, PL: power law, BB: black body, TT:a two-component thermal model.
Reference:
(1)\citet{Nobukawa2009};
(2) \citet{Reid1993};
(3) \citet{Hewitt2009}; 
(4) \citet{Kaspi2001};
(5)\citet{Verbiest2012};
(6) \citet{Funk2007};
(7) \citet{Kilpatrick2016};
(8) \citet{Reynolds2006};
(9) \citet{Petriella2013};
(10) \citet{Petriella2013};
(11) \citet{Safi-Harb2001};
(12) \citet{Tian2008};
(13) \citet{Bamba2003}; 
(14) \citet{Kumar2014};
(15) \citet{Tian2008b};
(16) \citet{Misanovic2010};
(17) \citet{Schwentker1994};
(18) \citet{Zhu2017};
(19) \citet{Verbiest2012};
(20) \citet{Su2009};
(21) \citet{Folgheraiter1997};
(22) \citet{Yamaguchi2004};
(23) \citet{Zhou2011};
(24) \citet{Safi-Harb2000};
(25) \citet{Jiang2010};
(26) \citet{Fox2001};
(27) \citet{Lorimer2000};
(28) \citet{Keohane2007};
(29) \citet{Zhu2014};
(30) \citet{Brogan2001};
(31) \citet{Karpova2015};
(32) \citet{Kothes2008};
(33) \citet{Matheson2013};
(34) \citet{Kothes2003}
(35) \citet{Arzoumanian2011}; 
(36) \citet{Kargaltsev2010}.
\end{flushleft}
\end{table*}

\label{lastpage}
\end{document}